\documentclass[floatfix]{emulateapj}

\usepackage{amsmath}
\usepackage[colorlinks = true,linkcolor = blue, citecolor = blue]{hyperref}
\usepackage{natbib}
\usepackage{multirow}
\usepackage{placeins}

\def\Vhrulefill{\leavevmode\leaders\hrule height 0.7ex depth \dimexpr0.4pt-0.7ex\hfill\kern0pt}

\shorttitle{Modelling the Molecular Gas in NGC\,6240}

\begin{document}
\title{Modelling the Molecular Gas in NGC\,6240}
\author{R. Tunnard\altaffilmark{1}, T.\,R. Greve\altaffilmark{1}, S. Garcia-Burillo\altaffilmark{2}, J. Graci\'{a} Carpio\altaffilmark{3}, A. Fuente\altaffilmark{2}, L. Tacconi\altaffilmark{3}, R. Neri\altaffilmark{4} and A. Usero\altaffilmark{2}.}
\altaffiltext{1}{Department of Physics and Astronomy, University College London, Gower Street, London WC1E 6BT, UK; \email{richard.tunnard.13@ucl.ac.uk}}
\altaffiltext{2}{Observatorio Astron\'{o}mico Nacional, Observatorio de Madrid, Alfonso XII, 3, 28014 Madrid, Spain}
\altaffiltext{3}{Max-Planck-Institute for Extraterrestrial Physics (MPE), Giessenbachstra\ss e 1, 85748 Garching, Germany}
\altaffiltext{4}{IRAM, 300 rue de la Piscine, Domaine Universitaire, 38406 Saint Martin d'H\`{e}res Cedex, France}

\begin{abstract}
\noindent We present the first observations of H$^{13}$CN$(1-0)$, H$^{13}$CO$^+(1-0)$ and SiO$(2-1)$ in NGC\,6240, obtained with the IRAM PdBI. Combining a Markov Chain Monte Carlo (MCMC) code with Large Velocity Gradient (LVG) modelling, and with additional data from the literature, we simultaneously fit three gas phases and six molecular species to constrain the physical condition of the molecular gas, including mass$-$luminosity conversion factors. We find $\sim10^{10}M_\odot$ of dense molecular gas in cold, dense clouds ($T_{\rm k}\sim10$\,K, $n_{{\rm H}_2}\sim10^6$\,cm$^{-3}$) with a volume filling factor $<0.002$, embedded in a shock heated molecular medium ($T_{\rm k}\sim2000$\,K, $n_{{\rm H}_2}\sim10^{3.6}$\,cm$^{-3}$), both surrounded by an extended diffuse phase ($T_{\rm k}\sim200$\,K, $n_{{\rm H}_2}\sim10^{2.5}$\,cm$^{-3}$). We derive a global $\alpha_{\rm CO}=1.5^{7.1}_{1.1}$ with gas masses $\log_{10}\left(M / [M_\odot]\right)=10.1_{10.0}^{10.8}$, dominated by the dense gas. We also find $\alpha_{\rm HCN} = 32^{89}_{13}$, which traces the cold, dense gas. The [$^{12}$C]/[$^{13}$C] ratio is only slightly elevated ($98^{230}_{65}$), contrary to the very high [CO]/[$^{13}$CO] ratio (300-500) reported in the literature. However, we find very high [HCN]/[H$^{13}$CN] and [HCO$^+$]/[H$^{13}$CO$^+$] abundance ratios $(300^{500}_{200})$ which we attribute to isotope fractionation in the cold, dense clouds.
\end{abstract}

\section{Introduction}

\noindent The term ``dense gas'' typically refers to molecular gas with $n_{{\rm H}_2} > 10^4\,{\rm cm^{-3}}$, which is the typical density of molecular cloud cores bound by their self-gravity \citep[e.g.,][]{Lada1992}. This phase offers significant shielding against photo-dissociation while being heated and partially ionised by cosmic rays, leading to chemistry distinct from elsewhere in the ISM. It is best traced by the rotational transitions of molecules such as HCN, HCO$^+$, and CS, which can reach high abundances relative to H$_2$ ($10^{-8}-10^{-6}$) under these conditions \citep{Nguyen1992,Solomon1992,Boger2005}. Due to both high densities and line trapping the low$-J$ transitions of these lines (in particular HCN) are almost exclusively optically thick, so that approaches to estimating gas mass developed for CO \citep{Young1982,Papadopoulos2012,Bolatto2013} become applicable, with appropriate calibrations, allowing us to trace specifically dense molecular gas.

Single-dish surveys of primarily the HCN, but also the HCO$^+$, $J=1-0$ line towards samples of local (ultra)luminous infrared galaxies ((U)LIRGs) suggest that a larger fraction of the molecular gas in these galaxies is in a dense state compared to that of normal star forming galaxies \citep{Solomon1992,Gao2004,GarciaBurillo2012}, where the HCN/CO $J=1-0$ luminosity ratio is adopted as a proxy of the dense gas mass fraction (typically $\sim 0.1$ in LIRGs/ULIRGs cf.\ $\sim 0.03$ in normal star forming galaxies, \citealp{Gao2004,Greve2014}). However, to date, only a handful of sources have been analysed using multi-line data-sets of dense gas tracers, in conjunction with radiative transfer analyses, to constrain the dense gas mass \citep{Krips2008,Greve2009,Papadopoulos2014}.

Gas mass conversion factors typically assume: {\bf 1)} something about the kinematical state of the gas; usually that it resides within virialized clouds (\citealp{Bolatto2013}, although see \citealp{Papadopoulos2012} for a formulation that allows for a generalised kinematic state): perhaps not unreasonable for dense gas tracers but certainly questionably for CO. {\bf 2)} that the lines are optically thick, which is generally a good assumption, at least for the $J=1-0$ lines. {\bf 3)} metallicity and/or molecular abundances. These are particularly uncertain, being strongly affected by the environment the emission is tracing, e.g., photon dominated regions (PDRs), X-ray dominated regions (XDRs), hot-cores or cold, dark clouds, and are usually a subject of interest in themselves. There are also possible contaminations from other sources, such as mid-IR pumping of HCN (and perhaps HNC and HCO$^+$), which increases the low$-J$ HCN line intensity \citep{Aalto2007,Aalto2012,Aalto2015}. As we move to higher redshifts the effect of the CMB becomes noticeably non-linear, with the greatest effects on the low$-J$ CO lines, and even in the local universe bright continuum backgrounds could be having significant effects on observed lines and line ratios \citep{Papadopoulos2000,Papadopoulos2010}. HCO$^+$ is subject to a complicated network of ion chemistry and photochemical reactions, making it particularly complicated molecular tracer \citep{Viti2002,Papadopoulos2007}.

Optically thin dense gas tracers, in particular isotopologues of the main species (e.g., H$^{13}$CN and H$^{13}$CO$^+$), can be especially powerful as they provide much tighter constraints on large velocity gradient (LVG) models when combined with their $^{12}$C isotopologues. Although usually $40-100\times$, and perhaps up to $\sim 500\times$, less abundant than their $^{12}$C isotopologues, optical depth effects mean that these lines are usually only $\sim 10-15\times$ fainter, and are detectable with the IRAM Plateau de Bure Interferometer (PdBI) and certainly with the Atacama Large Millimeter/submillimeter Array (ALMA) for the closest LIRGs and ULIRGs.

\subsection{Chemical Modelling and Isotopologue Ratios}\label{subsec:chem}

\noindent If it can be accurately measured, the elemental [$^{12}$C]/[$^{13}$C] abundance ratio can be a powerful diagnostic of the evolutionary state of a galaxy, with a high [$^{12}$C]/[$^{13}$C] being evidence for a young, unprocessed ISM \citep{Milam2005,Henkel2010}. Recently, it has been suggested by \citet{Papadopoulos2011} that a high [$^{12}$C]/[$^{13}$C] ratio might be evidence for a cosmic ray dominated star formation paradigm, leading to a top heavy stellar initial mass function and thereby an excess of massive, $^{12}$C producing, stars. Typical values of [$^{12}$C]/[$^{13}$C] in the local universe are $\sim25$ in the central Milky Way, 68 in the Milky Way as a whole and up to $\sim100$ in local starbursts and $\gtrsim100$ in high redshift ULIRGs \citep{Milam2005,Henkel2010,Henkel2014}. Typically, these ratios are measured using a combination of observations of CN, $^{13}$CN, CO, $^{13}$CO and C$^{18}$O lines. However, this measurement is non-trivial due to chemical effects leading to isotope fractionation under a variety of circumstances, as well as optical depth effects.

The theory of $^{12}$C/$^{13}$C isotope fractionation rests heavily, although not exclusively, upon isotope charge exchange (ICE) reactions, such as:

\begin{align}
\rm
^{13}C^+ + CO &\rightleftharpoons \rm {C^+} + {^{13}CO} + \rm 35\,K,\\
\rm ^{13}CO + HCO^+ &\rightleftharpoons \rm CO + H^{13}CO^+ +\rm 17\,K,\\
\rm ^{13}C^+ + CN &\rightleftharpoons \rm {C^+} + {^{13}CN} +\rm 31\,K,
\end{align}
\citep{Watson1976,Langer1978,Langer1984,Roueff2015}. Since these forward reactions are slightly exothermic, in cold molecular gas these reactions can lead isotope fractionation. However, there are large and complex chemical networks associated with and linking these and other molecules, necessitating sophisticated chemical modelling. There is also competition with selective photo-dissociation (SPD): for CO UV photo-dissociation is a line process \citep{vanDishoeck1988}, allowing for isotopologue specific self-shielding and thereby fractionating CO in the opposite direction to ICE. In contrast, CN photo-dissociation is a continuum process \citep{Lavendy1987}.

\citet{Langer1984} presented time dependent chemical models of molecular cloud chemistry for a range of densities ($5\times10^2 - 1\times10^5$\,cm$^{-3}$), temperatures ($6-80$\,K), metallicities and C/O abundances, with the aim of modelling $^{13}$C fractionation. They found large variations between species and with conditions, as well as multiple degeneracies. Three distinctly behaved carbon pools emerged: CO, HCO$^+$, and other carbon bearing species such as C$^+$, H$_2$CO, HCN, CS, CH etc., with each pool broadly following a given trend in fractionation. 

Their key findings were the confirmation that a small decrease in the [CO]/[$^{13}$CO] ratio from ICE can, due to the high relative abundance of CO, rapidly deplete the pool of available $^{13}$C, leading to very elevated ratios such as [HCN]/[H$^{13}$CN], which attained ratios as high as $6\times$ that of [CO]/[$^{13}$CO]. HCO$^+$ on the other hand exists in a complex equilibrium with both other pools and can be fractionated in either direction, although to a lesser extent. It is also very susceptible to variations in metallicity and the C/O ratio, but was never found to deviate from the [CO]/[$^{13}$CO] ratio by more than a factor of 1.5. Combining their results, they predicted a bracketing of the true [$^{12}$C]/[$^{13}$C] ratio:
\begin{equation}\label{eqn:chembracket}
[{\rm CO}]/[^{13}{\rm CO}] \leq [^{12}{\rm C}]/[^{13}{\rm C}] \leq [{\rm HCN}]/[\rm{H}^{13}{\rm CN}],
\end{equation}
while [HCO$^+$]/[H$^{13}$CO$^+$] is impossible to interpret in isolation, and extremely challenging even when combined with observations of multiple species and their isotopologue ratios.

\bigskip

\noindent Observational studies over the subsequent three decades have presented an even more complex picture. \citet{Milam2005} observed CN and $^{13}$CN along lines of sight to dense clouds in the Milky Way, finding no difference in the isotopologue ratios of CN, CO and H$_2$CO, instead finding that all three ratios increase from $\sim25$ in the Galactic Centre to $\sim130$ at a Galactic radius of 16.4\,kpc, ostensibly due to the ISM being more evolved (and hence enriched in $^{13}$C) near the Galactic centre.

On the other hand, \citet{Sheffer2007} used HST UV observations to measure [CO]/[$^{13}$CO] over 25 lines of sight, finding evidence of CO isotope fractionation in both positive and negative directions, as well as finding that the [CN]/[$^{13}$CN] ratio appeared to anti-correlate with [CO]/[$^{13}$CO], as predicted by \citet{Langer1984}. 

Similarly, \citet{Ritchey2011} found significant evidence for anti-correlated fractionation of CO and CN along lines of sight to diffuse molecular clouds, such as $\zeta$ Oph, which presents [CO]/[$^{13}$CO] = $167\pm15$ and [CN]/[$^{13}$CN] = $47.3^{+5.5}_{-4.4}$ \citep{Lambert1994,Crane1988}. \citet{Ritchey2011} identified CH$^+$ as a potential candidate to reliably constrain the true [$^{12}$C]/[$^{13}$C] ratio as it is produced only in non-thermal processes\footnote{CH$^+$ is formed via the reaction:
\begin{equation}
\rm C^+ + H_2 \rightarrow \rm CH^+ + H,
\end{equation} 
which is endothermic with $\Delta$E/$k_B=4640$\,K, so requires localised magnetohydrodynamic shocks to form in molecular clouds \citep{Elitzur1978,Elitzur1980,Ritchey2011}}, so is insusceptible to fractionation. Indeed, towards $\zeta$ Oph [CH$^+$]/[$^{13}$CH$^+$] = $67.5\pm4.5$, c.f.\ the ISM standard of 68 \citep{Crane1991,Milam2005}.

\bigskip

\noindent The most recent, state-of-the-art chemical modelling study of isotope fractionation was conducted by \citet{Roueff2015}, who aimed to refine the understanding of [$^{14}$N]/[$^{15}$N] fractionation by cementing the related [$^{12}$C]/[$^{13}$C] fractionation chemistry. Their detailed chemical model included some of the most up-to-date reactions and chemical networks and explored time dependent isotope fractionation in cold (10\,K), dense ($2\times10^4$ and $2\times10^5$\,cm$^{-3}$) molecular clouds, with an elemental [$^{12}$C]/[$^{13}$C] = 68. One of the significant changes is the addition of the reaction:
\begin{equation}
\rm HNC + C \rightarrow \rm HCN + C,
\end{equation}
\citep{Loison2014}, which couples HCN back into the chemical network post-production, allowing it to be fractionated at any time that the $^{13}$C pool is depleted. In addition, \citet{Roueff2015} use an updated $\Delta$E/$k_B$ for Equation 2 of 17\,K, c.f.\ 9\,K in \citet{Langer1984} \citep{Lohr1998,Maldenovic2014}.

The models of \citet{Roueff2015} have particularly interesting results for the isotopologue ratios of CN, HCN and HCO$^+$. Like \citet{Langer1984}, \citet{Roueff2015} found that in steady state [CO]/[$^{13}$CO] is only very slightly less than [$^{12}$C]/[$^{13}$C], at 67.4 and 68 in the lower and higher density models respectively. CN fractionates up in the first model, and down in the second. HCN on the other hand is elevated in all of their models, with steady state ratios consistently $2\sim3\times$ [CO]/[$^{13}$CO]. HCO$^+$ tends to equilibrate with [CO]/[$^{13}$CO] in all of the models. They also found that the H$_2$ ortho-para ratio has very significant effects on some species, such as CN, but not others, such as HCN.

\bigskip

\noindent The fundamental conclusion of the observational and theoretical studies is that isotope fractionation is possible, but that it is complex and hard to constrain due to the multitude of variables driving and opposing fractionation. Even state-of-the-art chemical models are limited in usefulness unless they have been run for the specific cloud conditions in question due to the strong sensitivity to temperature, density and metallicity. However, HCN appears to be reliably and strongly fractionated away from the elemental abundances, while HCO$^+$ can be fractionated in either direction and is very sensitive to the evolutionary stage of the molecular cloud. 

\bigskip


\noindent It is standard practice, when observing H$^{13}$CN, to assume a canonical [HCN]/[H$^{13}$CN] abundance ratio \citep[e.g.,][]{Pirogov1995, Nakajima2011, Tunnard2015}. In light of the chemical modelling discussed above this approach appears to risk being significantly flawed: since HCN preferentially traces colder, denser gas, it is especially sensitive to the effects of isotope fractionation. Also, there is a degeneracy between the kinetic temperature and the isotopologue ratio for HCN/H$^{13}$CN line ratios; assuming an incorrect [HCN]/[H$^{13}$CN] runs the risk of significantly biasing derived HCN column densities and estimates of the local kinetic temperature.

The chemical studies also suggest that any claims of elevated elemental [$^{12}$C]/[$^{13}$C] ratios, especially in extragalactic sources, must be interpreted with caution, and within not only the appropriate physical context, but also the appropriate chemical context \citep[e.g., as in][]{Henkel2014}. There is also the potential for strong optical depth effects \citep{Szucs2014}, although that particular challenge appears intractable with unresolved, single dish extragalactic observations.

\subsection{NGC\,6240}

\noindent NGC\,6240 is an early stage major merger galaxy in the constellation of Ophiuchus. Hosting two active supermassive black holes and possessing an infra-red luminosity $L_{8-1000\mu{\rm m}}=10^{11.73}$\,$L_\odot$ it is a luminous infrared galaxy (LIRG) and a LINER galaxy (low-ionization nuclear emission-line region) \citep{Komossa2003,Lutz2003,Sanders2003,Veron2006}. Studies by \citet{Tacconi1999} and \citet{Tecza2000} have shown that while star formation is concentrated in two nuclei, the molecular gas is predominantly in a turbulent, thick inter-nuclear disk. They also demonstrated that the extremely luminous infrared H$_2$ $S(0)$ and $S(1)$ emission ($L({\rm H}_2)\sim2\times10^9$\,$L_\odot$) is most likely due to slow C shocks between gas clouds in this dense, turbulent disk; a finding confirmed by the study of the CO spectral line energy distribution (SLED), complete up to and including $J=13-12$, by \citet{Meijerink2013}. 

As one of the closest LIRGs the molecular emission in NGC\,6240 has been extensively studied. The finding of \citet{Greve2009} that the CO SLED required at least a two component fit has been extended upon by \citet{Papadopoulos2014}, who fitted the CO SLED of CO from $J=1-0$ to $J=13-12$ with an ``inside out'' decomposition, where the results of HCN and HCO$^+$ $J=1-0$ to $J=4-3$ SLED fits are used to constrain the high$-J$ CO lines, and a third phase fitted to the remaining CO lines. Two of their key findings for NGC\,6240 were a very high [$^{12}$C]/[$^{13}$C] ratio of $300-500$ and that most of the H$_2$ gas mass is inconsistent with self-gravitating and photoelectrically heated gas, requiring either cosmic ray or turbulent heating. Indeed, \citet{Meijerink2013} model the CO SLED as being entirely due to shock heated gas. An independent two phase analysis was also conducted by \citet{Kamenetzky2014}, who used a nested sampling routine to model only the CO SLED, finding a hot, diffuse phase and a cold, dense phase.

\citet{Papadopoulos2014} argued that the high [$^{12}$C]/[$^{13}$C] (300-500) abundance ratio they found implies the existence of a very unusual environment in NGC\,6240. They suggest that it is either a sign of a prolonged period of star formation with a top heavy initial mass function (IMF) or an inflow of unprocessed gas from the outer edges of the galaxy. However, the massive, wide outflow in CO observed by \citet{Feruglio2013a} in NGC\,6240 suggests that an extensive inflow of gas would be difficult to maintain. That being said, \citet{GarciaBurillo2014} observed simultaneous inflows and outflows into/from the circumnuclear disk of NGC\,1068, so this possibility cannot be entirely ruled out.

As well as presenting a striking CO SLED, NGC\,6240 was shown to possess massive ($\sim120$\,$M_{\odot}$\,yr$^{-1}$) molecular outflows at $-600$, $+800$\,km\,s$^{-1}$ by \citet{Feruglio2013a} in CO$(1-0)$, while \citet{Wang2014} found that the extended (5\,kpc) hard X-ray spectrum ($\sim6$\,keV) is due to $\sim2200$\,km\,s$^{-1}$ outflows, which they ascribe to the nuclear starburst with a supernova rate of $\sim2$\,yr$^{-1}$.

This combination of extreme conditions make NGC\,6240 a unique laboratory for star formation physics under unusual conditions.

\bigskip

\noindent This paper is outlined as follows: we describe our observations and their analysis in Section \ref{sec:obs} and the line extraction in Section \ref{sec:lineExtract}. We introduce our Monte Carlo Markov Chain - Large Velocity Gradient (MCMC LVG) model and results in Section \ref{sec:lvg}. The results are discussed in Section \ref{sec:discuss} and we outline our conclusions in Section \ref{sec:conclusions}. We include further details of our model in the Appendix. We adopt cosmological parameters $h=0.70$, $\Omega_m=0.3$ and $\Omega_\Lambda=0.7$, giving a luminosity distance of 107\,Mpc and an angular scale of 0.494\,kpc/$''$ at $z=0.0245$ \citep{Wright2006,Papadopoulos2014}. To avoid confusion, we note that throughout this work, unless explicitly stated otherwise, we present results as $x^{a}_{b}$, where $a$ and $b$ denote the values at the upper and lower limits on the 68\% credible interval.

\section{Observations}\label{sec:obs}

\noindent We obtained observations of H$^{13}$CN$(1-0)$, H$^{13}$CO$^+(1-0)$ and SiO$(2-1)$ in NGC\,6240 with the Plateau de Bure Interferometer (PdBI) on 2008 February 21 and 2008 March 7 in the most extended configuration (PI: J.\ Graci\'{a}-Carpio), obtaining minimum and maximum baselines of 59.7\,m (16.8\,k$\lambda$) and 760\,m (215.8\,k$\lambda$) respectively, centred on the sky frequency of 84.6\,GHz. Using natural weighting to maximise sensitivity we obtained a synthesised beam $1.86''\times0.74''$.

The data were reduced in CLIC, part of the IRAM software package GILDAS\footnote{\url{http://www.iram.fr/IRAMFR/GILDAS/}.}, while for the UV analysis and imaging we used the GILDAS program, MAPPING.

\section{Spectrum Extraction and Data Analysis}\label{sec:lineExtract}

\noindent Due to the poor signal to noise of the UV visibilities we analysed the data independently in both the UV plane and in the final imaged datacube. These two methods are outlined below, and produced spectra consistent on a channel by channel basis, within the channel uncertainties.

\bigskip

\begin{figure*}
	\centering
	\includegraphics[width=\textwidth]{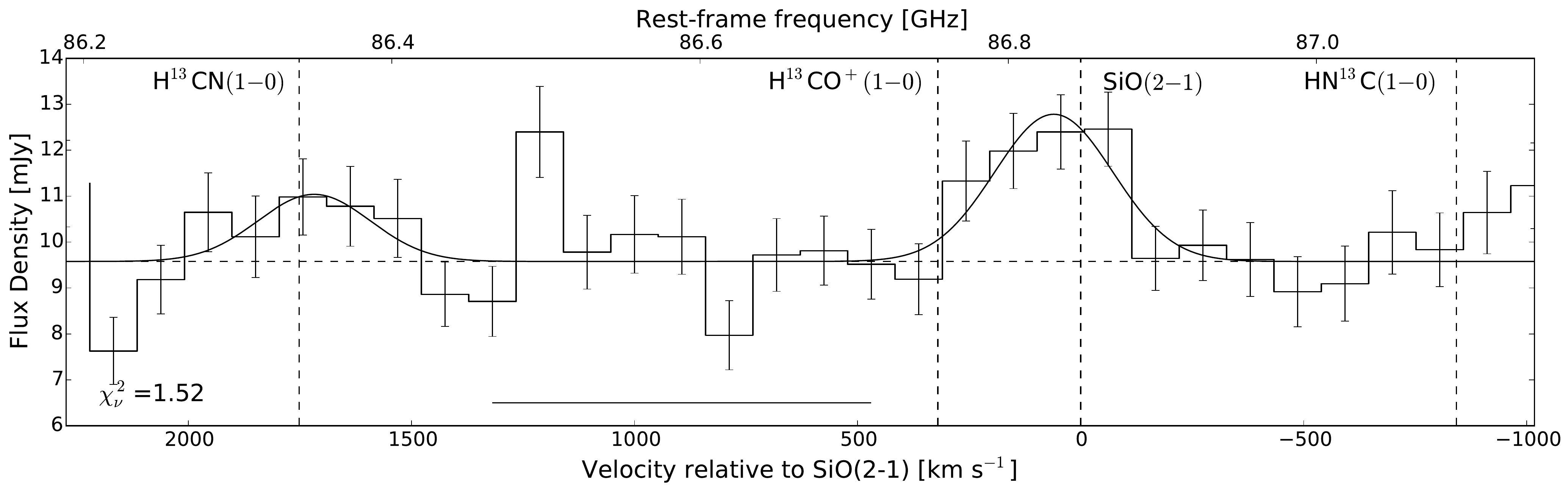}
	\caption{The UV fitted spectrum and the least squares two Gaussian plus baseline fit. Velocities are relative to the SiO$(2-1)$ line centre, and rest frame frequencies in GHz are given at the top of the plot. The $1\sigma$ channel uncertainties (median $\pm0.8$\,mJy) from the UV fitting are shown, and do not include the absolute flux calibration uncertainty. The horizontal bar indicates the region identified as line free and used for the continuum subtraction in the imaging step. The reduced $\chi^2$ of the fit is 1.52.}\label{fig:uvspec}
\end{figure*}

\noindent Elliptical Gaussian and disk models were fitted to the UV visibilities on a channel-by-channel basis, using the GILDAS task UV\_FIT and 30\,MHz channels. The fitted fluxes and fit rms's were identical for the two models and we adopted the elliptical Gaussian model. Fitting of channels narrower than 30\,MHz was attempted but produced pathological fits in multiple channels due to the poor S/N of the visibilities. The spectrum obtained is shown in Figure \ref{fig:uvspec}, where we fit two Gaussian profiles and a $0^{\rm th}$ order polynomial baseline\footnote{All reasonable fits with a 1$^{\rm st}$ order polynomial baseline had a gradient of almost zero. However, due to the randomised starting conditions we use to check for convergence, some solutions with a 1$^{\rm st}$ order polynomial were pathological, and we adopt instead the stable 0$^{\rm th}$ order.}. The SiO$(2-1)$ line is clearly detected and there is a possible detection of H$^{13}$CN$(1-0)$. The significance of this detection is quantified in Section \ref{subsec:speclines} below. Channel errors are shown at the $\pm1\sigma$ level and are unique for each channel, based on uncertainty in the fitted flux. The median channel error is $\pm0.8$\,mJy.

A concern with the UV fit to each channel is the possibility of fitting to a different spatial centre, and thereby unintentionally applying a spatial pseudo-averaging across the channels. The majority of the channels fitted a centre consistent within the uncertainties $(\sim\pm0.06'')$, while the full range is $0.3''$ in Dec and $0.1''$ in RA, dominated by four outlying channels. These offsets are $<25\%$ of the fitted FWHMs, and $<5\%$ of the aperture used for extracting the spectrum from the imaged spectrum (see below), and so this is not a significant effect here.

\bigskip

\noindent For imaging, we first continuum subtracted the data by extracting the visibilities from the 210\,MHz line free region from 86.47\,GHz to 86.67\,GHz (rest frame) to define the continuum, channel averaging the visibilities, and then subtracting the channel averaged visibilities in the UV plane. The visibilities were imaged in MAPPING and the spectrum extracted in CASA \citep{CASA} with an elliptical aperture $5.58''\times2.22''$, PA $=162^{\circ}$ (three times the clean beam), centred on the phase centre RA$=16^{\rm h}52^{\rm m}58^{\rm s}.890$, Dec$=+02^\circ24^\prime03^{\prime\prime}.90$. This includes conversions from mJy\,beam$^{-1}$ to mJy and allows for a slightly extended source.

\begin{figure*}
	\centering
	\includegraphics[width=\textwidth]{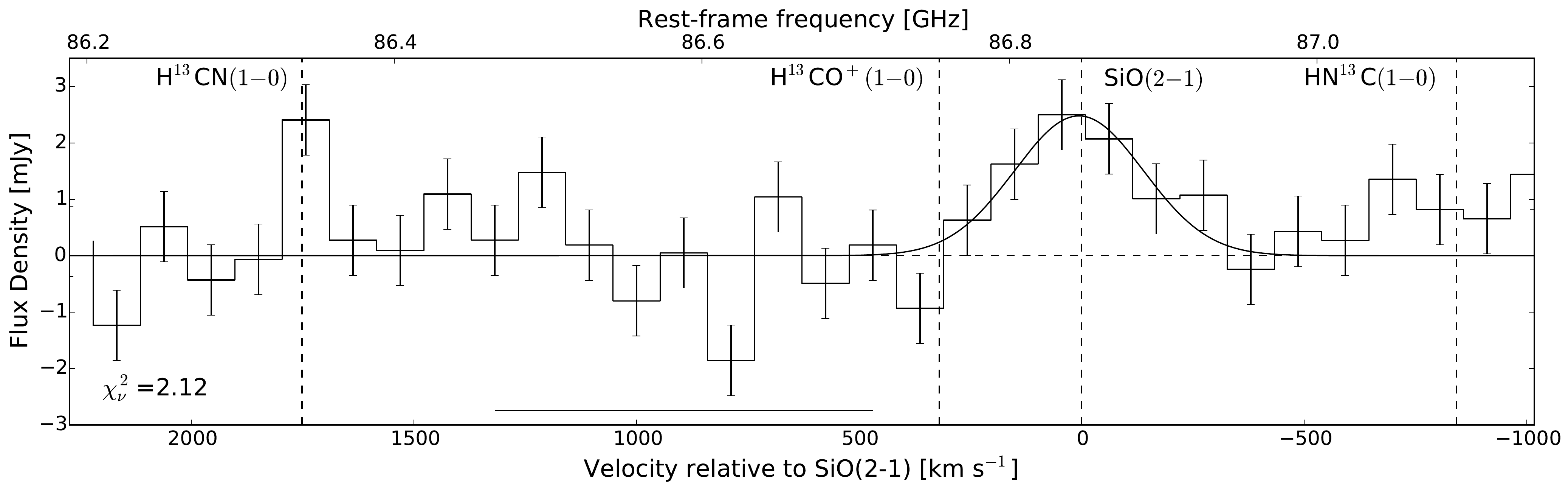}
	\caption{The imaged spectrum, extracted with an aperture as described in Section \ref{sec:lineExtract}, and the least squares Gaussian fit to the SiO$(2-1)$ line. Velocities are relative to the SiO$(2-1)$ line centre, and rest frame frequencies in GHz are given at the top of the plot. The $1\sigma$ rms channel uncertainties ($\pm0.6$\,mJy) are shown, and do not include the absolute flux calibration uncertainty. The horizontal bar indicates the region used for the continuum. The reduced $\chi^2$ of the fit is 2.12.}\label{fig:imspec}
\end{figure*}

The noise in the imaged cube was estimated with the GILDAS task NOISE to be 0.2\,mJy\,beam$^{-1}$. To estimate the noise in the spectrum we placed eight apertures, identical to that used for the spectrum extraction, arranged around the science aperture so as to create a regular $3\times3$ grid with the science spectrum in the centre. We then extracted the spectra from within these eight apertures. Since the cube is continuum subtracted, the signal in these apertures should be entirely noise (both thermal, and artificial from processing). For each spectrum the rms about zero was found, and then these eight rms's were averaged to produce a final rms for the imaged spectrum of 0.6\,mJy\,channel$^{-1}$. The spectrum is shown in Figure \ref{fig:imspec}. We fitted a single Gaussian to the SiO$(2-1)$ line and find a reduced $\chi^2$ of 2.12. Attempts to fit a second Gaussian component consistently lead to pathological results due to all of the putative H$^{13}$CN$(1-0)$ flux lying in a single channel.

\bigskip

\begin{figure*}
	\centering
	\includegraphics[width=\textwidth]{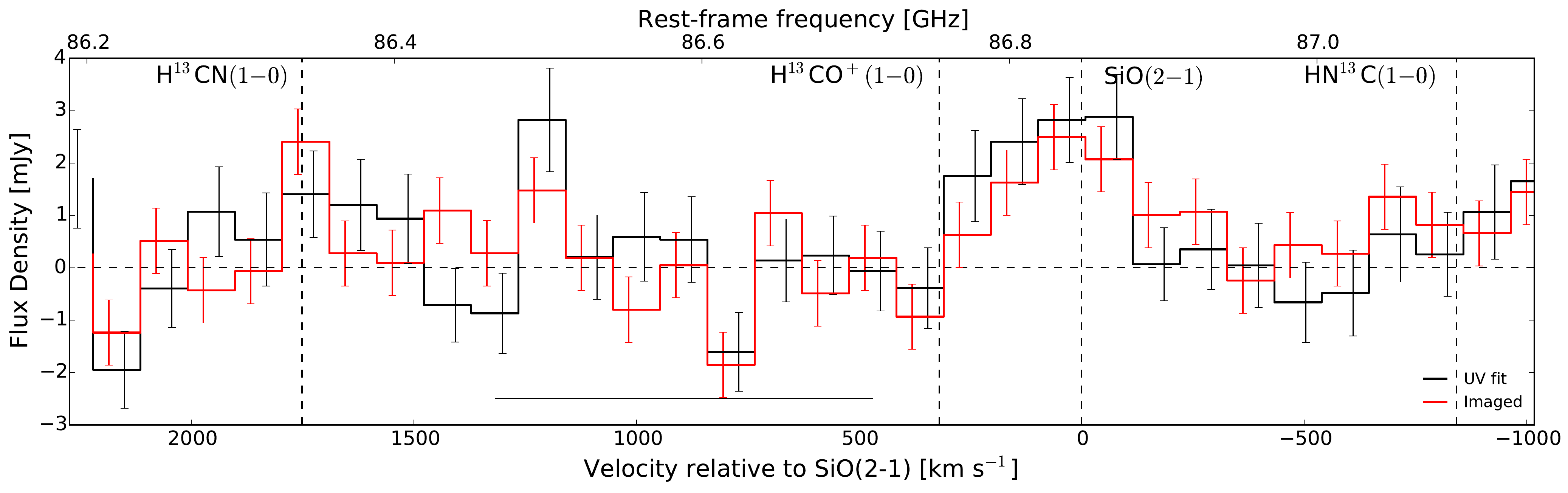}
	\caption{A comparison of the UV fitted and Imaged spectra. Axes are as in Figures \ref{fig:uvspec} and \ref{fig:imspec}. The UV fitted spectrum has been continuum subtracted using the baseline fitted simultaneously with the Gaussian lines. Channel uncertainties have been offset left and right for clarity.}\label{fig:spec_comp}
\end{figure*}

\noindent The UV fitted and imaged spectra are compared in Figure \ref{fig:spec_comp}. The channel errors have been offset horizontally for clarity. Every channel is consistent between the two spectra within $1\sigma$.


	\subsection{Continuum}\label{subsec:continuum}
	
\begin{figure}
	\centering
	\includegraphics[width=0.475\textwidth]{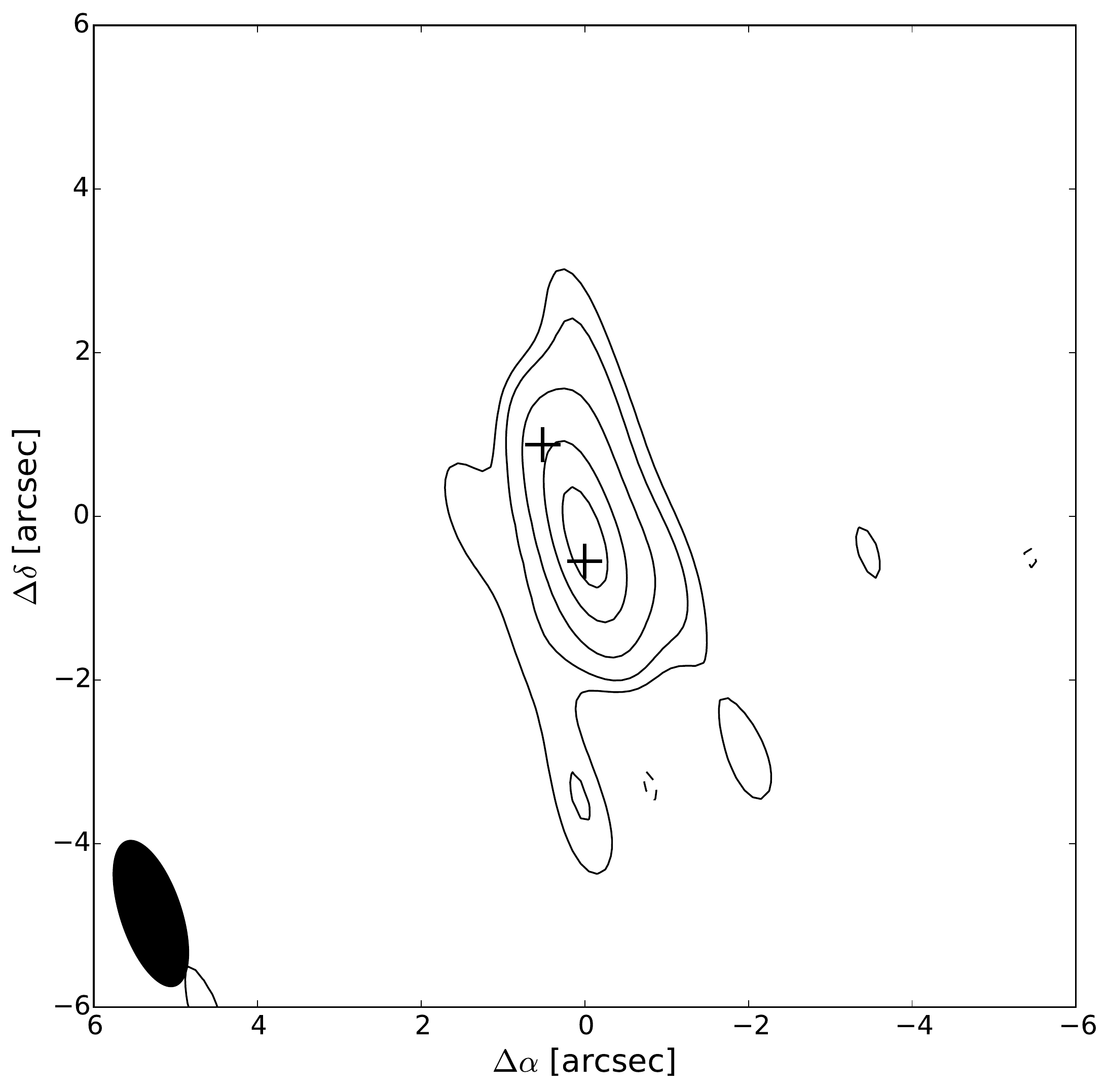}
	\caption{The NGC\,6240 3\,mm continuum. Black crosses mark the positions of the AGNs as reported in \citet{Hagiwara2011}. Contours are at -3, 3, 5, 10, 20 and 30$\sigma$, where $\sigma = 1.63\times10^{-4}$\,Jy\,beam$^{-1}$.}\label{fig:continuum}
\end{figure}

	\noindent We used the 210\,MHz region from 86.47\,GHz to 86.67\,GHz (rest frame) to define the continuum. This region, expected to be line free, was verified as such by an inspection of the UV fitted spectrum. The imaged continuum is shown in Figure \ref{fig:continuum} and we find an aperture extracted flux density of $11.3\pm1.7$\,mJy (using the same aperture and method as for the spectrum extraction). This is noticeably larger than the flux density of the elliptical Gaussian fitted to the continuum UV visibilities, which gives a flux density of $9.70\pm0.29$\,mJy ($9.7\pm1.5$\,mJy including the 15\% absolute flux calibration uncertainty), although it is still consistent within $1\sigma$. This is in part due to the extended irregularly shaped emission to the south-east of the pointing centre, which is not accounted for in the single component UV fit. Ideally we would fit further components to the residual visibilities \citep[see e.g., ][]{Feruglio2013a}, but the poor signal to noise ratio of our data prevents meaningfully fitting additional components.
	
	An estimate of the continuum was also provided by the baseline fitted to the UV fitted spectrum. This measured a continuum flux density of $9.58\pm0.19$\,mJy, or, including our 15\% absolute flux calibration uncertainty, $9.6\pm1.5$\,mJy.
	
	Our continuum flux density is slightly than expected: \citet{Nakanishi2005} found 16.6$\pm1.7$\,mJy at $\sim88.6$\,GHz rest frame and $10.8\pm1.1$\,mJy at $\sim110.2$\,GHz rest frame, with the Nobeyama Millimeter Array and RAINBOW interferometers. Their measurement is consistent with our imaged continuum at the $2\sigma$ level. Expanding the aperture used for the imaged continuum does not significantly change the recovered flux density. 
	
	\subsection{Spectral Lines}\label{subsec:speclines}
	
	\noindent Given the low signal to noise ratio of our observations we adopt an elementary statistical approach to line identification. A priori, we expect to detect H$^{13}$CN$(1-0)$ and H$^{13}$CO$^+(1-0)$ based on previous detections of HCN and HCO$^+$ in NGC\,6240, while we expect to detect SiO$(2-1)$ based on comparisons with other LIRGs and major mergers. We give parameters for these lines in Table \ref{tab:lines}. 
	
	Only SiO$(2-1)$ is clearly detected in both the UV fitted and imaged spectra. The H$^{13}$CN$(1-0)$ line appears to be present in the UV fitted spectrum, but not clearly so in the imaged spectrum, and there is no sign of H$^{13}$CO$^+(1-0)$ in either. While, if detected, it would be at least partially blended with the SiO$(2-1)$ line, there is absolutely no sign of emission redwards of the H$^{13}$CO$^+(1-0)$ line centre, or bluewards in the residuals of the SiO$(2-1)$ line fit.
	
	For our statistical test of the line detections we treat the channel uncertainties as Gaussian. We sum the channels within $\pm400$\,km\,s$^{-1}$ of the line centre and sum the channel rms's in quadrature, to create a single `detection' channel and quantify this signal in units of $\sigma_{\rm detect}$, the rms of the detection channel. We define our upper limit on the flux density, $S_{\rm max}$, as the flux density which has a 2\% chance of being measured at the value we see or less if the line is in fact present, i.e.,
	
	\begin{equation}
	S_{\rm max} = \left[S_{\rm detect} + \sqrt{2\sigma^2}{\rm erf}^{-1}(2{\rm p} -1)\right],
	\end{equation}
	where $S_{\rm detect}$ is the measured signal in the detection channel, p is our 0.02 probability and erf is the Gauss error function. For ${\rm p}=0.02$ this gives $S_{\rm max} = S_{\rm detect} + 2.054\sigma$ (Masci 2011)\footnote{\scriptsize \url{wise2.ipac.caltech.edu/staff/fmasci/UpperLimits_FM2011.pdf}}.

	\begin{table}[h!]
		\begin{center}
		\caption{Observed Line Parameters}\label{tab:lines}
		\begin{tabular}{l c c c c}\hline
		 & $\nu_0$ & E$_{\rm u}$/k & $S_{\rm line}$ (UV) & $S_{\rm line}$ (imaged)\\
		 Line & [GHz] & [K] & [Jy\,km\,s$^{-1}$] & [Jy\,km\,s$^{-1}$] \\ \hline\hline
		 H$^{13}$CN$(1-0)$\dotfill\footnote{Spectral Line Atlas of Interstellar Molecules \citep[SLAIM,][]{SLAIM}.} & 86.340 & 4.14 & $0.46\pm0.07$ & $<0.8$\\
		 H$^{13}$CO$^+(1-0)\dotfill^{\rm a}$ & 86.754 & 4.16 & $<0.4$ & $<0.4$ \\
		 SiO$(2-1)\dotfill$\footnote{The Cologne Database for Molecular Spectroscopy \citep[CDMS,][]{CDMS}.} & 86.847 & 6.25 & $1.10\pm0.15$ & $0.90\pm0.09$ \\ \hline
		\end{tabular}
		\end{center}
		The uncertainties quoted here do not include the 15\% absolute calibration uncertainty.
	\end{table}

	We use two measures to test whether the line is actually detected. Firstly, we test against the null hypothesis that we see a signal at least as bright as in the data, given that there is no line present - i.e., we assume that there is no chance of absorption. This is a one-tailed test with probability
	\begin{equation}
	\alpha = \frac{1}{2}\left[1-{\rm erf}\left(\frac{x-\mu}{\sqrt{2\sigma^2}}\right)\right],
	\end{equation}
	and a value consistent with no line is 0.5.
	
	Secondly, we test against the null hypothesis that we see a signal with an absolute value at least as great as is seen, given that there is no line present - i.e, we allow for absorption. This is a two-tailed test with probability
	\begin{equation}
	\alpha = 1-{\rm erf}\left(\frac{|x-\mu|}{\sqrt{2\sigma^2}}\right),
	\end{equation}
	and a value consistent with no line is 1. While the one-tailed test specifically searches for emission lines, the two-tailed test makes no such distinction. 
	
	\bigskip
	
	\noindent For the aperture extracted SiO$(2-1)$ line and continuum region we found one-tailed values of $7.0\times10^{-7}$ and 0.53 respectively, precisely as expected for a clear line detection and a flat continuum. The two-tailed values are $1.4\times10^{-6}$ and 0.95, again consistent with a clear line detection and a flat continuum. 
	
	For the imaged H$^{13}$CN$(1-0)$ line the one and two-tailed tests gave 0.014 and 0.028 respectively, while the UV fitted line gave 0.014 and 0.004. While not entirely conclusive, these tests argue for a detection of the H$^{13}$CN$(1-0)$ line in the UV fitted spectrum. For H$^{13}$CO$^+(1-0)$ we subtracted the SiO$(2-1)$ line fit from the spectrum before running the tests. In all cases we find values completely consistent with no line detection with one and two-tailed tests giving 0.51 and 0.98 respectively.

	For the H$^{13}$CN$(1-0)$ line we find an upper limit $S_\nu$d$v < 0.8$\,Jy\,km\,s$^{-1}$ from the imaged spectrum and $S_\nu$d$v < 0.9$\,Jy\,km\,s$^{-1}$ from the UV fitting. The imaged spectrum cannot be well fitted with a Gaussian due to almost all of the flux residing in one channel, but for the UV fitted spectrum the Gaussian fit to the line gives $S_\nu$d$v = 0.46 \pm 0.07 $\,Jy\,km\,s$^{-1}$. The SiO$(2-1)$ line was clearly detected in both spectra, and was fitted with line fluxes of $S_\nu$d$v = 1.10 \pm 0.15 $\,Jy\,km\,s$^{-1}$ and $S_\nu$d$v = 0.90 \pm 0.09 $\,Jy\,km\,s$^{-1}$ in the UV fitted and imaged spectra respectively (not including the 15\% absolute calibration uncertainty). For H$^{13}$CO$^+(1-0)$ we first subtracted the SiO$(2-1)$ Gaussian line fit before summing the channel fluxes. We obtain an upper limit of $S_\nu$d$v < 0.4$\,Jy\,km\,s$^{-1}$ for both spectra. Attempts to simultaneously fit two Gaussians to the SiO$(2-1)$ and H$^{13}$CO$^+(1-0)$ line centres consistently reported zero line flux for H$^{13}$CO$^+(1-0)$, consistent with a non-detection/upper limit. The measured fluxes and upper limits are recorded in Table \ref{tab:lines}.
	
\begin{figure*}
\centering
\includegraphics[width=\textwidth]{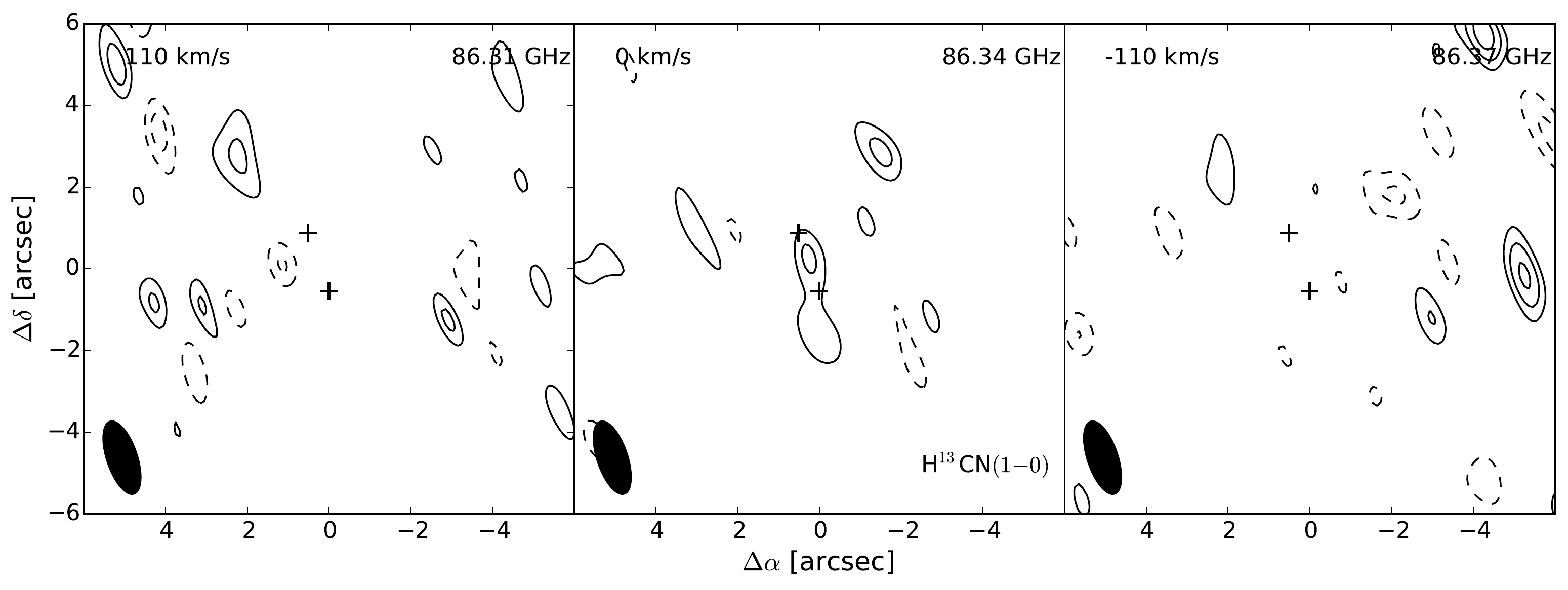}\\
\includegraphics[width=\textwidth]{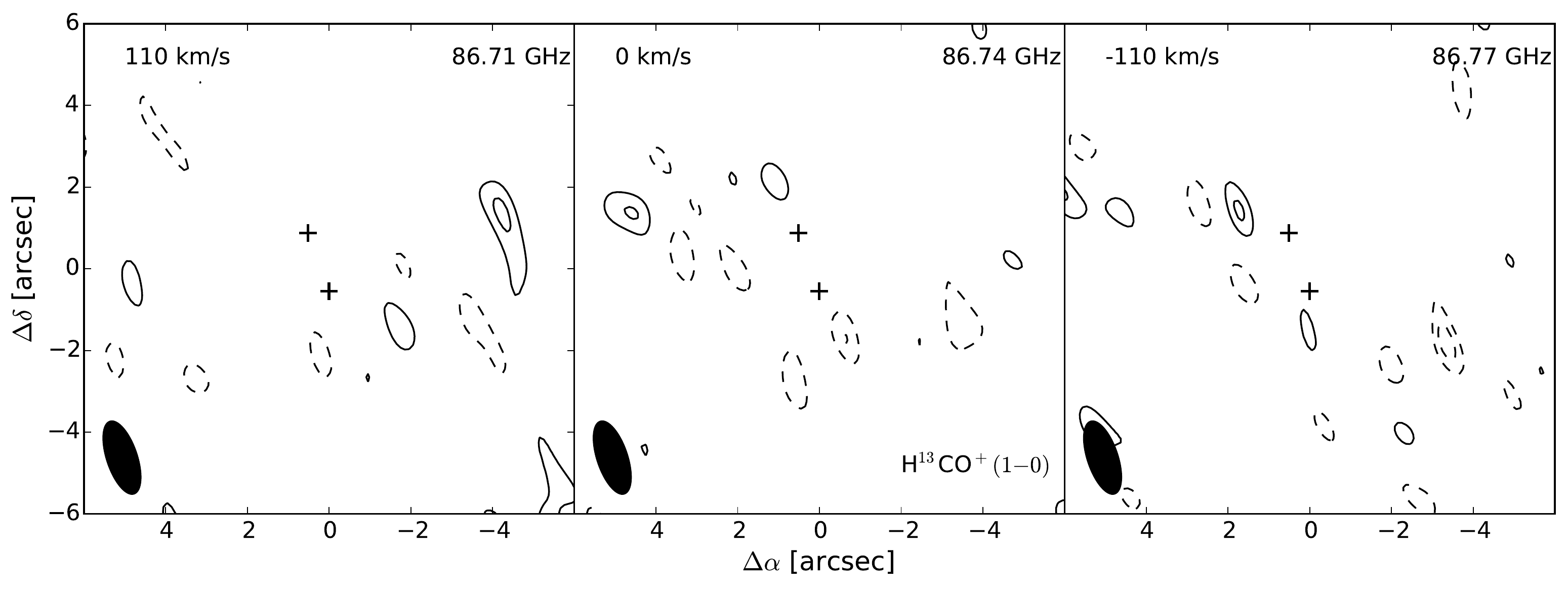}\\
\includegraphics[width=\textwidth]{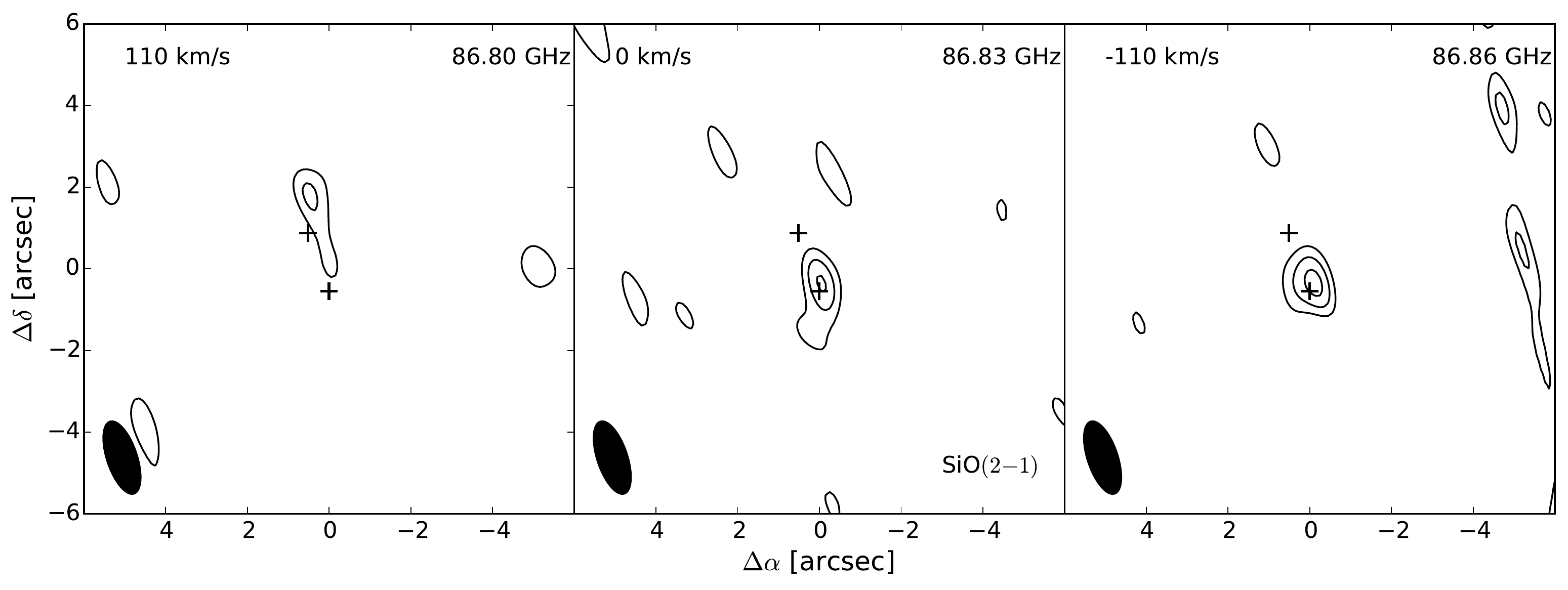}
\caption{Continuum subtracted channel maps for H$^{13}$CN$(1-0)$, H$^{13}$CO$^+(1-0)$ and SiO$(2-1)$. Black crosses mark the AGN as reported in \citet{Hagiwara2011} Contours are at $-$3, $-$2, 2, 3 and 4\,$\sigma$, where $\sigma = 3.9\times10^{-4}$\,Jy\,beam$^{-1}$. H$^{13}$CN$(1-0)$ appears to be located between the two nuclei, co-located with the the HCN$(1-0)$ \citep{Nakanishi2005}. SiO$(2-1)$ on the other hand appears to be concentrated around the more active southern nucleus.}\label{fig:channel_maps}
\end{figure*}

	We present 30\,MHz channels maps for H$^{13}$CN$(1-0)$, H$^{13}$CO$^+(1-0)$ and SiO$(2-1)$ in Figure \ref{fig:channel_maps}. H$^{13}$CN$(1-0)$ is seen only in the channel containing the line centre, at a peak level of $3\sigma$. It is concentrated between the two nuclei, as was seen for HCN$(1-0)$ by \citet{Nakanishi2005}. There is no sign of H$^{13}$CO$^+(1-0)$ in any of the channels, consistent with the non-detection in the UV fitted and imaged spectra. Finally, SiO$(2-1)$ appears at the 3 and $4\sigma$ level in the 0 and $-110$\.km\,s$^{-1}$ channels centred on the southern nucleus, with a possible $3\sigma$ detection in the $+110$\,km\,s$^{-1}$ channel about the northern nucleus.

	\smallskip

	\noindent We briefly note than there is no sign of HN$^{13}$C$(1-0)$ in either of the spectra. Also, if the line were detected it would extend beyond the edge of our spectrum, hindering its analysis. We mark the line frequency in Figures \ref{fig:uvspec} and \ref{fig:imspec} for reference.

	\subsection{Spatial Filtering and Missing Flux}
	
	\noindent Interferometers sample discrete spatial frequencies, and as such may underestimate the true flux of a source. As there are no single dish observations of any of our targeted lines we have no direct means of measuring the fraction of recovered flux and instead use comparisons with other line and continuum measurements from the literature. Nominally, our observations should only resolve out emission on scales larger than $12''$, three times larger than the largest line and continuum structures seen by \citet{Nakanishi2005}, \citet{Iono2007} and \citet{Scoville2015}, so it is unlikely that we have any significant loss of flux. However, given the detection of CO on scales $\sim25''$ by \citet{Feruglio2013a} we proceed with caution, despite not expecting HCN or HCO$^+$ emission to be similarly extended.
	
	\citet{Nakanishi2005} found the HCN$(1-0)$ line to be more compact than both the continuum and HCO$^+(1-0)$. The HCO$^+(1-0)$ line was slightly more extended, but still only out to scales $\sim4''$. It is of course possible that the observations of \citet{Nakanishi2005} are also filtering a significant fraction of the flux, and indeed their single dish flux to interferometric flux ratio was 0.68 for HCN$(1-0)$, using the single dish data of \citet{Solomon1992}. However, the single dish fluxes are extremely variable, and comparisons with more recent single dish observations give flux ratios of $1.1\pm0.4$ \citep{Greve2009} and $0.9\pm0.4$ \citep{Krips2008}.
	
	The high spatial resolution ($0.5''$) interferometric (ALMA) HCN$(4-3)$ observations of \citet{Scoville2015} have APEX single dish counterparts from \citet{Papadopoulos2014} and a flux ratio $0.90\pm0.16$, i.e., consistent with no spatial filtering/almost complete flux recovery. While extrapolating from the $J=4-3$ transition to our $J=1-0$ transitions is unacceptably unreliable, combined with the very compact HCN$(1-0)$ emission and the comparisons in the previous paragraph this comparison does argue for our observations recovering most of the H$^{13}$CN$(1-0)$ flux.
	
	We note that when comparing ALMA observations with JCMT fluxes \citet{Scoville2015} raised significant concerns regarding the calibration accuracy of the single dish measurements.

	\citet{Wang2013} observed SiO$(2-1)$ in a six local galaxies with the IRAM 30m, including NGC\,6240. They found a double peaked line profile, with a combined flux of $0.72\pm0.30$\,Jy\,km\,s$^{-1}$\footnote{Converted from the $T_A^*$ fluxes in their Table 1, using their quoted $F_{eff}=95\%$ and the 30m conversion factor $S/T_A^*=3.06 F_{eff}/A_{eff}$\,Jy\,K$^{-1}$, with $A_{eff}=0.63$ from \url{http://www.iram.es/IRAMES/mainWiki/Iram30mEfficiencies}.}, consistent with our observations. \citet{Wang2013} also observe what may be a marginal H$^{13}$CO$^+$ detection, with a peak $T_A^*\simeq0.55$\,mK. We approximate this as corresponding to $\simeq0.08$\,Jy\,km\,s$^{-1}$, consistent with our upper limit on the line of $0.4$\,Jy\,km\,s$^{-1}$.

	\subsection{The LTE limit}\label{sec:ltelim}
	
	\noindent \citet{Papadopoulos2014} reported the discovery of an extremely high [$^{12}$C]/[$^{13}$C] abundance ratio in NGC\,6240, evidenced by the [CO]/[$^{13}$CO] ratio of 300 - 500. Here we test whether this ratio is compatible with our observed H$^{13}$CN$(1-0)$ line flux under the assumption of local thermodynamic equilibrium (LTE).
	
	We compare our H$^{13}$CN$(1-0)$ line flux to an aggregate of literature measurements of the HCN$(1-0)$ line using
	
	\begin{align}
	S_{\nu} &= \frac{2kT_A\nu^2}{c^2}\Omega_A,\\
	T_A{\rm d}v &= \frac{hc^3}{8\pi k\nu^2}N_uA_{ul}\left(\frac{\Omega_s}{\Omega_A}\right)\left(\frac{1-e^{-\tau}}{\tau}\right),
	\end{align}
	where $\Omega_{A,S}$ is the solid angle of the primary beam and source respectively, $N_u$ is the column density of the species in level $u$ in cm$^{-2}$ and $A_{ul}$ is the Einstein A coefficient for the transition between levels $u$ and $l$ \citep{Goldsmith1999}. Combining these equations and assuming that HCN and H$^{13}$CN share the same excitation temperature, $T_{\rm ex}$, so that $N_{{\rm HCN, }J}/N_{\text{H$^{13}$CN,}J}$ = [HCN]/[H$^{13}$CN], we obtain for the $J = 1-0$ transition:
	
	\begin{equation}
	\frac{S_1{\rm d}v_1}{S_2{\rm d}v_2} = \frac{[{\rm HCN}]}{[{\rm H}^{13}{\rm CN}]} \frac{A_{1,1-0}}{A_{2,1-0}} \frac{\tau_2\left(1-e^{-\tau_1}\right)}{\tau_1\left(1-e^{-\tau_2}\right)},
	\end{equation}
	where the subscripts 1 and 2 refer to HCN and H$^{13}$CN respectively, Therefore, the lower limit on [HCN]/[H$^{13}$CN] for a given flux ratio is found when both lines are optically thin\footnote{Assuming that the optical depth of HCN$(1-0)$ is always greater than that of H$^{13}$CN$(1-0)$ - a safe assumption.}. In this optically thin LTE case and using the upper limit on our H$^{13}$CN$(1-0)$ line flux, we find [HCN]/[H$^{13}$CN]$\geq12$. However, exploring a range of optical depths we find that constrained by the $J=1-0$ lines alone the [HCN]/[H$^{13}$CN] could be as large as 1000, consistent with the high values found by \citet{Papadopoulos2014}. We further constrain the isotopologue ratio in Section \ref{sec:lvg} using LVG modelling. 
	
	\bigskip
	
	\noindent A high [$^{12}$C]/[$^{13}$C] ratio, leading to high [HCN]/[H$^{13}$CN] and [HCO$^+$]/[H$^{13}$CO$^+$] ratios, is also consistent with SiO$(2-1)$ being brighter than both H$^{13}$CN$(1-0)$ and H$^{13}$CO$^+(1-0)$ (SiO$(2-1)$/H$^{13}$CO$^+(1-0) > 3)$; SiO$(2-1)$ is not over-luminous, rather the low abundances of the $^{13}$C isotopologues leads to unusually faint H$^{13}$CN$(1-0)$ and H$^{13}$CO$^+(1-0)$ lines. \citet{Usero2004} found SiO$(2-1)$/H$^{13}$CO$^+(1-0)$ intensity ratios of about 3 in some regions of NGC\,1068, but this galaxy is unusually bright in SiO.

\section{LVG Modelling}\label{sec:lvg}

\noindent We use the publicly available code RADEX\footnote{\url{http://home.strw.leidenuniv.nl/~moldata/radex.html}} \citep{Radex} with collisional data from the Leiden LAMBDA database \citep{Green1974,Flower1999,Dayou2006, Lique2006,Dumouchel2010,Yang2010} to model spectral line ratios as a function of the gas phase conditions. We employ an MCMC script using the Metropolis-Hastings algorithm \citep{Metropolis1953,hastings1970} to explore the parameter space. Details of our MCMC LVG model are included in Appendix \ref{app:lvg}. We use this code to fit three gas phases and six species simultaneously.

We combine our PdBI data with literature data for HCN and HCO$^+$ shown in Table \ref{tab:lvgInputs} and adopt CO and $^{13}$CO data from \citet{Papadopoulos2014}. We have available to us the complete HCN and HCO$^+$ ladders from $J=1-0$ to $J=4-3$ except for HCO$^+(2-1)$, the complete CO SLED up to and including $J=13-12$, and the $^{13}$CO $J=1-0$, $2-1$ and $6-5$ lines.

\bigskip

\noindent In collecting data from a range of telescopes and over a range of frequencies it is essential to consider whether there is a common putative source size and whether the single dish beam coupling to the source varies significantly between observations. We first convert the line fluxes from our observations and the literature into line luminosities:

\begin{equation}
L' = 3.25\times10^7\left(\frac{\nu_{\rm obs}}{{\rm GHz}}\right)^{-2}\left(\frac{D_{\rm L}}{\rm Mpc}\right)^2\left(1+z\right)^{-3}\left(\frac{\int_{\Delta v}S_v{\rm d}v}{{\rm Jy\,km\,s}^{-1}}\right),
\end{equation}
where $D_{\rm L}$ is the luminosity distance of the source and $L'$ has units of K\,km\,s$^{-1}$\,pc$^2$ \citep{Solomon1992}. Ideally we would have a complete, spatially resolved sample of lines so that we could precisely compare the line brightnesses at specific positions in the source, ensuring that we were comparing cospatial regions of equal size. Working instead with almost entirely unresolved single dish observations we make the following approximations.

\smallskip

\noindent \textbf{Beam coupling:}
Interferometric observations of the dense gas tracers exist for only the $J=1-0$ and $J=4-3$ HCN and HCO$^+$ lines. Convolved source sizes range from $4''$ to $1^{\prime\prime}.5$ as frequency increases, with the deconvolved low$-J$ source sizes being quite uncertain, but with very compact high$-J$ deconvolved sizes: $1''.1\times0''.6$ for HCN$(4-3)$ \citep{Scoville2015}. Even if the low-J emission is more widespread it will still be much smaller than the single dish beam sizes at the associated sky frequencies, so that the beam coupling corrections are $\lesssim1.05$ and as such are much less significant than the single dish calibration uncertainties $(>10\%$). Also, this would be a systematic underestimate of the flux, so is accounted for to first order when using line ratios.

\smallskip

\noindent \textbf{Source size:}
In all of the observations, except for the HCN$(4-3)$ of \citet{Scoville2015}, the source size is so uncertain that we cannot make any convincing estimates of the source size beyond the flat assumption that there is a common size to all $J$ levels. This could potentially introduce a bias into the models, with the lower$-J$ lines being integrated over a larger area than the high$-J$. However, this acts in the opposite sense to any potential beam coupling errors, which reduce the measured low$-J$ fluxes if the low$-J$ lines are significantly more extended. To some extent these opposing errors correct for any inaccuracies introduced by assuming a common source size and not attempting to account for changes in the beam coupling.

\bigskip

\noindent In addition to the observed line ratios we use the dynamical parameter $K_{vir}$ to constrain our models \citep[e.g.,][]{Greve2009,Papadopoulos2012,Papadopoulos2014}, where
\begin{equation}
K_{vir} = \frac{{\rm d}v/{\rm d}r}{\left({\rm d}v/{\rm d}r\right)_{\rm vir}} =\frac{1.54}{\sqrt{\alpha}}\frac{{\rm d}v}{{\rm d}r}\left(\frac{n_{{\rm H}_2}}{1000{\rm \,cm}^{-3}}\right)^{-0.5}.
\end{equation}
The geometric coefficient $\alpha$ ranges from $1-2.5$ with an expectation value $\left<\sqrt{\alpha}\right>=1.312$. The velocity gradient, ${\rm d}v/{\rm d}r$, describes the change in line-of-sight velocity through the cloud in km\,s$^{-1}$\,pc$^{-1}$. In galactic GMCs in virial equilibrium $K_{vir}\sim1$. However, as was pointed out by \citet{Papadopoulos2014}, in the extreme environments of LIRGs and ULIRGS a wider $K_{vir}$ range is necessary to capture the true environment due to unusual GMCs and unbound gas. This is especially true given that we also model the shocked CO gas. We therefore follow \citet{Papadopoulos2014} and adopt limits of 0.5 and 20.0 on the allowed $K_{vir}$ values for our models.

	\begin{table}
	\centering
	\caption{LVG Line Inputs}\label{tab:lvgInputs}
	\begin{tabular}{l c c c}\hline
	& $\nu_0$ & $S_{\rm line}$ & References\footnote{1 = \citet{Nakanishi2005}, 2 = \citet{Greve2009}, 3 = \citet{Krips2008}, 4 = \citet{GraciaCarpio2008}, 5 = \citet{Papadopoulos2014}, 6 = \citet{Scoville2015}, x = this work.}\\
	Line & [GHz] & [Jy\,km\,s$^{-1}$] &\\ \hline\hline
	HCN$(1-0)$ & 88.63  & $14\pm2$ & 1, 2, 3, 4\\
	HCN$(2-1)$ & 177.3 & $44\pm7$& 3\\
	HCN$(3-2)$ & 265.9 & $74\pm7$ & 3, 4, 5\\
	HCN$(4-3)$ & 354.5 & $41\pm6$ & 5, 6\\
	H$^{13}$CN$(1-0)$ & 86.34 & $0.46\pm0.07$ & x\\
	HCO$^+(1-0)$ & 89.19 & $21\pm3$ & 2\\
	H$^{13}$CO$^+(1-0)$ & 86.75 & $<0.44$ & x\\ 
	HCO$^+(3-2)$ & 267.6 & $141\pm21$ & 5\\
	HCO$^+(4-3)$ & 356.7 & $74\pm9$ & 2, 5\\ \hline
	\end{tabular}
	\end{table}
	
	\bigskip
	
	\noindent The modelling consisted of runs of increasing complexity to check for consistency both with existing data and between models. Here, we present only the final model. The precursors are included in the Appendix. 
		
		The limits on the $[X_{^{12}{\rm C}}]/[X_{^{13}{\rm C}}]$ range were taken from the analysis of the HCN$(1-0)$/H$^{13}$CN$(1-0)$ line ratio in Section \ref{sec:ltelim} to be 10 and 1000, while the limits on $T_{\rm k}$, $n_{{\rm H}_2}$ and ${\rm d}v/{\rm d}r$ are chosen to be consistent with \citet{Papadopoulos2014}, although we extend the lower limit on d$v/$d$r$ to 0.1\,km\,s$^{-1}$\,pc$^{-1}$ to better accommodate diffuse CO. The temperature of the background blackbody, $T_{\rm bg}$, is set to 3\,K. For models with free abundances we adopted $-12 < \log_{10}\left(X_{\rm mol}\right) < -4$ for HCN and HCO$^+$ and $-8 < \log_{10}\left(X_{\rm CO}\right) < -2$ for CO.

	\subsection{Three Phase Modelling}\label{subsec:threephasemodel}
	
	\noindent A concern with simultaneously modelling CO, HCN and HCO$^+$ is that about 30\% of the CO$(1-0)$ line flux is far more extended that any other molecular species, tracing an outflow out to almost $30''$, while the HCN and HCO$^+$ emission only extends out to $\lesssim 2''$. \citep{Feruglio2013a, Nakanishi2005}, potentially leading to significant errors and bias if we force it into the same gas phase as HCN and HCO$^+$. We therefore adopt a three phase model consisting of a diffuse, extended CO phase, a hot and shocked CO phase and a cold, dense multi-species phase.
	
	The parameter ranges are given in Table \ref{tab:3phaseinputs}. The temperature and density ranges are chosen so as to force the phases to be hot, diffuse or dense respectively, while still allowing a large range for each parameter to explore.
	
	\begin{table}
	\centering
	\caption{Three Phase Input Ranges. All values $\log_{10}$.}\label{tab:3phaseinputs}
	\begin{tabular}{l c c c}\hline
	Parameter & Hot, shocked & Diffuse & Cold, dense\\ \hline\hline
	$T_{\rm k}$ & $2.0-3.5$ & $1.0-3.0$ & $0.5-1.5$ \\
	$n_{{\rm H}_2}$ & $1-6$ & $1-6$ & $4-8$ \\
	d$v/$d$r$ & \multicolumn{3}{c}{\Vhrulefill\ $-1 - 3$\ \Vhrulefill}\\
	$[$CO$]/[^{13}$CO$]$ & \multicolumn{3}{c}{\Vhrulefill\ $1 - 3$\ \Vhrulefill}\\
	$\frac{[{\rm HC(N/O}^+)]}{[{\rm H}^{13}{\rm C(N/O}^+)]}$ & $-$ & $-$ & $1-3$ \\
	$X_{\rm CO}$ & \multicolumn{3}{c}{\Vhrulefill\ $-8 - -2$\ \Vhrulefill}\\
	$X_{\rm HCN}$ & $-$ & $-$ & $-12--4$\\
	$X_{{\rm HCO}^+}$ & $-$ & $-$ & $-12--4$\\
	$f$ & $-2-0$ & $-$ & $-2-0$\\ \hline
	\end{tabular}
	\end{table}

	HCN and HCO$^+$ are only included in the cold, dense phase. This is motivated by the absence of any HCN emission from the shocked phase in the two phase models (Appendix \ref{sec:precursors}). There is a potential issue with excluding HCO$^+$ from the shocked and/or diffuse phase as it shows evidence in the precursor models of having a small but significant contribution from hotter and more diffuse gas. The choice to restrict HCO$^+$ to the dense phase is a pragmatic one: we simply do not have enough observed lines to constrain a more accurate model of HCO$^+$. This should not significantly affect the results as the CO lines are the primary determinants of the conditions in the diffuse and shocked phases, and the warmer phase HCO$^+$ emission is less than $10\%$ of the flux in any of the lines. Furthermore, it is very unlikely to be present in significant quantities in the hot, shocked gas, nor in the fully extended diffuse phase; rather it is expected on the UV illuminated edges of molecular clouds, which we do not model at all.
	
	CO is present in all three phases, and in each phase there is a free [CO]/[$^{13}$CO] ratio. In the cold, dense phase there is also an [HCN]/[H$^{13}$CN] ratio, which is shared with HCO$^+$. Inspired by \citet{Zhang2014} and \citet{Kamenetzky2014} we relate the fluxes from the three phases through two free scaling factors $f_1$ and $f_2$, such that:
	
	\begin{equation}
	S = S_{\rm shocked}f_1 + S_{\rm diffuse} + S_{\rm dense}f_2.
	\end{equation}
	
	It is important to appreciate that the $f$'s are \emph{NOT} the fractional flux densities emerging from the phases, as the fluxes from each phase may intrinsically be much greater (or smaller) than $S_{\rm diffuse}$. They are merely free parameters to allow the MCMC to scale the significance of the two phases. Physically, they are indirectly related to the beam filling factors of the phases, whereas the formulation of \citet{Kamenetzky2014}, who fitted line intensities instead of line ratios, used two beam filling factors directly as free parameters. 
	
	The contrast factors allow us to abstract away the issue that different molecular species present different beam filling factors. For the unresolved single dish observations this is essential, but even for high resolution interferometric observations this is a desirable property, allowing for that fact that the ISM can vary significantly over parsec scales (barely resolvable with ALMA in even the closest galaxies).

	\bigskip

\begin{figure*}
	\centering
	\includegraphics[width=\textwidth]{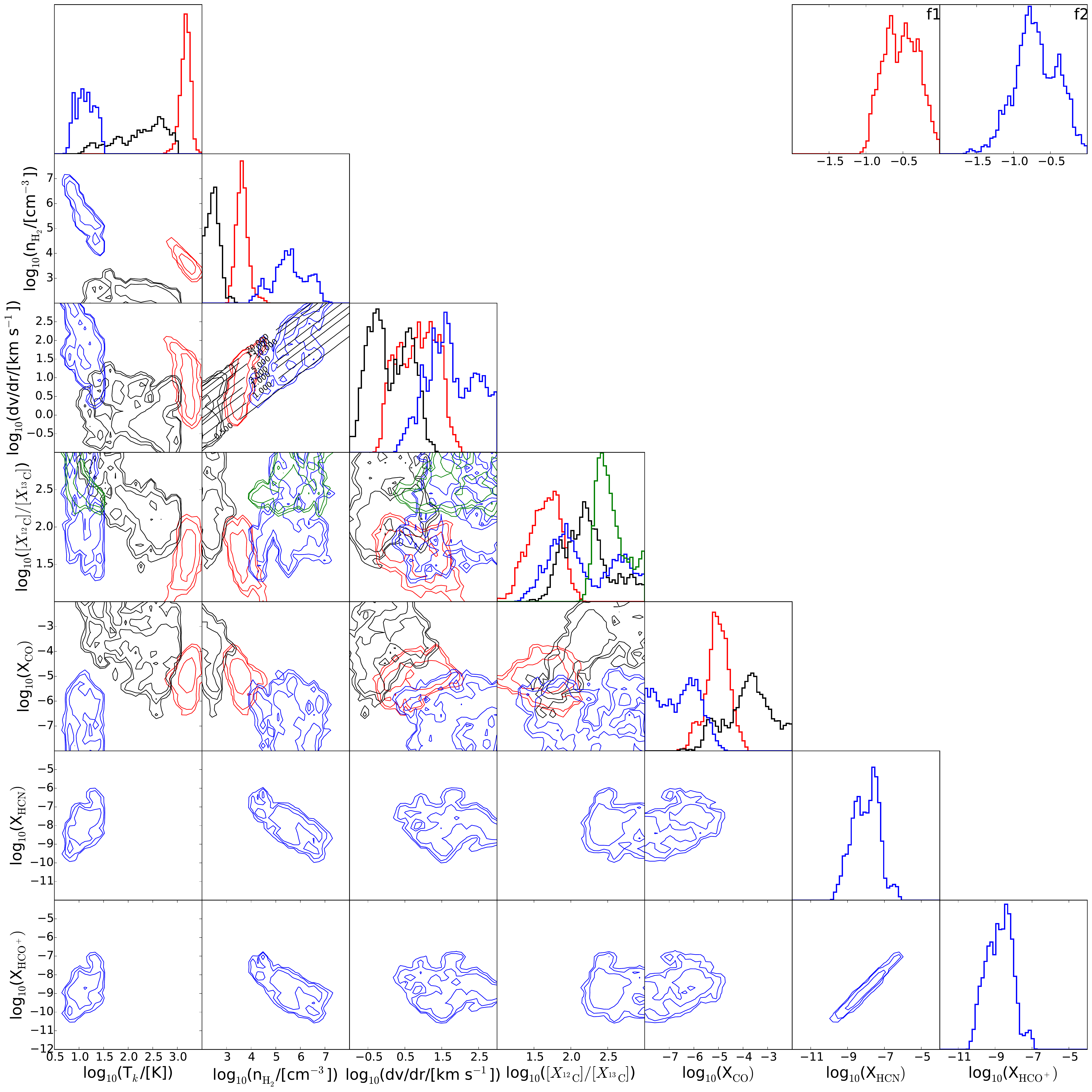}
	\caption{The step plot for the three phase HCN$+$HCO$^+$+CO model results. Contours are at the 68\%, 95\% and 99\% credible intervals. Red, black and blue contours are the shocked, diffuse and dense gas phases respectively. In the $[X_{^{12}{\rm C}}]/[X_{^{13}{\rm C}}]$ plots the green contours correspond to the [HCN]/[H$^{13}$CN] abundance ratio (shared with HCO$^+$), and the blue the [CO]/[$^{13}$CO] ratio, both in the cold, dense phase.}\label{fig:3phaseCO}
	\end{figure*}

\begin{figure*}
	\centering
	\includegraphics[width=0.475\textwidth]{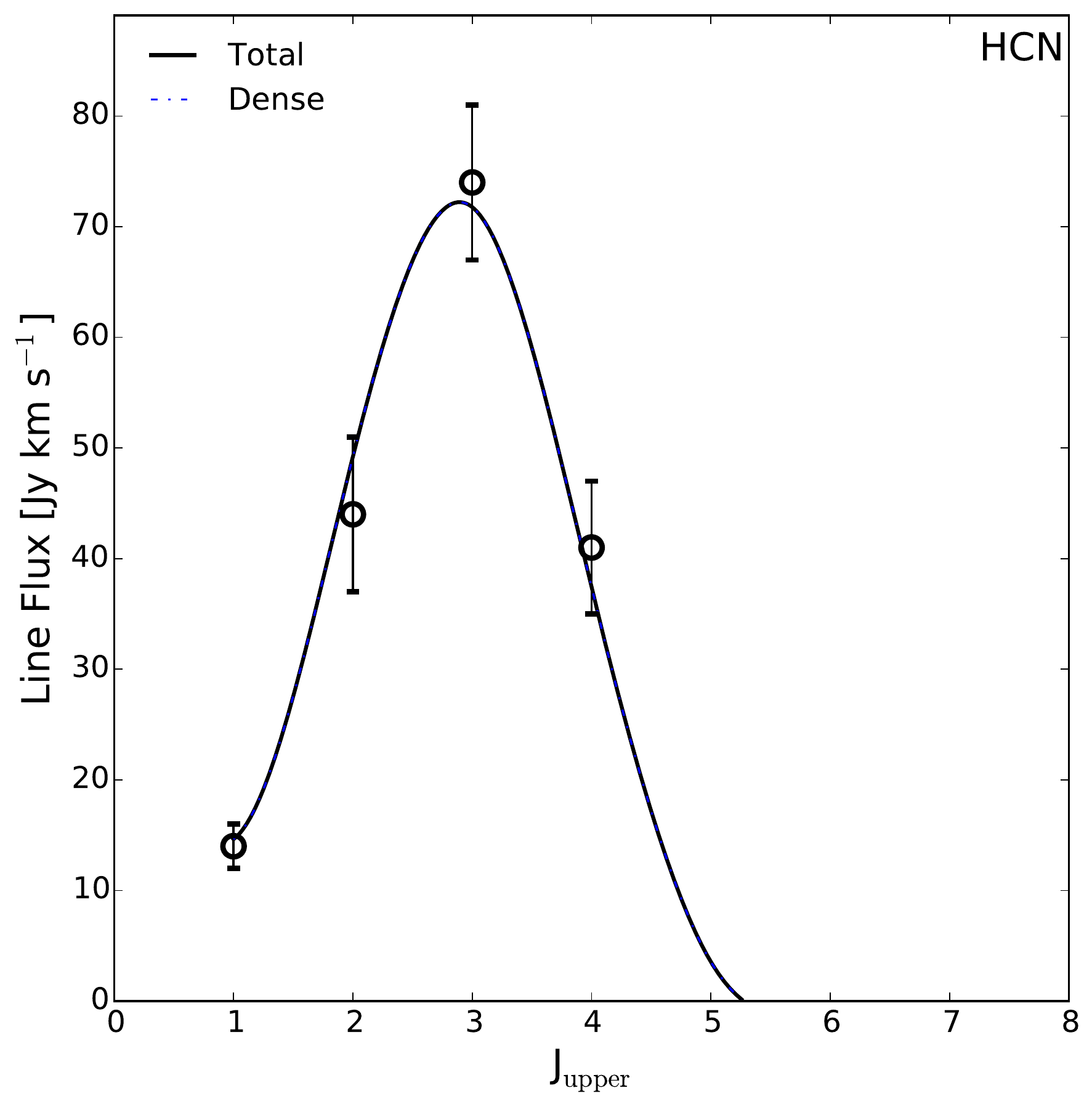}
	\hfill
	\includegraphics[width=0.475\textwidth]{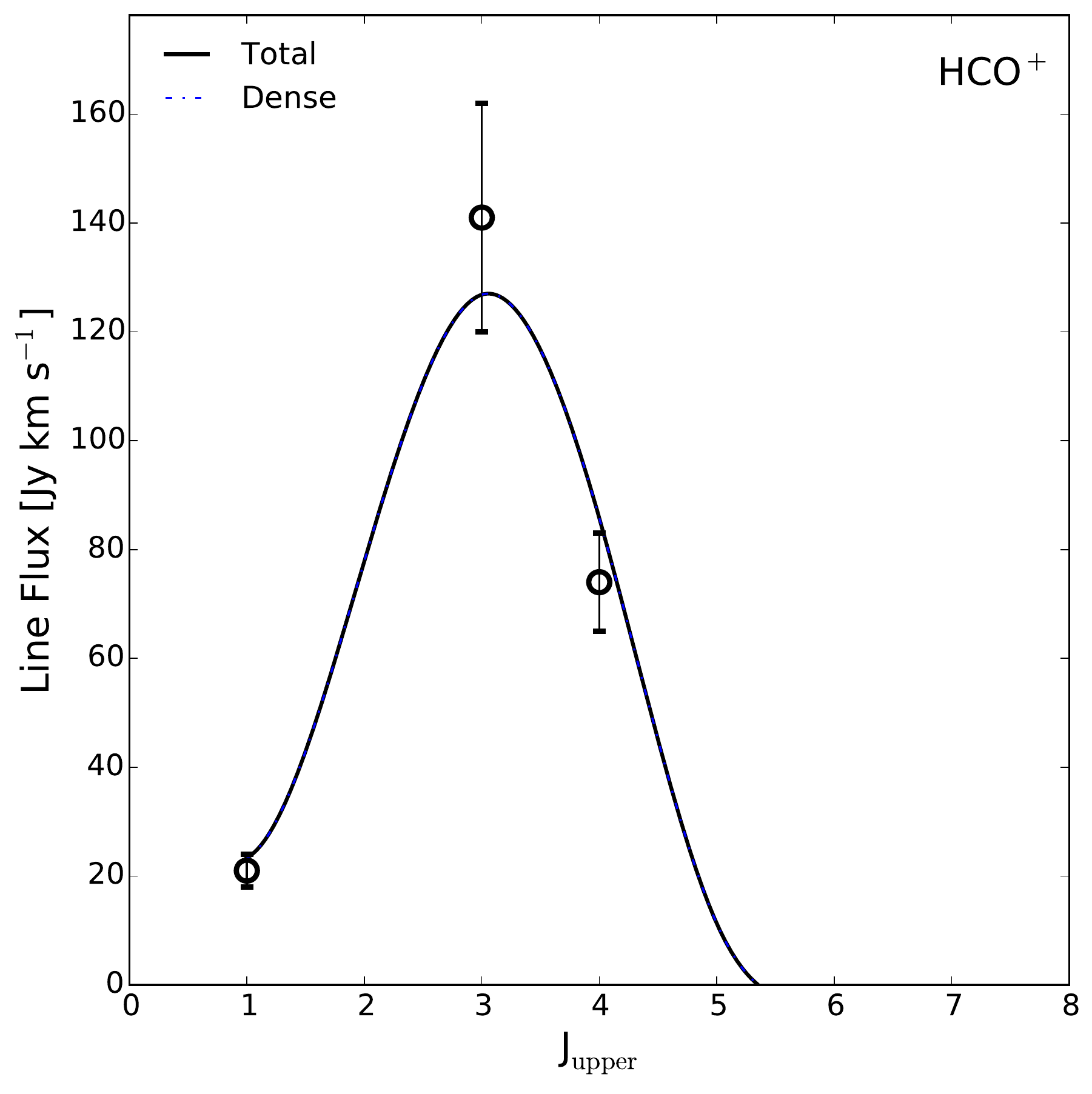}\\
	\includegraphics[width=0.475\textwidth]{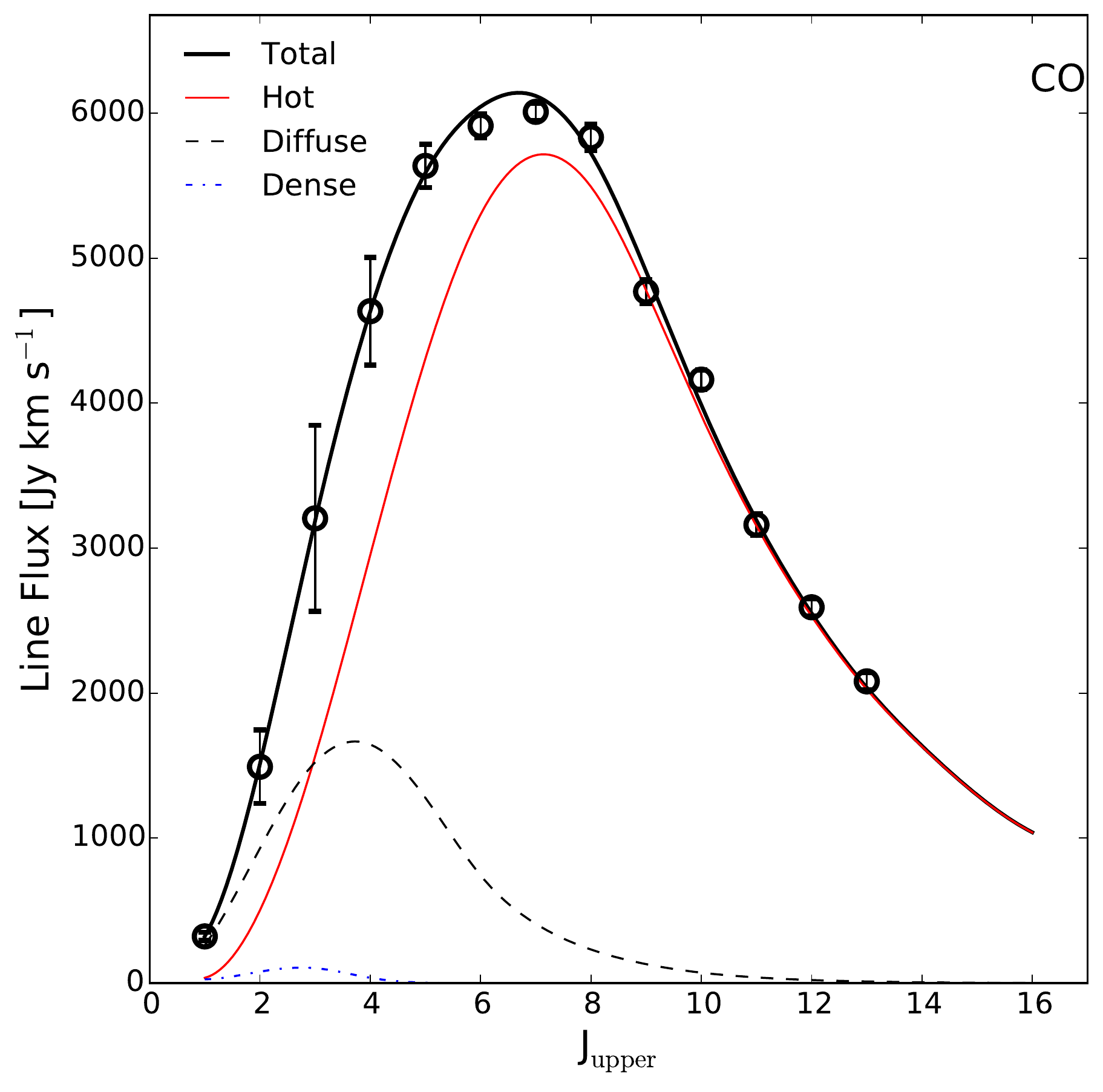}
	\hfill
	\includegraphics[width=0.475\textwidth]{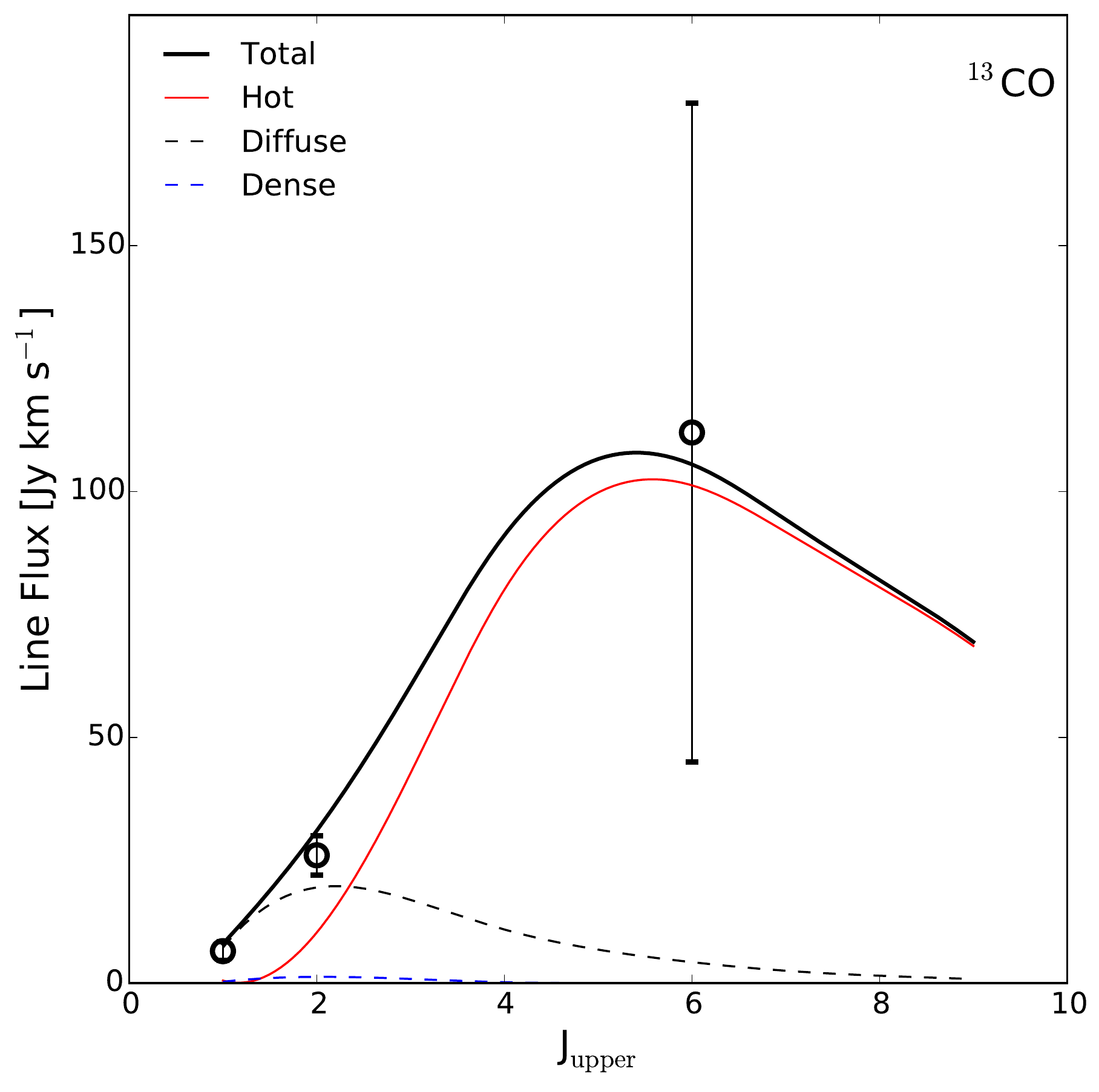}
	\caption{SLED fits for the three phase HCN$+$HCO$^+$+CO model, showing the relative contribution of each gas phase. The same normalisation has been used for the CO and $^{13}$CO SLEDs.}\label{fig:3phaseCOSLED}
	\end{figure*}

\begin{figure}
	\centering
	\includegraphics[width=0.45\textwidth]{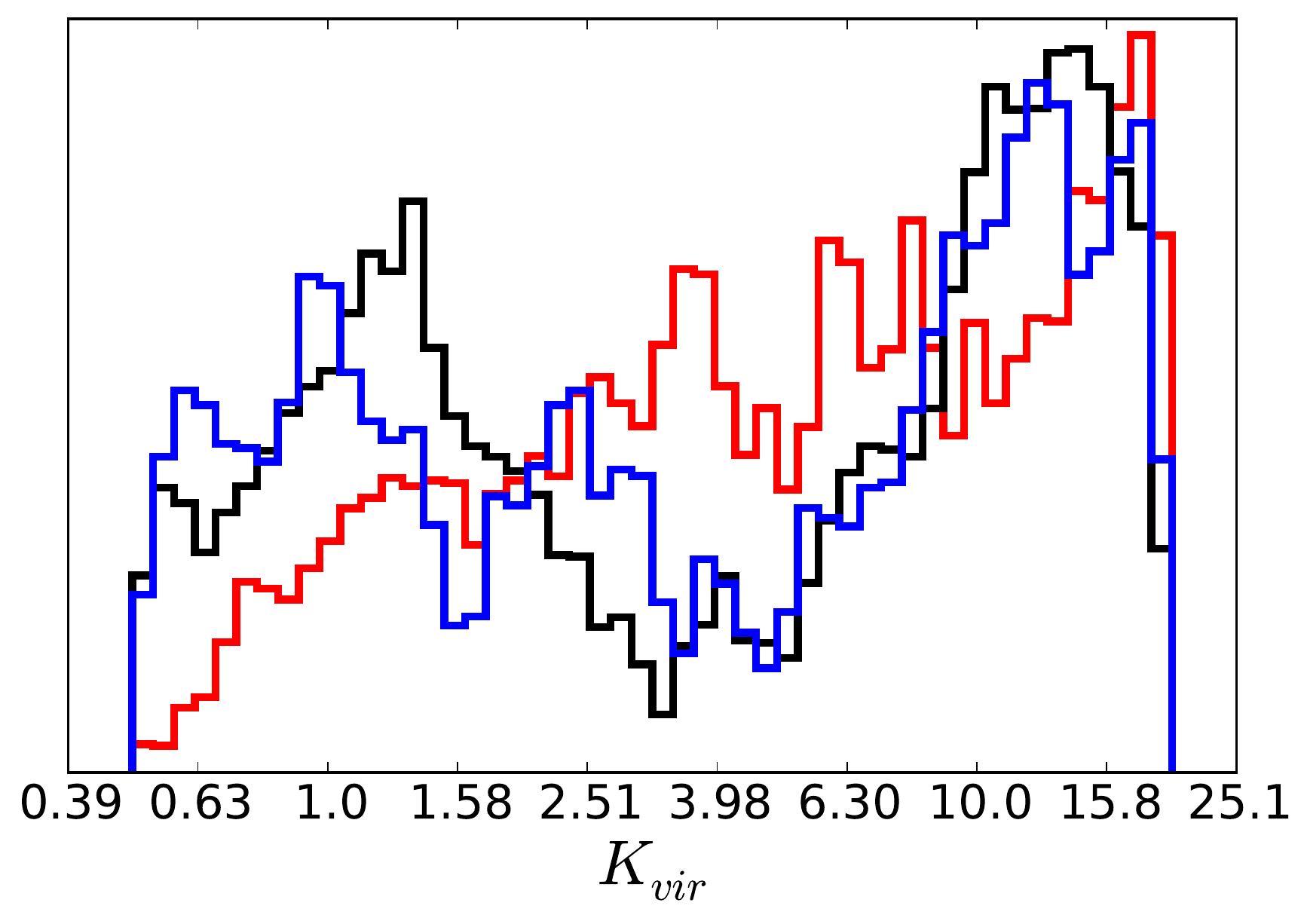}
	\caption{The $K_{vir}$ pdfs for the three phase model. Colours are as in Figure \ref{fig:3phaseCO}, with red, black and blue correspond to the shocked, diffuse and dense phases respectively. Both the diffuse and dense phases present doubly peaked pdfs, in particular with a peak about $K_{vir}=1$, while the shocked phase pdf steadily increases towards the upper limit of $K_{vir}=20$.}\label{fig:3phasekvir}
	\end{figure}

	\noindent The results of the three phase modelling are shown in Figures \ref{fig:3phaseCO}, \ref{fig:3phaseCOSLED} and \ref{fig:3phasekvir}, while the numerical values are tabulated in Table \ref{tab:3phaseResults}. The model successfully splits into three clear phases, without any tension against the parameter ranges. $X_{\rm CO}$ varies greatly between the gas phases, being greatest in the diffuse phase lying near the canonical value of $2\times10^{-4}$, before dropping slightly to around $1\times10^{-5}$ in the shocked phase and then dropping rapidly to around $10^{-6} - 10^{-7}$ in the cold, dense phase; strongly suggestive of freeze-out of CO onto dust grains in the cold, dense phase.
	
	The isotopologue abundance ratios are also largely as we would expect: both the shocked and the dense gas phases have relatively low [CO]/[$^{13}$CO] ratios of $30-100$, likely representative of the true [$^{12}$C]/[$^{13}$C] ratio in the galaxy. The diffuse phase present a slightly elevated ratio of $\sim 100-200$, consistent with this ubiquitous gas phase experiencing selective photodissociation due to the nuclear starbursts. Finally, the [HCN]/[H$^{13}$CN] ratio (recall that this is shared with HCO$^+$) is the highest, at around 300, consistent with isotope fractionation in cold, dense molecular cores.
	
	The SLEDs in Figure \ref{fig:3phaseCOSLED} show exceptionally good fits to the observed lines. The cold, dense phase has very little contribution to the CO or $^{13}$CO SLEDs, and in both of these the shocked phase dominates the emission at $J \gtrsim 4$, while the low$-J$ lines are set by the ubiquitous diffuse phase.
	
	For each point in the MCMC trace we can calculate $K_{vir}$: the $K_{vir}$ pdfs are shown in Figure \ref{fig:3phasekvir}. Both the diffuse and dense phases present doubly peaked pdfs, in particular with a peak about $K_{vir}=1$, while the shocked phase pdf steadily increases towards the upper limit of $K_{vir}=20$. $K_{vir}$ is generally a poorly constrained parameter in our models, but these results are suggestive of approximately virialised gas dominating the diffuse and dense gas phases, while the shocked phase is highly turbulent. 
	
\begin{figure}
	\centering
	\includegraphics[width=0.45\textwidth]{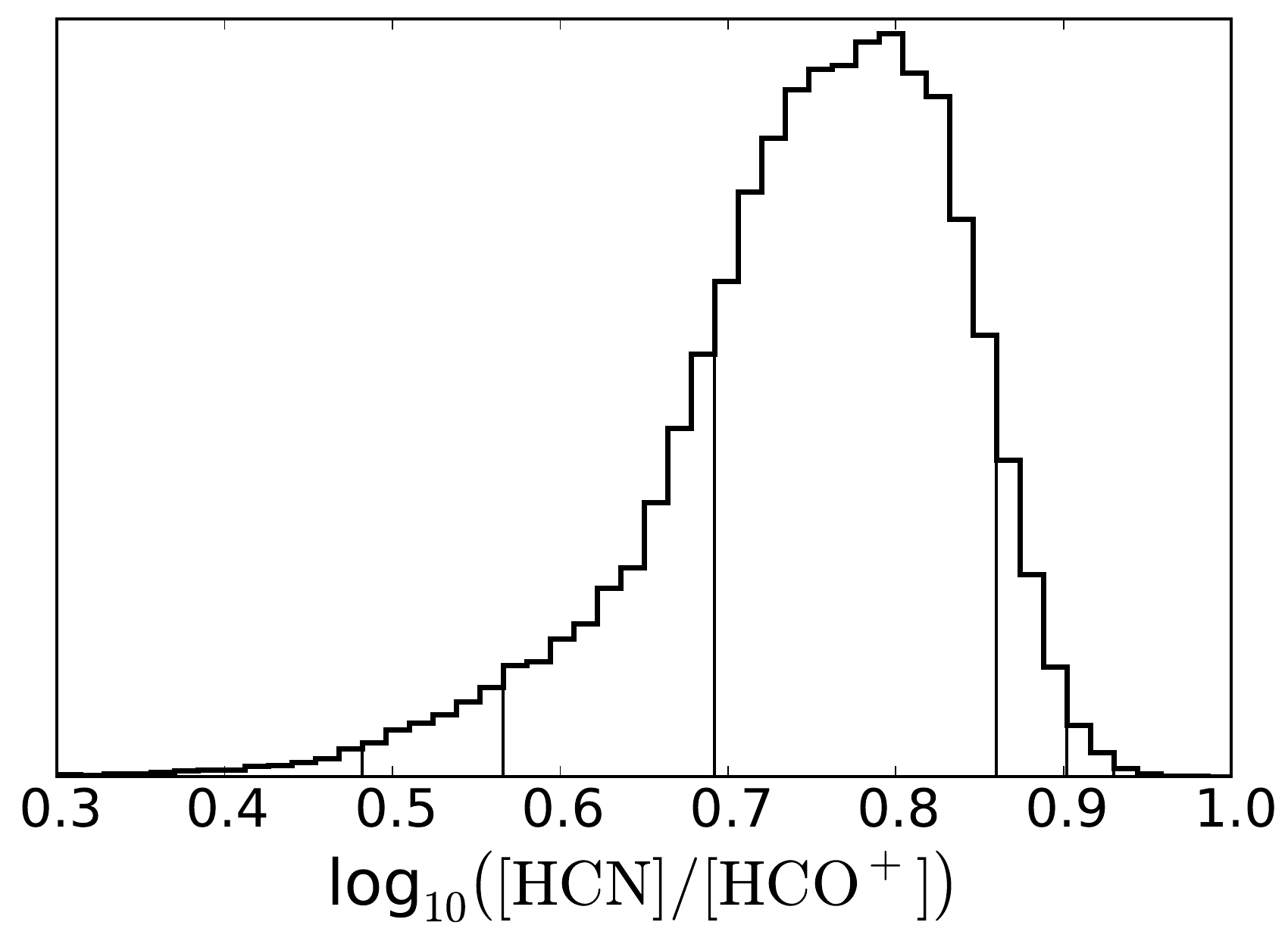}
	\caption{The [HCN]/[HCO$^+$] abundance ratio pdf from the three phase model. The value is tightly constrained as $6.2_{4.3}^{7.9}$.}\label{fig:3phasehcnhcoratio}
	\end{figure}

	We also show the [HCN]/[HCO$^+$] abundance ratio pdf in Figure \ref{fig:3phasehcnhcoratio}. This ratio is very tightly constrained as $6.2_{4.3}^{7.9}$.

	\begin{table}
	\centering
	\caption{Three Phase Results. All values $\log_{10}$. Values in parenthesis are the best fit values.}\label{tab:3phaseResults}
	\begin{tabular}{l c c c}\hline
	Parameter & Hot, shocked & Diffuse & Cold, dense\\ \hline\hline
	$T_{\rm k}$ & $3.2^{3.3}_{3.1}(3.2)$ & $2.3^{2.9}_{1.8}(2.4)$ & $1.1^{1.3}_{1.0}(0.9)$ \\[1mm]
	$n_{{\rm H}_2}$ & $3.6^{3.9}_{3.4}(3.5)$ & $2.5^{2.7}_{2.2}(2.4)$ & $5.6^{6.5}_{5.0}(6.6)$ \\[1mm]
	d$v/$d$r$ & $0.8^{1.0}_{0.1}(0.2)$ & $0.1^{0.4}_{-0.6}(0.2)$ & $1.8^{1.9}_{1.1}(1.5)$ \\[1mm]
	$\frac{[{\rm CO}]}{[^{13}{\rm CO}]}$ & $1.6^{1.9}_{1.5}(1.8)$ & $2.2^{2.4}_{1.9}(2.2)$ & $2.2^{2.8}_{1.6}(2.5)$\\[1mm]
	$\frac{[{\rm HC(N/O}^+)]}{[{\rm H}^{13}{\rm C(N/O}^+)]}$ & $-$ & $-$ & $2.5^{2.7}_{2.3}(2.6)$ \\[1mm]
	$X_{\rm CO}$ & $-5.0^{-4.5}_{-5.4}(-5.1)$ & $-3.8^{-2.7}_{-5.4}(-3.7)$ & $-6.6^{-5.6}_{-7.5}(-7.4)$\\[1mm]
	$X_{\rm HCN}$ & $-$ & $-$ & $-8.0^{-7.2}_{-7.9}(-9.6)$\\[1mm]
	$X_{{\rm HCO}^+}$ & $-$ & $-$ & $-8.8^{-8.0}_{-9.0}(-10.2)$\\[1mm]
	$f$ & $-0.5^{-0.3}_{-0.7}(-0.6)$ & $-$ & $-0.7^{-0.5}_{-1.0}(-0.2)$\\[1mm]
	Mass $[M_\odot]$\footnote{CO estimated mass from calculated $\alpha_{\rm CO}$ and the observed CO$(1-0)$ line.} & $9.4^{9.8}_{8.9}(9.1)$ & $9.3^{9.9}_{8.6}(8.7)$ & $10.3^{10.8}_{9.6}(10.0)$ \\[1mm] \hline
	\end{tabular}
	\end{table}
	
	\bigskip
	
\begin{figure*}
	\centering
	\includegraphics[width=0.495\textwidth]{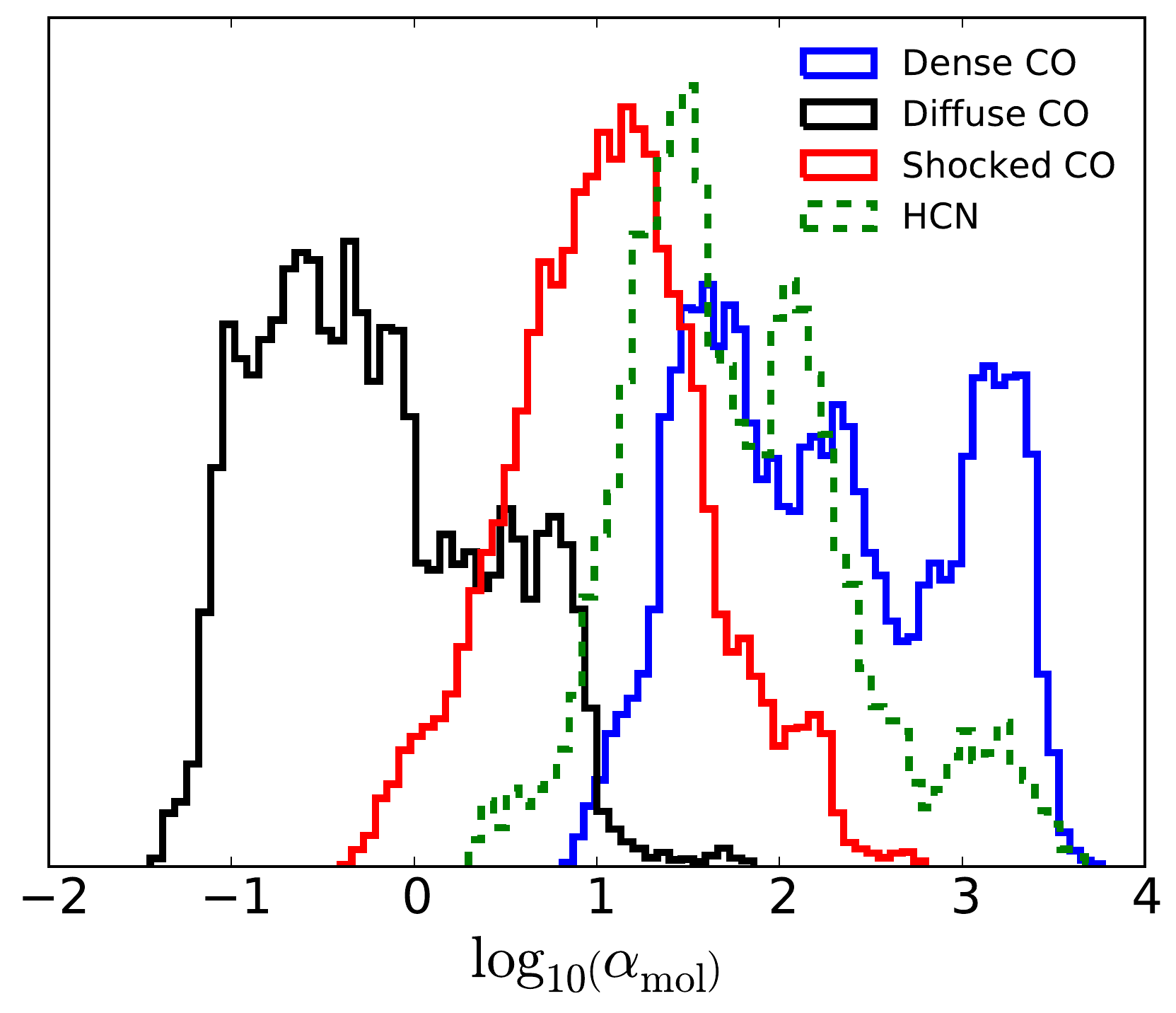}
	\hfill
	\includegraphics[width=0.495\textwidth]{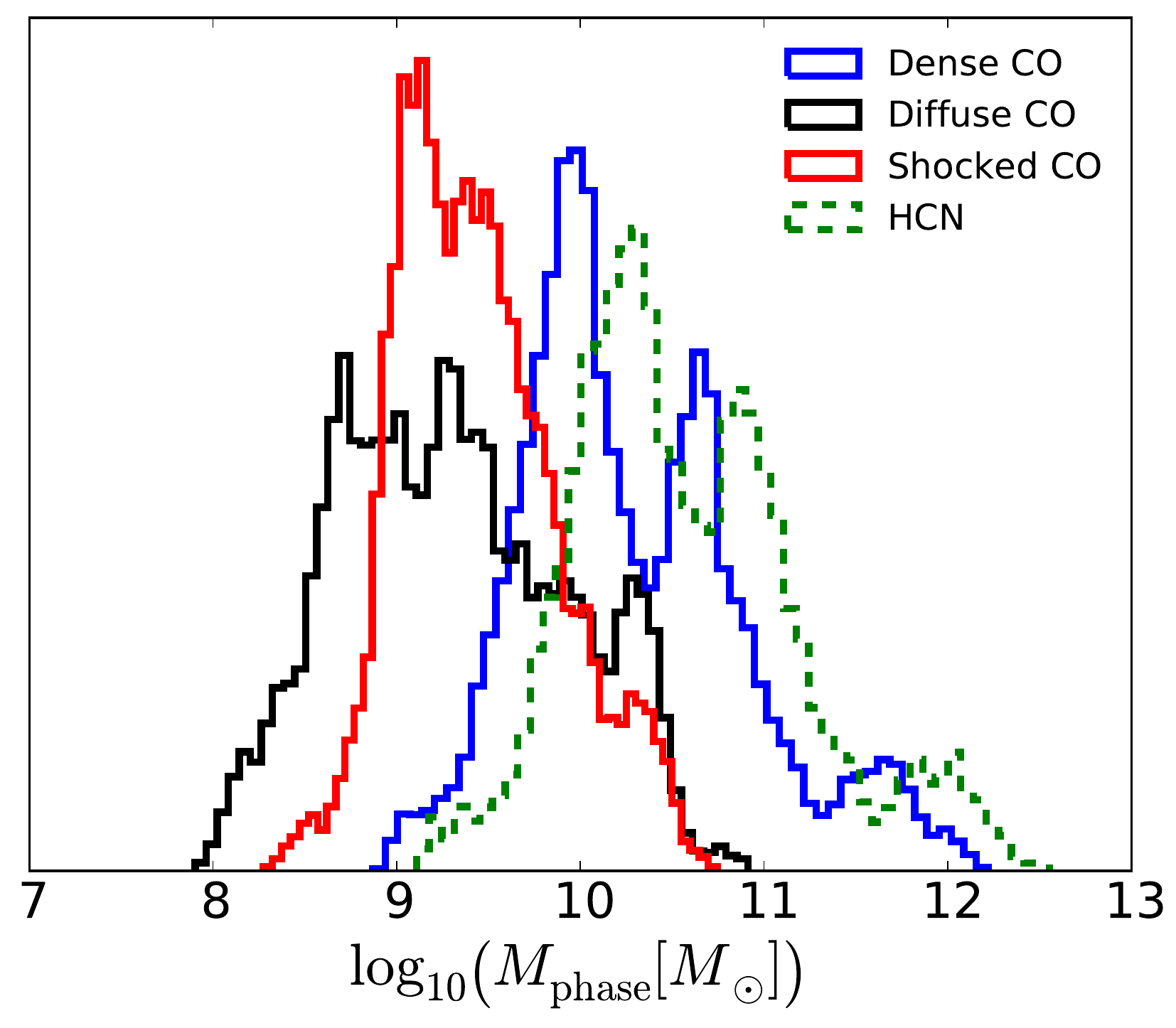}
	\caption{The $\alpha_{\rm CO}$ factor for each of the gas phases of the three phase model and the $\alpha_{\rm HCN}$ (encompassing the dense phase) (left), and the corresponding mass of the gas phase (right). The diffuse and shocked phases are consistent with the ``standard'' $\alpha_{\rm CO}$ values of between $0.5$ and 4, while the dense gas phase is considerably higher. While contributing a very small fraction of the CO$(1-0)$ line flux the dense gas phase dominates the molecular gas mass in the galaxy. The CO estimate and the HCN estimate of the mass of the dense phase are highly consistent.}\label{fig:3phase_alpha}
	\end{figure*}

	\noindent We can also calculate $\alpha_{{\rm mol}(1-0)}$, where $\alpha_{{\rm mol}(1-0)}= M_{\rm dense}/L^{\prime}_{{\rm mol}(1-0)}$ is the ratio of the gas mass of the phase to the $J=1-0$ line luminosity of the species from that phase\footnote{Units of $M_{\odot}\left({\rm K\,km\,s}^{-1}{\rm\,pc}^{2}\right)^{-1}$.}. We use equation A4 for optically thick gas from \citet{Papadopoulos2012}\footnote{$\alpha_{{\rm mol}(1-0)} = \frac{3.25}{\sqrt{\alpha}}\frac{\sqrt{\left<n_{{\rm H}_2}\right>}}{\left<T_{\rm b,1-0}\right>}K_{vir}^{-1}$} to derive $\alpha_{{\rm mol}(1-0)}$ for CO and HCN and present the pdfs, along with the corresponding masses, in Figure \ref{fig:3phase_alpha}. The spread in each of the $\alpha_{\rm CO}$ pdfs is between 1.5 and 2 dex, with the centres shifting to increasing $\alpha_{\rm CO}$ as we move from the diffuse phase, through the shocked phase and onto the dense phase. There is less spread in the masses, and in particular the HCN derived and CO derived dense gas masses are very consistent with almost identical pdfs but for a systematic shift of $\sim -0.2$\,dex on the CO mass estimates. These are discussed in Section \ref{subsec:alphas}.
	
	\bigskip
	
	\noindent Interestingly, the thermal pressure of the shocked and dense gas phases is $P/k\sim10^7$\,K\,km\,s$^{-1}$ - the same value that \citet{Dopita2005} found necessary to fit the SED in NGC\,6240 and the value found for the hot, diffuse phase in \citet{Kamenetzky2014}. This makes sense in light of the three phase model: the diffuse CO and shocked CO dominate the CO SLED, so when \citet{Kamenetzky2014} fitted to the CO SLED only they recovered these phases. However, the diffuse phase is only partially co-located with the shocked phase so is not in pressure balance, whereas the dense phase, identified here by fitting HCN, HCO$^+$ and CO simultaneously, \emph{is} co-located with the shocked phase, and approximately in pressure balance with it. On the other hand, we would expect a significant fraction of the pressure balance to be due to macroscopic turbulence, not just thermal pressure.
	
	\subsection{Comparison to the Literature}
	
	\noindent The dense, HCN bearing molecular gas in NGC\,6240 has been modelled previously, by \citet{Krips2008}, \citet{Greve2009} and \citet{Papadopoulos2014} (\citealp{Kamenetzky2014} only modelled the CO SLED). While \citet{Krips2008} modelled a single gas phase for both HCN and HCO$^+$, \citet{Greve2009} and \citet{Papadopoulos2014} fitted HCN and HCO$^+$ independently, each with their own gas phase parameters.
	
	While \citet{Krips2008} and \citet{Greve2009} both found similar gas kinetic temperature ranges ($20-120$\,K and $60-120$\,K respctively), \citet{Greve2009} found much larger $n_{{\rm H}_2}$ in the HCN phase of $\sim10^5$\,cm$^{-3}$, cf.\ $10^{3.5}-10^{4.5}$\,cm$^{-3}$ found by \citet{Krips2008}. On the other hand, the HCO$^+$ phase had an $n_{{\rm H}_2}$ of $\sim10^4$\,cm$^{-3}$, consistent with the results of \citet{Krips2008}. 
	
	\citet{Papadopoulos2014} present their results as 2D pdfs, finding degenerate ``bananas'' ranging from $n_{{\rm H}_2}\simeq10^6$\,cm$^{-3}$/$T_{\rm k}\simeq8$\,K  to $n_{{\rm H}_2}\simeq10^4$\,cm$^{-3}$/$T_{\rm k}\simeq1000$\,K, although the HCN pdf appears to peak at the high-density, low-$T_{\rm k}$ end of the ``banana''.
	
	Our results are broadly consistent with these previous works. Both our precursor models and the full three phase model argue for relatively cool ($8-30$\,K), dense ($10^{4.5}-10^{6.5}$\,cm$^{-3}$) gas dominating the HCN and HCO$^+$ emission. These models echo the trend seen by \citet{Greve2009} and \citet{Papadopoulos2014} of HCO$^+$ preferring a $\sim0.5$\,dex lower $n_{{\rm H}_2}$ than HCN while sharing a similar kinetic temperature. Our [HCN]/[HCO$^+$] abundance ratio of $5.3^{7.0}_{4.7}$ is consistent with \citet{Krips2008} who found ``around 10''.
	
\bigskip

	\noindent The multiple phases of the molecular gas in NGC\,6240 have been characterised previously, including by \citet{Armus2006} with Spitzer observations of H$_2$ vibrational lines, \citet{Papadopoulos2014} with the aforementioned spectral decomposition of the CO SLED and \citet{Kamenetzky2014} with a two phase MCMC model of the CO SLED. Our models are the first to simultaneously fit multiple species in multiple gas phases. As was pointed out by \citet{Kamenetzky2014}, and as we found here comparing to the results of \citet{Papadopoulos2014}, it is important to simultaneously fit multiple components to SLEDs to prevent the biasing of results.
	
	From modelling of H$_2$ vibrational lines \citet{Armus2006} found two highly excited phases of the molecular gas, with $6.7\times10^6\,M_\odot$ at $T_{\rm k}=957$\,K and $1.6\times10^9\,M_\odot$ at $T_{\rm k}=164\,$K respectively. \citet{Papadopoulos2014} first fitted the HCN and HCO$^+$ SLEDs, before using these results to fit high$-J$ components of the CO SLED, then fitting the residual CO SLED. This produces three gas phases with densities $\log_{10}\left(n_{{\rm H}_2}\right) = 5.0$, 4.3 and 3.0 and kinetic temperatures $\log_{10}\left(T_{\rm k}\right) = 1.5$, 2.6 and 2.0. The model of \citet{Kamenetzky2014} used a nested sampling algorithm to fit two gas phases to the CO SLED, finding a hot and a cold phase with best fit densities $\log_{10}\left(n_{{\rm H}_2}\right) =$ 4.1 and 5.8 and kinetic temperatures $\log_{10}\left(T_{\rm k}\right) =$ 3.1 and 1.2. These are reassuringly similar to the hot, shocked and cold, dense phase results of our multi-species, multi-phase analysis. It is curious however that the cold phase of their two phase model was closer to the fainter cold, dense phase than the brighter diffuse CO phase. There is no clear explanation for this, but it could very possibly be an artefact of the change from two to three phases.

\section{Discussion}\label{sec:discuss}

	\subsection{Validity of Multi-species Models}\label{subsec:mulitSpeciesValid}
	
	\noindent An important question of our analysis is whether or not it is appropriate to model multiple species as sharing the same gas phase. In reality there is a continuum of conditions within any telescope beam and any phase based model is only an approximation of the conditions dominating the emission, which will vary from molecule to molecule. Nevertheless, combining molecular species which are expected to trace similar regions appears to be a powerful tool for estimating molecular abundance ratios and gas conditions.
	
	The combination of two or more species in a single model is a compromise, with the extent of the assumptions dependent upon how similar the regions traced by the different species are likely to be. In our three phase model, the dense phase is shared by HCN, HCO$^+$ and CO, as well as their $^{13}$C isotopologues. That CO is present alongside other molecules goes without question, even if its abundance is reduced due to freeze-out onto ice grains, and the other two gas phases prevent CO from being heavily biased by the dense conditions; the colocation of HCN and HCO$^+$ is less concrete. While HCN and HCO$^+$ are almost certainly colocated, HCO$^+$ is quite possibly more extended, tracing warmer and less UV shielded regions \citep{Fuente2005}. On the other hand, it is not expected to be found in the very diffuse and extended CO gas phase we use here. Unfortunately, while an additional phase would be desirable we would not be able to place meaningful constraints upon the additional phase due to the paucity of lines. Two phase HCN and HCO$^+$ models (Appendix \ref{subsubsec:twophase}) suggest that the contribution of HCO$^+$ in a slightly hotter and less dense phase (physically this would most likely correspond to the transition region between the hot, diffuse phase and the cold, dense cores) is less than $10\%$ of any HCO$^+$ line. We believe that the colocation of HCN and HCO$^+$ in the cold, dense phase is therefore an undesirable but necessary, and acceptable, compromise.
		
	\subsection{The [$^{12}$C]/[$^{13}$C] Abundance Ratio in NGC\,6240}
	
	\noindent \citet{Papadopoulos2014} constrained the high$-J$ CO lines using the LVG solutions for HCN and HCO$^+$ before fitting a third phase to the remaining CO lines. When they then applied these three phases to their $^{13}$CO data, a very high [CO]/[$^{13}$CO] abundance ratio of $300-500$ was required for consistency and they argued for that this was evidence for a similarly elevated [$^{12}$C]/[$^{13}$C] ratio. Here, we have used our new observations of H$^{13}$CN and H$^{13}$CO$^+$ in combination with the CO and $^{13}$CO lines from the literature to place new constraints on the [$^{12}$C]/[$^{13}$C] abundance ratio. We have found that instead of arising in the HCN and HCO$^+$ bearing dense gas, the high$-J$ CO lines are due to the hot, shocked gas phase, which in turn changes the implied [CO]/[$^{13}$CO] ratios.
	
	A na\"{i}ve approach, simply taking the 68\% credible intervals on [CO]/[$^{13}$CO] and [HCN]/[H$^{13}$CN] as the brackets on the [$^{12}$C]/[$^{13}$C] ratio, places the ratio somewhere between 100 and 1000: however, this does not account for the much greater abundance of CO wrt HCN nor the dominance by mass of the dense phase. 
	
\begin{figure}
	\centering
	\includegraphics[width=0.495\textwidth]{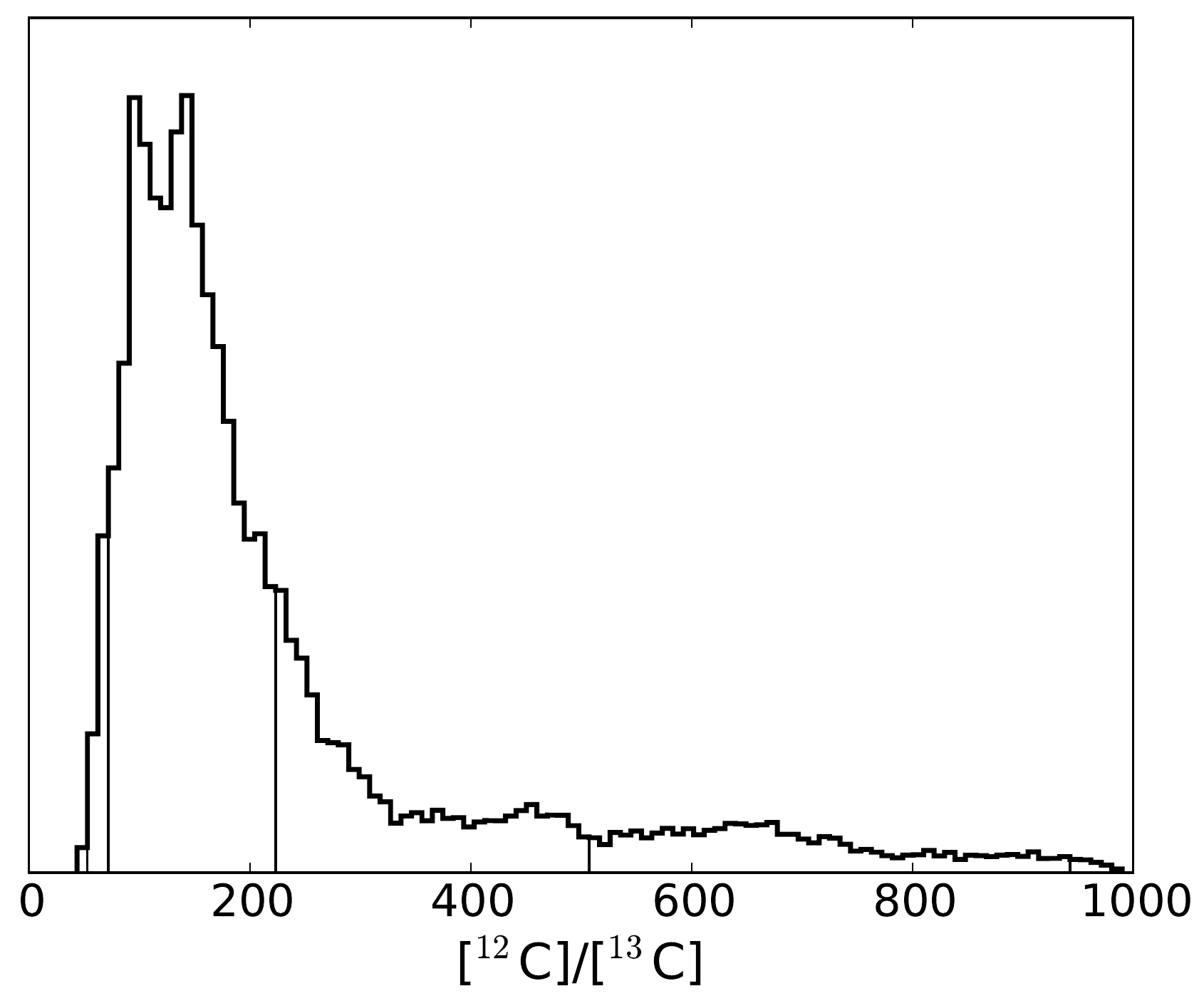}
	\caption{The pdf for the [$^{12}$C]/[$^{13}$C] ratio in NGC\,6240, derived from the three phase model in Section \ref{subsubsec:12c13c} using a mass and abundance weighted average. The ratio shows signs of being slightly elevated, but the upper 68\% credible interval is less than half the value found by \citet{Papadopoulos2014} and the pdf is most consistent with the ``normal'' ULIRG abundance ratio.}\label{fig:c12c13_wratio}
	\end{figure}

	We adopt a mass and abundance weighted average and find [$^{12}$C]/[$^{13}$C] = $98^{230}_{65}$. The pdf of the ratio is shown in Figure \ref{fig:c12c13_wratio}. The ratio is strongly peaked at ``normal'' ULIRG values of about 100, although there is an extended tail out to 1000. The model is compatible within $-1\sigma$ with a Galactic value, while the $+3\sigma$ boundary is at 550. We therefore argue against a super-ULIRG elevated [$^{12}$C]/[$^{13}$C] abundance ratio, but cannot conclusively exclude the range found by \citet{Papadopoulos2014}. The ideal solution, a full chemical-hydrodynamical model with radiation transfer, is beyond the scope of this paper. Nevertheless, it appears that the [$^{12}$C]/[$^{13}$C] ratio in NGC\,6240 is most similar to other ULIRGs.
	
	\bigskip
	
	\noindent Are the derived isotopologue abundance ratios consistent with the chemical models presented in Section \ref{subsec:chem}? Given the wide range of results from the chemical models, and their strong dependences on temperature, density, metallicity, the C/O ratio and the H$_2$ ortho-para ratio it is hard to say with certainty. We find [HCN]/[H$^{13}$CN] $\sim 2\times$ [CO]/[$^{13}$CO] in the dense and diffuse gas phases, consistent with \citet{Roueff2015}. The rather high [HCO$^+$]/[H$^{13}$CO$^+$] (which in the three phase model is fixed to be the same as [HCN]/[H$^{13}$CN]) is surprising, but not so extreme as to cause concern; \citet{Langer1984} found [HCO$^+$]/[H$^{13}$CO$^+$] $\sim 1.5\times$ [CO]/[$^{13}$CO] in 10\,K, $10^4$\,cm$^{-3}$, low metallicity gas. Furthermore, since this ratio is fixed to [HCN]/[H$^{13}$CN] this could be artificially elevating it (see Figure \ref{fig:fixXfree13C} in the Appendix).

	The [CO]/[$^{13}$CO] ratio appears to increase from $40^{80}_{30}$ in the hot phase to $160^{630}_{40}$ in the cold, dense phase. There is slight tension between these results and our interpretation of the model results as being evidence for ICE: in the case of ICE fractionation in the cold, dense phase should lower the [CO]/[$^{13}$CO] ratio; it is therefore contradictory if the hot, shocked phase, where there is no isotope fractionation, presents a lower [CO]/[$^{13}$CO] ratio than the dense phase. However, there is still a region of parameter space within the 68\% credible interval about the mean values where the [CO]/[$^{13}$CO] ratio is lower in the dense phase, so while there is tension it is not that great, and the model is still consistent with isotope fractionation. Furthermore, the hot phase [CO]/[$^{13}$CO] is almost entirely determined by the highly uncertain $^{13}$CO $J=6-5$ line, while the dense phase CO and $^{13}$CO emission is almost zero (Figure \ref{fig:3phaseCOSLED}), so that the [CO]/[$^{13}$CO] ratio is highly susceptible to small errors. The [CO]/[$^{13}$CO] vs $T_{\rm k}$ panel of Figure \ref{fig:3phaseCO} shows both that there is a large spread in all of the [CO]/[$^{13}$CO] values, and that there is a very large overlapping region within the 68\% credible intervals. Interestingly, within this spread in [CO]/[$^{13}$CO] all three phases are consistent with a value of $\sim80$, closer to the $\sim70$ of the Milky Way than to a globally elevated elemental [$^{12}$C]/[$^{13}$C] ratio.
	
	\bigskip
	
	\noindent It is possible that due to insufficient UV-plane sampling we have lost some of the continuum and line fluxes, as was mentioned in Section \ref{subsec:continuum}. The effect of this would be to bias our [HCN]/[H$^{13}$CN] and [HCO$^+$]/[H$^{13}$CO$^+$] to higher values; our finding that the elemental [$^{12}$C]/[$^{13}$C] ratio is not globally elevated in NGC\,6240 is therefore robust against this potential source of error, as the error would push the results towards an elevated value.

	\subsection{Line Luminosity to Gas Mass Conversion Factors}\label{subsec:alphas}	
	
	\noindent The ratio between line luminosity and gas mass, $\alpha$, is of great interest as it provides a means of estimating galaxy gas fractions from observations of a single molecular line, usually CO. However, the determination of a general $\alpha_{\rm CO}$ is particularly complicated, with dependences on luminosity \citep{Solomon1987} and metallicity, leading to a range of values from 0.3 to almost 300, and a typical Milky Way value of 4.8 dropping to about 1 in $z=2$ SMGs and 0.8 in (U)LIRGs\footnote{\citet{Downes1998} found a range for $\alpha_{\rm CO}$ from 0.3 to 1.3 in ULIRGs.} \citep{Downes1998,Bolatto2013}. Nevertheless, it has recently been argued that $\alpha_{\rm CO}$ may be underestimated in (U)LIRGs, and a Galactic value may be more accurate \citep{Papadopoulos2012}, due to gas over-densities arising from over-pressurisation due to turbulence.
	
	There is interest in $\alpha_{\rm HCN}$ as a possible estimator of the \emph{dense} molecular gas mass, which may be more robust over a wide range of luminosities and may more accurately trace the specifically star forming gas \citep{Gao2004,Greve2009,GarciaBurillo2012}. This in turn has been complicated by high resolution studies of ULIRGs which suggest that HCN emission may be dominated by shocked, X-ray heated or infra-red pumped emission \citep{Aalto2007,Sakamoto2010,Aalto2012,GarciaBurillo2014,Viti2014,Aalto2015,Martin2015,Tunnard2015}.
	
	We calculate $\alpha_{\rm mol}$ and corresponding mass pdfs for each gas phase. The values for our three phase model are shown in Figure \ref{fig:3phase_alpha} where we plot $\alpha_{\rm CO}$ for each of the three gas phases, and $\alpha_{\rm HCN}$ for the dense phase, as well as the corresponding masses. The $\alpha_{\rm CO}$ pdfs are broad, with spans of about $1.5-2$ dex. The diffuse and shocked phases both cover the range $\alpha_{\rm CO}=1-10$, while the diffuse phase extends down to 0.1. The dense phase $\alpha_{\rm CO}$ is much higher, due to a lower $X_{\rm CO}$, higher $n_{{\rm H}_2}$ and higher d$v/$d$r$ in this phase, all of which contribute to an increased $\alpha_{\rm CO}$. Despite contributing $<5\%$ of the CO$(1-0)$ line flux, the dense phase accounts for almost all of the molecular gas mass. The dense phase gas masses derived using HCN and using CO are consistent but for an unexplained systematic offset of $-0.2$\,dex on the CO mass estimate. $\alpha_{\rm HCN} = 32^{89}_{13}$: higher than the canonical value of 10 but similar to the $13-31.1$ found by \citet{Papadopoulos2014}, although some regions of their Figure 14 do extend to $\alpha_{\rm HCN}=60$. \citet{Gao2004} predicted $\alpha_{\rm HCN}\sim25$ for $X_{\rm HCN}=2\times10^{-8}$ and d$v/$d$r=5$\,km\,s$^{-1}$\,pc$^{-1}$: our lower $X_{\rm HCN}$ and higher d$v/$d$r$ are responsible for our slightly higher $\alpha_{\rm HCN}$. While the $+1\sigma$ level is very high, the best fit and $-1\sigma$ level are consistent with \citet{Greve2009}, who found $\alpha_{\rm HCN} = 17-37$ in NGC\,6240.
	
	\bigskip
	
	\begin{figure}
	\centering
	\includegraphics[width=0.495\textwidth]{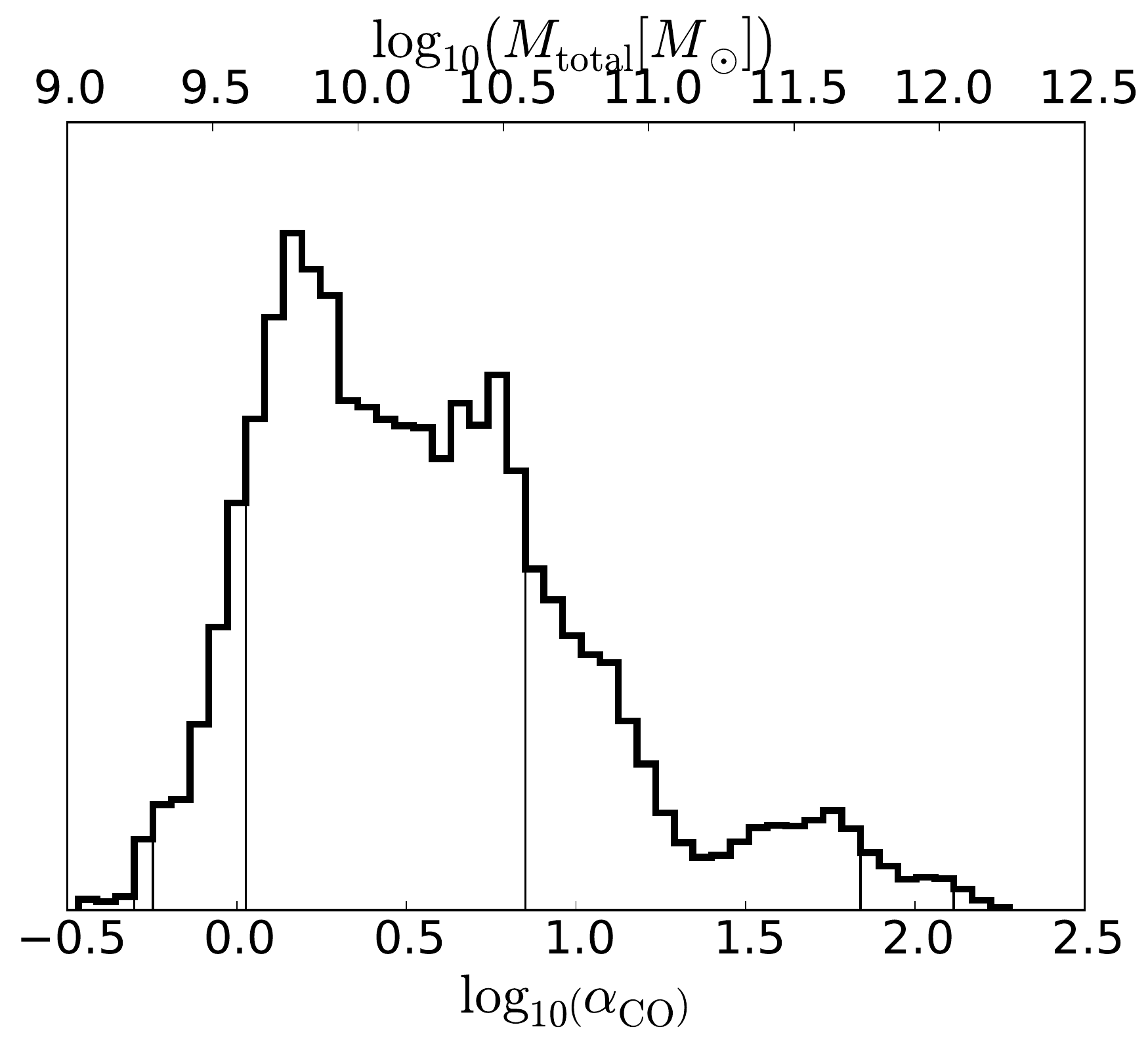}
	\caption{The global $\alpha_{\rm CO}$ for NGC\,6240 derived from the three phase model with the corresponding total molecular gas mass.}\label{fig:3ptotal_alpha}
	\end{figure}

	\noindent We derive a global $\alpha_{\rm CO}$ by calculating the sum of the masses of each gas phase, and dividing by the total CO$(1-0)$ line luminosity. The resultant $\alpha_{\rm CO}$ pdf and the corresponding mass pdf for NGC\,6240 is shown in Figure \ref{fig:3ptotal_alpha}. This global $\alpha_{\rm CO}$ is much more tightly constrained than the those of the individual phases, and we find $\alpha_{\rm CO}=1.5^{7.1}_{1.1}$ $(\log_{10}\left(M/[M_\odot]\right) = 10.1_{10.0}^{10.8})$: consistent with a Galactic $\alpha_{\rm CO}$, as predicted by \citet{Papadopoulos2012}, but perhaps more consistent with the canonical $\alpha_{\rm CO}$ of starbursts \citep[0.8,][]{Downes1998}. Notably, the very low (CO) luminosity dense phase contributes the majority of the gas mass.
	
	The high $\alpha_{\rm HCN}$ we derive is contrary to the findings of \citet{GarciaBurillo2012}, who argued for a lower $\alpha_{\rm HCN}$ in (U)LIRGs, matching the lower $\alpha_{\rm CO}$. It is very possible that this contradiction is simply due to the extremely unusual properties of NGC\,6240; NGC\,6240 is perhaps not even representative of other (U)LIRGs, existing in a class of its own.

	\subsection{Dense Gas and Star Formation in NGC\,6240}\label{subsec:densegasSF}

	\noindent The results of our LVG models present an intriguing picture of the dense molecular gas in NGC\,6240. We appear to be finding large quantities of very cold $(\sim 8$\,K) gas within the internuclear molecular disk, which is also host to extensively shocked and very hot $(\sim 2000$\,K) molecular gas. The question is how can such a large quantity of cold, dense gas coexist with the hot phase. Furthermore, star formation in NGC\,6240 is not found in the internuclear molecular disk; it is instead concentrated in the two nuclei. If we truly have almost $10^{10}\,M_\odot$ of cold, dense gas then we must be able to explain why this gas is not actively forming stars.
	
	An insight is offered if we apply an upper limit to the H$_2$ column density of the dense phase. There is an inherent degeneracy within the LVG model between a single, large cloud and multiple, small clouds distributed in velocity space across the large scale velocity field of the galaxy. However, for the dense gas phase a single cloud implies column densities in excess of $10^{26}$\,cm$^{-2}$, which are extremely unlikely. If instead we assume that the column densities of the clouds are $\simeq10^{23}$\,cm$^{-2}$ then we are instead seeing multiple small ($0.1 - 0.01$\,pc) dense clouds.
	
	\smallskip
	
	\noindent We therefore propose the following situation. Within the turbulent internuclear disk instabilities lead to rapid, localised cooling of pockets of gas, leading to multiple small, cold, dense clouds embedded in pressure equilibrium with the hot, diffuse phase but radiatively decoupled due to turbulent Doppler shifting of the lines. This rapid cooling leads to an evolving chemistry, potentially with molecules less stable than CO destroyed in the high temperatures of the shocks reforming as the gas cools. These clouds are still turbulent however, and may indeed be transient, explaining the lack of star formation. 
	
	\smallskip
	
	\noindent In the three phase model the CO abundance is significantly lower in the cold, dense phase than in the hot phase or diffuse phase: we have almost a continuum of $X_{\rm CO}$ from the the diffuse phase value of $\sim 10^{-4}$, falling to $\sim 10^{-5}$ in the shocked phase and then to $\sim 10^{-6}$ in the dense phase. This is consistent with the many cold clouds interpretation above, with freeze out of CO onto dust grain mantles depleting the gas phase CO in cold, dense clouds - as observed in Milky Way cold cores \citep{Kramer1999,Tafalla2002,Liu2013,Ripple2013}. This might also explain the very strangely hot and diffuse CS model solutions from the precursor models (Table \ref{tab:modelsResults}): CS is depleted along with CO \citep{Tafalla2002}, so the only CS we observe is in the hot gas phase. Unfortunately this is highly speculative, and cannot be confirmed without much higher spatial resolution observations able to resolve the high density molecular clouds.

    We note that with $n_{{\rm H}_2}\sim1\times10^5$, a uniform density profile and spherical gas distribution the dynamical mass implies an upper limit on the dense gas volume filling factor of 0.002 within the central 600\,pc of NGC\,6240. This is consistent with the image, presented above, of the molecular disk being almost entirely filled with hot, shocked gas, but with the majority (by mass) of the molecular gas surviving in cold, dense clouds embedded within the hot phase. 
	
	\bigskip
	
	\noindent Since the majority of the dense gas and the ongoing star formation are separate in NGC\,6240 \citep{Tacconi1999} we cannot calculate reliable local gas depletion times and instead we explore the galaxy averaged gas depletion time. Combining the star formation rate (SFR) of $60\pm30$\,$M_\odot$\,yr$^{-1}$ \citep{Yun2002,Feruglio2013a} with our $\alpha_{{\rm HCN}(1-0)}$ and $L'_{{\rm HCN}(1-0)}$ we can obtain estimates of the gas depletion time in NGC\,6240, and we find a depletion time $6_4^{13}\times 10^8$\,yr.
	
	However, the dynamical mass in the central 600\,pc of $\sim10^{10}\,M_\odot$ is significantly less than the dense gas mass implied by HCN and the dense phase CO. If we exclude results predicting gas masses in excess of the dynamical mass the depletion time is reduced to $0.9_{0.6}^{1.9}\times10^8$\,yr.
    
     We do however question whether the dynamical mass is in fact particularly meaningful in the central regions of NGC\,6240. This galaxy is an ongoing major merger, so it is not clear that the velocity dispersion should actually correlate with the mass. Estimating the mass from the velocity dispersion assumes either virialisation or stable orbital rotation, neither of which is necessarily true for the central gas in NGC\,6240, which is dissipating turbulent energy from the merger \citep{Tacconi1999} and is most likely a transient feature itself. It is very possible that this explains the super-dynamical masses suggested by some regions of our results and a large region of the parameter space in \citet{Papadopoulos2014}.

	\subsection{Implications of the [HCN]/[HCO$^+$] Abundance Ratio}
    
    \noindent We have found [HCN]/[HCO$^+$] = $6.2^{7.9}_{4.3}$ in NGC\,6240. This compares favourably to the value of ``around 10'' found by \citet{Krips2008}. NGC\,6240 is frequently compared to Arp\,220, where \citet{Tunnard2015} found [HCN]/[HCO$^+$] $>18$ in the western nucleus (WN) and [HCN]/[HCO$^+$] $0.5-5$ in the eastern nucleus (EN). The abundance ratio for NGC\,6240 appears to be most consistent with the less luminous EN, which is dominated by star formation (whereas Arp\,220 WN may contain an energetically significant AGN). 
    
    However, in NGC\,6240 the comparison is probably less meaningful. Unlike in most of the galaxies of \citet{Krips2008}, the majority of the molecular gas is spatially separated from the starbursts and putative AGNs in the two nuclei. As \citet{Krips2008} pointed out, the extreme conditions in NGC\,6240 make it quite unique, and the abundance ratios may not carry the same meaning as they do in other galaxies and Arp\,220 (which may itself be in a subset of galaxies possessing Compact Obscured Nuclei (CONs), which generate unique chemistries \citealp{Costagliola2015}). In the context of \citet{Krips2008}, our results place NGC\,6240 in the intermediate region between a starburst and AGN dominated galaxy. The derived $n_{{\rm H}_2}$ is higher than the usual $<10^{4.5}$\,cm$^{-3}$ for AGNs, while the [HCN]/[HCO$^+$] ratio is higher than the $0.01-1$ of starburst but lower than the $>10$ of AGN. Given the separation of the starburst and AGN nuclei from the bulk of the molecular gas (which lies between the two nuclei) the fact that NGC\,6240 does not fit clearly into any specific category is not surprising.

\section{Conclusions}\label{sec:conclusions}

\noindent We have used new observations of the $^{13}$C isotopologue lines of HCN and HCO$^+$ in NGC\,6240 to model the physical conditions and abundance ratios of the molecular gas. We have:

\begin{itemize}
\item presented the first observations of SiO$(2-1)$ and of the isotopologue line H$^{13}$CN$(1-0)$ in NGC\,6240, and an upper limit on H$^{13}$CO$^+(1-0)$. 

\item combined these observations with literature data as inputs for an MCMC wrapper for RADEX, modelling three gas phases with six species simultaneously to account for shocked molecular gas, extended diffuse molecular gas and cold, dense molecular gas, to find the conditions of the gas in NGC\,6240. These models suggest that the cold, dense gas exists in multiple, small (0.01-0.1\,pc) clouds embedded in the hot, shocked inter-nuclear gas disk. The cold gas is particularly cold with $T_{\rm k}=13^{20}_{10}$, while the hot phase reaches temperatures of almost 2000\,K.

\item used our MCMC code and the new $^{13}$C observations to demonstrate that the very high [$^{12}$C]/[$^{13}$C] ratio in NGC\,6240 found by \citet{Papadopoulos2014} was most likely due to assuming that the high$-J$ CO lines originate in the dense gas traced by HCN, whereas in NGC\,6240 they originate in shocks \citep{Meijerink2013} and the HCN lines are cold, dense gas dominated, isolated by using our new H$^{13}$CN observations. Our models suggest instead that [$^{12}$C]/[$^{13}$C] $=98^{230}_{65}$ and is ``standard'' for ULIRGs, and even consistent with the Milky Way average value of 68. The high [HCN]/[H$^{13}$CN] and [HCO$^+$]/[H$^{13}$CO$^+$] ratios $(300^{500}_{200})$ are due to isotope fractionation in cold, dense gas, with the [HCO$^+$]/[H$^{13}$CO$^+$] ratio being particularly uncertain.

\item derived $\alpha_{\rm CO}$ values for the diffuse, shocked and dense gas phases in NGC\,6240, and combined these to produce a global $\alpha_{\rm CO}=1.5^{7.1}_{1.1}$, in between the ULIRG value of 0.8 and 4.2 appropriate for Milky Way GMCs. We also calculate $\alpha_{\rm HCN} = 32^{89}_{13}$, where $\alpha_{\rm HCN}$ is tracing the cold, dense gas.

\bigskip

\noindent These results demonstrate the extreme value of isotopologue lines in LVG modelling, with their powerful ability to break degeneracies even when the isotopologue abundance ratios are unknown. Similarly, we have highlighted the dangers of fitting multiple species gas phase components by eye, as was quantified for multiple CO gas phases by \citet{Kamenetzky2014}. 

Our model produces chemically feasible molecular abundances and isotopologue abundance ratios, despite the significant simplifications. Nevertheless, the high [HCO$^+$]/[H$^{13}$CO$^+$] ratio is unexpected, and warrants further investigation. A distinct possibility, evidenced by our slightly low continuum flux density, is that we are missing a significant fraction of the molecular line flux. This would present as elevated [HCN]/[H$^{13}$CN] and [HCO$^+$]/[H$^{13}$CO$^+$] abundance ratios. Importantly, even if this is the case the main conclusions of this work, that the elemental [$^{12}$C]/[$^{13}$C] ratio is not elevated in NGC\,6240, are not affected; they are instead reinforced if we are missing line flux. 

In any case, further observations of $^{13}$CO, H$^{13}$CN and H$^{13}$CO$^+$ are warranted, both in NGC\,6240 and other galaxies with CO, HCN and HCO$^+$ SLEDs. These isotopologue lines provide added value and, as we have demonstrated here, even a single line can simultaneously improve temperature constraints while constraining the isotopologue abundance ratio, without the need to assume a, potentially erroneous, ratio.

\end{itemize}
\FloatBarrier
\acknowledgements 
{\small Acknowledgements: RT would like to thank the organisers and sponsors of the ICIC Data Analysis Workshop. This research is supported by an STFC PhD studentship. TRG acknowledges support from an STFC Advanced Fellowship. SGB acknowledges support from Spanish grants AYA2010-15169 and AYA2012-32295 and from the Junta de Andalucia through TIC-114 and the Excellence Project P08-TIC-03531. AU acknowledges support from Spanish grants AYA2012-32295 and FIS2012-32096. AF and SGB acknowledge support from Spanish MICIN program CONSOLIDER IMAGENIO 2010 under grant ASTROMOL (ref. CSD2009-00038). We thank the anonymous referee for their constructive comments, which greatly helped in improving the structure and readability of this paper. This work was based on observations carried out with the IRAM PdBI, supported by INSU/CNRS(France), MPG(Germany), and IGN(Spain).}

\bibliography{ngc6240.bib}

\begin{thebibliography}{}
\expandafter\ifx\csname natexlab\endcsname\relax\def\natexlab#1{#1}\fi

\bibitem[{{Aalto} {et~al.}(2012){Aalto}, {Garcia-Burillo}, {Muller}, {Winters},
  {van der Werf}, {Henkel}, {Costagliola}, \& {Neri}}]{Aalto2012}
{Aalto}, S., {Garcia-Burillo}, S., {Muller}, S., {et~al.} 2012, \aap, 537, A44

\bibitem[{{Aalto} {et~al.}(2002){Aalto}, {Polatidis}, {H{\"u}ttemeister}, \&
  {Curran}}]{Aalto2002}
{Aalto}, S., {Polatidis}, A.~G., {H{\"u}ttemeister}, S., \& {Curran}, S.~J.
  2002, \aap, 381, 783

\bibitem[{{Aalto} {et~al.}(2007){Aalto}, {Spaans}, {Wiedner}, \&
  {H{\"u}ttemeister}}]{Aalto2007}
{Aalto}, S., {Spaans}, M., {Wiedner}, M.~C., \& {H{\"u}ttemeister}, S. 2007,
  \aap, 464, 193

\bibitem[{{Aalto} {et~al.}(2015){Aalto}, {Garcia-Burillo}, {Muller}, {Winters},
  {Gonzalez-Alfonso}, {van der Werf}, {Henkel}, {Costagliola}, \&
  {Neri}}]{Aalto2015}
{Aalto}, S., {Garcia-Burillo}, S., {Muller}, S., {et~al.} 2015, \aap, 574, A85

\bibitem[{{Armus} {et~al.}(2006){Armus}, {Bernard-Salas}, {Spoon}, {Marshall},
  {Charmandaris}, {Higdon}, {Desai}, {Hao}, {Teplitz}, {Devost}, {Brandl},
  {Soifer}, \& {Houck}}]{Armus2006}
{Armus}, L., {Bernard-Salas}, J., {Spoon}, H.~W.~W., {et~al.} 2006, \apj, 640,
  204

\bibitem[{{Boger} \& {Sternberg}(2005)}]{Boger2005}
{Boger}, G.~I., \& {Sternberg}, A. 2005, \apj, 632, 302

\bibitem[{{Bolatto} {et~al.}(2013){Bolatto}, {Wolfire}, \&
  {Leroy}}]{Bolatto2013}
{Bolatto}, A.~D., {Wolfire}, M., \& {Leroy}, A.~K. 2013, \araa, 51, 207

\bibitem[{{Costagliola} {et~al.}(2015){Costagliola}, {Sakamoto}, {Muller},
  {Mart{\'{\i}}n}, {Aalto}, {Harada}, {van der Werf}, {Viti}, {Garcia-Burillo},
  \& {Spaans}}]{Costagliola2015}
{Costagliola}, F., {Sakamoto}, K., {Muller}, S., {et~al.} 2015, ArXiv e-prints,
  arXiv:1506.09027

\bibitem[{{Crane} \& {Hegyi}(1988)}]{Crane1988}
{Crane}, P., \& {Hegyi}, D.~J. 1988, \apjl, 326, L35

\bibitem[{{Crane} {et~al.}(1991){Crane}, {Hegyi}, \& {Lambert}}]{Crane1991}
{Crane}, P., {Hegyi}, D.~J., \& {Lambert}, D.~L. 1991, \apj, 378, 181

\bibitem[{{Dayou} \& {Balan{\c c}a}(2006)}]{Dayou2006}
{Dayou}, F., \& {Balan{\c c}a}, C. 2006, \aap, 459, 297

\bibitem[{{Dopita} {et~al.}(2005){Dopita}, {Groves}, {Fischera}, {Sutherland},
  {Tuffs}, {Popescu}, {Kewley}, {Reuland}, \& {Leitherer}}]{Dopita2005}
{Dopita}, M.~A., {Groves}, B.~A., {Fischera}, J., {et~al.} 2005, \apj, 619, 755

\bibitem[{{Downes} \& {Solomon}(1998)}]{Downes1998}
{Downes}, D., \& {Solomon}, P.~M. 1998, \apj, 507, 615

\bibitem[{{Dumouchel} {et~al.}(2010){Dumouchel}, {Faure}, \&
  {Lique}}]{Dumouchel2010}
{Dumouchel}, F., {Faure}, A., \& {Lique}, F. 2010, \mnras, 406, 2488

\bibitem[{{Elitzur} \& {Watson}(1978)}]{Elitzur1978}
{Elitzur}, M., \& {Watson}, W.~D. 1978, \apjl, 222, L141

\bibitem[{{Elitzur} \& {Watson}(1980)}]{Elitzur1980}
---. 1980, \apj, 236, 172

\bibitem[{{Feruglio} {et~al.}(2013){Feruglio}, {Fiore}, {Maiolino},
  {Piconcelli}, {Aussel}, {Elbaz}, {Le Floc'h}, {Sturm}, {Davies}, \&
  {Cicone}}]{Feruglio2013a}
{Feruglio}, C., {Fiore}, F., {Maiolino}, R., {et~al.} 2013, \aap, 549, A51

\bibitem[{{Flower}(1999)}]{Flower1999}
{Flower}, D.~R. 1999, \mnras, 305, 651

\bibitem[{{Fuente} {et~al.}(2005){Fuente}, {Garc{\'{\i}}a-Burillo}, {Gerin},
  {Teyssier}, {Usero}, {Rizzo}, \& {de Vicente}}]{Fuente2005}
{Fuente}, A., {Garc{\'{\i}}a-Burillo}, S., {Gerin}, M., {et~al.} 2005, \apjl,
  619, L155

\bibitem[{{Fuente} {et~al.}(1995){Fuente}, {Martin-Pintado}, \&
  {Gaume}}]{Fuente1995}
{Fuente}, A., {Martin-Pintado}, J., \& {Gaume}, R. 1995, \apjl, 442, L33

\bibitem[{{Gao} \& {Solomon}(2004)}]{Gao2004}
{Gao}, Y., \& {Solomon}, P.~M. 2004, \apjs, 152, 63

\bibitem[{{Garc{\'{\i}}a-Burillo} {et~al.}(2012){Garc{\'{\i}}a-Burillo},
  {Usero}, {Alonso-Herrero}, {Graci{\'a}-Carpio}, {Pereira-Santaella},
  {Colina}, {Planesas}, \& {Arribas}}]{GarciaBurillo2012}
{Garc{\'{\i}}a-Burillo}, S., {Usero}, A., {Alonso-Herrero}, A., {et~al.} 2012,
  \aap, 539, A8

\bibitem[{{Garc{\'{\i}}a-Burillo} {et~al.}(2014){Garc{\'{\i}}a-Burillo},
  {Combes}, {Usero}, {Aalto}, {Krips}, {Viti}, {Alonso-Herrero}, {Hunt},
  {Schinnerer}, {Baker}, {Boone}, {Casasola}, {Colina}, {Costagliola},
  {Eckart}, {Fuente}, {Henkel}, {Labiano}, {Mart{\'{\i}}n}, {M{\'a}rquez},
  {Muller}, {Planesas}, {Ramos Almeida}, {Spaans}, {Tacconi}, \& {van der
  Werf}}]{GarciaBurillo2014}
{Garc{\'{\i}}a-Burillo}, S., {Combes}, F., {Usero}, A., {et~al.} 2014, \aap,
  567, A125

\bibitem[{{Goldsmith} \& {Langer}(1999)}]{Goldsmith1999}
{Goldsmith}, P.~F., \& {Langer}, W.~D. 1999, \apj, 517, 209

\bibitem[{{Graci{\'a}-Carpio} {et~al.}(2008){Graci{\'a}-Carpio},
  {Garc{\'{\i}}a-Burillo}, {Planesas}, {Fuente}, \& {Usero}}]{GraciaCarpio2008}
{Graci{\'a}-Carpio}, J., {Garc{\'{\i}}a-Burillo}, S., {Planesas}, P., {Fuente},
  A., \& {Usero}, A. 2008, \aap, 479, 703

\bibitem[{{Greaves} \& {Church}(1996)}]{Greaves1996}
{Greaves}, J.~S., \& {Church}, S.~E. 1996, \mnras, 283, 1179

\bibitem[{{Green} \& {Thaddeus}(1974)}]{Green1974}
{Green}, S., \& {Thaddeus}, P. 1974, \apj, 191, 653

\bibitem[{{Greve} {et~al.}(2009){Greve}, {Papadopoulos}, {Gao}, \&
  {Radford}}]{Greve2009}
{Greve}, T.~R., {Papadopoulos}, P.~P., {Gao}, Y., \& {Radford}, S.~J.~E. 2009,
  \apj, 692, 1432

\bibitem[{{Greve} {et~al.}(2014){Greve}, {Leonidaki}, {Xilouris}, {Wei{\ss}},
  {Zhang}, {van der Werf}, {Aalto}, {Armus}, {D{\'{\i}}az-Santos}, {Evans},
  {Fischer}, {Gao}, {Gonz{\'a}lez-Alfonso}, {Harris}, {Henkel}, {Meijerink},
  {Naylor}, {Smith}, {Spaans}, {Stacey}, {Veilleux}, \& {Walter}}]{Greve2014}
{Greve}, T.~R., {Leonidaki}, I., {Xilouris}, E.~M., {et~al.} 2014, \apj, 794,
  142

\bibitem[{{Hagiwara} {et~al.}(2011){Hagiwara}, {Baan}, \&
  {Kl{\"o}ckner}}]{Hagiwara2011}
{Hagiwara}, Y., {Baan}, W.~A., \& {Kl{\"o}ckner}, H.-R. 2011, \aj, 142, 17

\bibitem[{Hastings(1970)}]{hastings1970}
Hastings, W.~K. 1970, Biometrika, 57, 97

\bibitem[{{Henkel} {et~al.}(2010){Henkel}, {Downes}, {Wei{\ss}}, {Riechers}, \&
  {Walter}}]{Henkel2010}
{Henkel}, C., {Downes}, D., {Wei{\ss}}, A., {Riechers}, D., \& {Walter}, F.
  2010, \aap, 516, A111

\bibitem[{{Henkel} {et~al.}(2014){Henkel}, {Asiri}, {Ao}, {Aalto}, {Danielson},
  {Papadopoulos}, {Garc{\'{\i}}a-Burillo}, {Aladro}, {Impellizzeri},
  {Mauersberger}, {Mart{\'{\i}}n}, \& {Harada}}]{Henkel2014}
{Henkel}, C., {Asiri}, H., {Ao}, Y., {et~al.} 2014, \aap, 565, A3

\bibitem[{{Iono} {et~al.}(2007){Iono}, {Wilson}, {Takakuwa}, {Yun}, {Petitpas},
  {Peck}, {Ho}, {Matsushita}, {Pihlstrom}, \& {Wang}}]{Iono2007}
{Iono}, D., {Wilson}, C.~D., {Takakuwa}, S., {et~al.} 2007, \apj, 659, 283

\bibitem[{{Kamenetzky} {et~al.}(2014){Kamenetzky}, {Rangwala}, {Glenn},
  {Maloney}, \& {Conley}}]{Kamenetzky2014}
{Kamenetzky}, J., {Rangwala}, N., {Glenn}, J., {Maloney}, P.~R., \& {Conley},
  A. 2014, \apj, 795, 174

\bibitem[{{Komossa} {et~al.}(2003){Komossa}, {Burwitz}, {Hasinger}, {Predehl},
  {Kaastra}, \& {Ikebe}}]{Komossa2003}
{Komossa}, S., {Burwitz}, V., {Hasinger}, G., {et~al.} 2003, \apjl, 582, L15

\bibitem[{{Kramer} {et~al.}(1999){Kramer}, {Alves}, {Lada}, {Lada}, {Sievers},
  {Ungerechts}, \& {Walmsley}}]{Kramer1999}
{Kramer}, C., {Alves}, J., {Lada}, C.~J., {et~al.} 1999, \aap, 342, 257

\bibitem[{{Krips} {et~al.}(2008){Krips}, {Neri}, {Garc{\'{\i}}a-Burillo},
  {Mart{\'{\i}}n}, {Combes}, {Graci{\'a}-Carpio}, \& {Eckart}}]{Krips2008}
{Krips}, M., {Neri}, R., {Garc{\'{\i}}a-Burillo}, S., {et~al.} 2008, \apj, 677,
  262

\bibitem[{{Lada}(1992)}]{Lada1992}
{Lada}, E.~A. 1992, \apjl, 393, L25

\bibitem[{{Lambert} {et~al.}(1994){Lambert}, {Sheffer}, {Gilliland}, \&
  {Federman}}]{Lambert1994}
{Lambert}, D.~L., {Sheffer}, Y., {Gilliland}, R.~L., \& {Federman}, S.~R. 1994,
  \apj, 420, 756

\bibitem[{{Langer} {et~al.}(1984){Langer}, {Graedel}, {Frerking}, \&
  {Armentrout}}]{Langer1984}
{Langer}, W.~D., {Graedel}, T.~E., {Frerking}, M.~A., \& {Armentrout}, P.~B.
  1984, \apj, 277, 581

\bibitem[{{Langer} {et~al.}(1978){Langer}, {Wilson}, {Henry}, \&
  {Guelin}}]{Langer1978}
{Langer}, W.~D., {Wilson}, R.~W., {Henry}, P.~S., \& {Guelin}, M. 1978, \apjl,
  225, L139

\bibitem[{{Lavendy} {et~al.}(1987){Lavendy}, {Robbe}, \&
  {Gandara}}]{Lavendy1987}
{Lavendy}, H., {Robbe}, J.~M., \& {Gandara}, G. 1987, Journal of Physics B
  Atomic Molecular Physics, 20, 3067

\bibitem[{{Lique} {et~al.}(2006){Lique}, {Spielfiedel}, \&
  {Cernicharo}}]{Lique2006}
{Lique}, F., {Spielfiedel}, A., \& {Cernicharo}, J. 2006, \aap, 451, 1125

\bibitem[{{Liu} {et~al.}(2013){Liu}, {Wu}, \& {Zhang}}]{Liu2013}
{Liu}, T., {Wu}, Y., \& {Zhang}, H. 2013, \apjl, 775, L2

\bibitem[{{Lohr}(1998)}]{Lohr1998}
{Lohr}, L.~L. 1998, \jcp, 108, 8012

\bibitem[{{Loison} {et~al.}(2014){Loison}, {Wakelam}, \&
  {Hickson}}]{Loison2014}
{Loison}, J.-C., {Wakelam}, V., \& {Hickson}, K.~M. 2014, \mnras, 443, 398

\bibitem[{{Lutz} {et~al.}(2003){Lutz}, {Sturm}, {Genzel}, {Spoon}, {Moorwood},
  {Netzer}, \& {Sternberg}}]{Lutz2003}
{Lutz}, D., {Sturm}, E., {Genzel}, R., {et~al.} 2003, \aap, 409, 867

\bibitem[{{Mart{\'{\i}}n} {et~al.}(2015){Mart{\'{\i}}n}, {Kohno}, {Izumi},
  {Krips}, {Meier}, {Aladro}, {Matsushita}, {Takano}, {Turner}, {Espada},
  {Nakajima}, {Terashima}, {Fathi}, {Hsieh}, {Imanishi}, {Lundgren}, {Nakai},
  {Schinnerer}, {Sheth}, \& {Wiklind}}]{Martin2015}
{Mart{\'{\i}}n}, S., {Kohno}, K., {Izumi}, T., {et~al.} 2015, \aap, 573, A116

\bibitem[{{McMullin} {et~al.}(2007){McMullin}, {Waters}, {Schiebel}, {Young},
  \& {Golap}}]{CASA}
{McMullin}, J.~P., {Waters}, B., {Schiebel}, D., {Young}, W., \& {Golap}, K.
  2007, in Astronomical Society of the Pacific Conference Series, Vol. 376,
  Astronomical Data Analysis Software and Systems XVI, ed. R.~A. {Shaw},
  F.~{Hill}, \& D.~J. {Bell}, 127

\bibitem[{{Meijerink} {et~al.}(2013){Meijerink}, {Kristensen}, {Wei{\ss}}, {van
  der Werf}, {Walter}, {Spaans}, {Loenen}, {Fischer}, {Israel}, {Isaak},
  {Papadopoulos}, {Aalto}, {Armus}, {Charmandaris}, {Dasyra}, {Diaz-Santos},
  {Evans}, {Gao}, {Gonz{\'a}lez-Alfonso}, {G{\"u}sten}, {Henkel}, {Kramer},
  {Lord}, {Mart{\'{\i}}n-Pintado}, {Naylor}, {Sanders}, {Smith}, {Spinoglio},
  {Stacey}, {Veilleux}, \& {Wiedner}}]{Meijerink2013}
{Meijerink}, R., {Kristensen}, L.~E., {Wei{\ss}}, A., {et~al.} 2013, \apjl,
  762, L16

\bibitem[{{Metropolis} {et~al.}(1953){Metropolis}, {Rosenbluth}, {Rosenbluth},
  {Teller}, \& {Teller}}]{Metropolis1953}
{Metropolis}, N., {Rosenbluth}, A.~W., {Rosenbluth}, M.~N., {Teller}, A.~H., \&
  {Teller}, E. 1953, \jcp, 21, 1087

\bibitem[{{Milam} {et~al.}(2005){Milam}, {Savage}, {Brewster}, {Ziurys}, \&
  {Wyckoff}}]{Milam2005}
{Milam}, S.~N., {Savage}, C., {Brewster}, M.~A., {Ziurys}, L.~M., \& {Wyckoff},
  S. 2005, \apj, 634, 1126

\bibitem[{{Mladenovi{\'c}} \& {Roueff}(2014)}]{Maldenovic2014}
{Mladenovi{\'c}}, M., \& {Roueff}, E. 2014, \aap, 566, A144

\bibitem[{M{\"u}ller {et~al.}(2005)M{\"u}ller, Schl{\"o}der, Stutzki, \&
  Winnewisser}]{CDMS}
M{\"u}ller, H.~S., Schl{\"o}der, F., Stutzki, J., \& Winnewisser, G. 2005,
  Journal of Molecular Structure, 742, 215

\bibitem[{Nakajima {et~al.}(2011)Nakajima, Takano, Kohno, \&
  Inoue}]{Nakajima2011}
Nakajima, T., Takano, S., Kohno, K., \& Inoue, H. 2011, The Astrophysical
  Journal Letters, 728, L38

\bibitem[{{Nakanishi} {et~al.}(2005){Nakanishi}, {Okumura}, {Kohno}, {Kawabe},
  \& {Nakagawa}}]{Nakanishi2005}
{Nakanishi}, K., {Okumura}, S.~K., {Kohno}, K., {Kawabe}, R., \& {Nakagawa}, T.
  2005, \pasj, 57, 575

\bibitem[{{Nguyen} {et~al.}(1992){Nguyen}, {Jackson}, {Henkel}, {Truong}, \&
  {Mauersberger}}]{Nguyen1992}
{Nguyen}, Q.-R., {Jackson}, J.~M., {Henkel}, C., {Truong}, B., \&
  {Mauersberger}, R. 1992, \apj, 399, 521

\bibitem[{{Papadopoulos}(2007)}]{Papadopoulos2007}
{Papadopoulos}, P.~P. 2007, \apj, 656, 792

\bibitem[{{Papadopoulos} {et~al.}(2010){Papadopoulos}, {Isaak}, \& {van der
  Werf}}]{Papadopoulos2010}
{Papadopoulos}, P.~P., {Isaak}, K., \& {van der Werf}, P. 2010, \apj, 711, 757

\bibitem[{{Papadopoulos} {et~al.}(2000){Papadopoulos}, {R{\"o}ttgering}, {van
  der Werf}, {Guilloteau}, {Omont}, {van Breugel}, \&
  {Tilanus}}]{Papadopoulos2000}
{Papadopoulos}, P.~P., {R{\"o}ttgering}, H.~J.~A., {van der Werf}, P.~P.,
  {et~al.} 2000, \apj, 528, 626

\bibitem[{{Papadopoulos} {et~al.}(2011){Papadopoulos}, {Thi}, {Miniati}, \&
  {Viti}}]{Papadopoulos2011}
{Papadopoulos}, P.~P., {Thi}, W.-F., {Miniati}, F., \& {Viti}, S. 2011, \mnras,
  414, 1705

\bibitem[{{Papadopoulos} {et~al.}(2012){Papadopoulos}, {van der Werf},
  {Xilouris}, {Isaak}, \& {Gao}}]{Papadopoulos2012}
{Papadopoulos}, P.~P., {van der Werf}, P., {Xilouris}, E., {Isaak}, K.~G., \&
  {Gao}, Y. 2012, \apj, 751, 10

\bibitem[{{Papadopoulos} {et~al.}(2014){Papadopoulos}, {Zhang}, {Xilouris},
  {Weiss}, {van der Werf}, {Israel}, {Greve}, {Isaak}, \&
  {Gao}}]{Papadopoulos2014}
{Papadopoulos}, P.~P., {Zhang}, Z.-Y., {Xilouris}, E.~M., {et~al.} 2014, \apj,
  788, 153

\bibitem[{Pirogov {et~al.}(1995)Pirogov, Zinchenko, Lapinov, Myshenko, \&
  Shul'Ga}]{Pirogov1995}
Pirogov, L., Zinchenko, I., Lapinov, A., Myshenko, V., \& Shul'Ga, V. 1995,
  Astronomy and Astrophysics Supplement Series, 109, 333

\bibitem[{{Remijan} {et~al.}(2007){Remijan}, {Markwick-Kemper}, \& {ALMA
  Working Group on Spectral Line Frequencies}}]{SLAIM}
{Remijan}, A.~J., {Markwick-Kemper}, A., \& {ALMA Working Group on Spectral
  Line Frequencies}. 2007, in Bulletin of the American Astronomical Society,
  Vol.~39, American Astronomical Society Meeting Abstracts, 132.11

\bibitem[{{Ripple} {et~al.}(2013){Ripple}, {Heyer}, {Gutermuth}, {Snell}, \&
  {Brunt}}]{Ripple2013}
{Ripple}, F., {Heyer}, M.~H., {Gutermuth}, R., {Snell}, R.~L., \& {Brunt},
  C.~M. 2013, \mnras, 431, 1296

\bibitem[{{Ritchey} {et~al.}(2011){Ritchey}, {Federman}, \&
  {Lambert}}]{Ritchey2011}
{Ritchey}, A.~M., {Federman}, S.~R., \& {Lambert}, D.~L. 2011, \apj, 728, 36

\bibitem[{{Roueff} {et~al.}(2015){Roueff}, {Loison}, \& {Hickson}}]{Roueff2015}
{Roueff}, E., {Loison}, J.~C., \& {Hickson}, K.~M. 2015, \aap, 576, A99

\bibitem[{{Sakamoto} {et~al.}(2010){Sakamoto}, {Aalto}, {Evans}, {Wiedner}, \&
  {Wilner}}]{Sakamoto2010}
{Sakamoto}, K., {Aalto}, S., {Evans}, A.~S., {Wiedner}, M.~C., \& {Wilner},
  D.~J. 2010, \apjl, 725, L228

\bibitem[{{Sanders} {et~al.}(2003){Sanders}, {Mazzarella}, {Kim}, {Surace}, \&
  {Soifer}}]{Sanders2003}
{Sanders}, D.~B., {Mazzarella}, J.~M., {Kim}, D.-C., {Surace}, J.~A., \&
  {Soifer}, B.~T. 2003, \aj, 126, 1607

\bibitem[{{Scoville} {et~al.}(2015){Scoville}, {Sheth}, {Walter}, {Manohar},
  {Zschaechner}, {Yun}, {Koda}, {Sanders}, {Murchikova}, {Thompson},
  {Robertson}, {Genzel}, {Hernquist}, {Tacconi}, {Brown}, {Narayanan},
  {Hayward}, {Barnes}, {Kartaltepe}, {Davies}, {van der Werf}, \&
  {Fomalont}}]{Scoville2015}
{Scoville}, N., {Sheth}, K., {Walter}, F., {et~al.} 2015, \apj, 800, 70

\bibitem[{{Sheffer} {et~al.}(2007){Sheffer}, {Rogers}, {Federman}, {Lambert},
  \& {Gredel}}]{Sheffer2007}
{Sheffer}, Y., {Rogers}, M., {Federman}, S.~R., {Lambert}, D.~L., \& {Gredel},
  R. 2007, \apj, 667, 1002

\bibitem[{{Solomon} {et~al.}(1992){Solomon}, {Downes}, \&
  {Radford}}]{Solomon1992}
{Solomon}, P.~M., {Downes}, D., \& {Radford}, S.~J.~E. 1992, \apjl, 387, L55

\bibitem[{{Solomon} {et~al.}(1987){Solomon}, {Rivolo}, {Barrett}, \&
  {Yahil}}]{Solomon1987}
{Solomon}, P.~M., {Rivolo}, A.~R., {Barrett}, J., \& {Yahil}, A. 1987, \apj,
  319, 730

\bibitem[{Sz{\H{u}}cs {et~al.}(2014)Sz{\H{u}}cs, Glover, \&
  Klessen}]{Szucs2014}
Sz{\H{u}}cs, L., Glover, S.~C., \& Klessen, R.~S. 2014, Monthly Notices of the
  Royal Astronomical Society, 445, 4055

\bibitem[{{Tacconi} {et~al.}(1999){Tacconi}, {Genzel}, {Tecza}, {Gallimore},
  {Downes}, \& {Scoville}}]{Tacconi1999}
{Tacconi}, L.~J., {Genzel}, R., {Tecza}, M., {et~al.} 1999, \apj, 524, 732

\bibitem[{{Tafalla} {et~al.}(2002){Tafalla}, {Myers}, {Caselli}, {Walmsley}, \&
  {Comito}}]{Tafalla2002}
{Tafalla}, M., {Myers}, P.~C., {Caselli}, P., {Walmsley}, C.~M., \& {Comito},
  C. 2002, \apj, 569, 815

\bibitem[{{Tecza} {et~al.}(2000){Tecza}, {Genzel}, {Tacconi}, {Anders},
  {Tacconi-Garman}, \& {Thatte}}]{Tecza2000}
{Tecza}, M., {Genzel}, R., {Tacconi}, L.~J., {et~al.} 2000, \apj, 537, 178

\bibitem[{Tunnard {et~al.}(2015)Tunnard, Greve, Garcia-Burillo, Carpio,
  Fischer, Fuente, Gonz{\'a}lez-Alfonso, Hailey-Dunsheath, Neri, Sturm, Usero,
  \& Planesas}]{Tunnard2015}
Tunnard, R., Greve, T.~R., Garcia-Burillo, S., {et~al.} 2015, ApJ, 800, 25

\bibitem[{{Usero} {et~al.}(2004){Usero}, {Garc{\'{\i}}a-Burillo}, {Fuente},
  {Mart{\'{\i}}n-Pintado}, \& {Rodr{\'{\i}}guez-Fern{\'a}ndez}}]{Usero2004}
{Usero}, A., {Garc{\'{\i}}a-Burillo}, S., {Fuente}, A.,
  {Mart{\'{\i}}n-Pintado}, J., \& {Rodr{\'{\i}}guez-Fern{\'a}ndez}, N.~J. 2004,
  \aap, 419, 897

\bibitem[{{van der Tak} {et~al.}(2007){van der Tak}, {Black}, {Sch{\"o}ier},
  {Jansen}, \& {van Dishoeck}}]{Radex}
{van der Tak}, F.~F.~S., {Black}, J.~H., {Sch{\"o}ier}, F.~L., {Jansen}, D.~J.,
  \& {van Dishoeck}, E.~F. 2007, \aap, 468, 627

\bibitem[{{van Dishoeck} \& {Black}(1988)}]{vanDishoeck1988}
{van Dishoeck}, E.~F., \& {Black}, J.~H. 1988, \apj, 334, 771

\bibitem[{{V{\'a}zquez-Semadeni} {et~al.}(1997){V{\'a}zquez-Semadeni},
  {Ballesteros-Paredes}, \& {Rodr{\'{\i}}guez}}]{VazquezSemadeni1997}
{V{\'a}zquez-Semadeni}, E., {Ballesteros-Paredes}, J., \& {Rodr{\'{\i}}guez},
  L.~F. 1997, \apj, 474, 292

\bibitem[{{V{\'e}ron-Cetty} \& {V{\'e}ron}(2006)}]{Veron2006}
{V{\'e}ron-Cetty}, M.-P., \& {V{\'e}ron}, P. 2006, \aap, 455, 773

\bibitem[{{Viti} {et~al.}(2002){Viti}, {Natarajan}, \& {Williams}}]{Viti2002}
{Viti}, S., {Natarajan}, S., \& {Williams}, D.~A. 2002, \mnras, 336, 797

\bibitem[{{Viti} {et~al.}(2014){Viti}, {Garc{\'{\i}}a-Burillo}, {Fuente},
  {Hunt}, {Usero}, {Henkel}, {Eckart}, {Martin}, {Spaans}, {Muller}, {Combes},
  {Krips}, {Schinnerer}, {Casasola}, {Costagliola}, {Marquez}, {Planesas}, {van
  der Werf}, {Aalto}, {Baker}, {Boone}, \& {Tacconi}}]{Viti2014}
{Viti}, S., {Garc{\'{\i}}a-Burillo}, S., {Fuente}, A., {et~al.} 2014, \aap,
  570, A28

\bibitem[{{Wang} {et~al.}(2013){Wang}, {Zhang}, {Shi}, \& {Zhang}}]{Wang2013}
{Wang}, J., {Zhang}, J., {Shi}, Y., \& {Zhang}, Z. 2013, \apjl, 778, L39

\bibitem[{Wang {et~al.}(2011)Wang, Zhang, \& Shi}]{Wang2011}
Wang, J., Zhang, Z., \& Shi, Y. 2011, MNRASL, 416, L21

\bibitem[{{Wang} {et~al.}(2014){Wang}, {Nardini}, {Fabbiano}, {Karovska},
  {Elvis}, {Pellegrini}, {Max}, {Risaliti}, {U}, \& {Zezas}}]{Wang2014}
{Wang}, J., {Nardini}, E., {Fabbiano}, G., {et~al.} 2014, \apj, 781, 55

\bibitem[{{Watson} {et~al.}(1976){Watson}, {Anicich}, \&
  {Huntress}}]{Watson1976}
{Watson}, W.~D., {Anicich}, V.~G., \& {Huntress}, Jr., W.~T. 1976, \apjl, 205,
  L165

\bibitem[{{Wright}(2006)}]{Wright2006}
{Wright}, E.~L. 2006, \pasp, 118, 1711

\bibitem[{{Yang} {et~al.}(2010){Yang}, {Stancil}, {Balakrishnan}, \&
  {Forrey}}]{Yang2010}
{Yang}, B., {Stancil}, P.~C., {Balakrishnan}, N., \& {Forrey}, R.~C. 2010,
  \apj, 718, 1062

\bibitem[{{Young} \& {Scoville}(1982)}]{Young1982}
{Young}, J.~S., \& {Scoville}, N. 1982, \apj, 258, 467

\bibitem[{{Yun} \& {Carilli}(2002)}]{Yun2002}
{Yun}, M.~S., \& {Carilli}, C.~L. 2002, \apj, 568, 88

\bibitem[{{Zhang} {et~al.}(2014){Zhang}, {Henkel}, {Gao}, {G{\"u}sten},
  {Menten}, {Papadopoulos}, {Zhao}, {Ao}, \& {Kaminski}}]{Zhang2014}
{Zhang}, Z.-Y., {Henkel}, C., {Gao}, Y., {et~al.} 2014, \aap, 568, A122

\end{thebibliography}

\FloatBarrier
\appendix
\section{A. LVG Models}\label{app:lvg}

	We wrote an MCMC wrapper for the LVG code RADEX to allow a rapid and efficient exploration of high dimension parameter spaces and easy generation of posterior pdfs. The code allows us to fit multiple species simultaneously, as well as fitting multiple gas phases simultaneously. The code is written in Python, with the RADEX calculations in Fortran presenting the rate limiting step: $\gtrsim99.8\%$ of the runtime is RADEX IO and calculations.
	
	We adopt a Metropolis-Hastings (MH) sampler for its simplicity and robustness. The likelihoods from the LVG models are generally simple, contiguous shapes, and as such are well suited to the MH algorithm. We do not include covariance terms between the new parameters when sampling: each is sampled from a 1D Gaussian distribution. We adopt a $\chi^2$ likelihood function:
	
	\begin{equation}
	\mathcal{L} \propto \exp\left(-\chi^2/2\right).
	\end{equation}
	Constraints on the parameter ranges are implemented using uniform priors in log-space. The MCMC is conducted in log-space, with values converted to linear space for input to RADEX. This provides simple scale invariance and prevents biasing towards high values of the parameters.
	
	We use stock RADEX, and as such must be wary of non-convergences, bad fluxes and high optical depths. These are checked for when reading the RADEX results, and if identified the point is rejected and resampled. The alternative would be to treat this as a point of zero likelihood - test runs of these two showed no differences in the results. Since the MCMC naturally tends towards regions of greater likelihood the RADEX errors are not usually significant except for when using randomised initial positions. In the rare instance that the code becomes trapped in a region of errors new initial parameters are randomly generated. 
	
	\subsection{LVG Parameters}
	
	Our aim was to extend the results of \citet{Papadopoulos2014} with our new data and an MCMC model. We therefore adopted a similar parameter space and $K_{vir}$ constraints: $0.5 < K_{vir} < 20$. See Table \ref{tab:3phaseinputs} for the three phase parameter ranges and Table \ref{tab:models} for the precursor model inputs.
			 
	While we model the abundance and velocity gradient $X_{\rm mol}$ and ${\rm d}v/{\rm d}r$, RADEX requires as inputs the column density $N_{mol}$ and line width $\Delta v$. As only the ratio $N_{\rm mol}/\Delta v$ is relevant\footnote{For a given product $N_{\rm mol}/\Delta v$, changing $\Delta v$ only changes the quoted flux, proportional to the change in $\Delta v$, which cancels in any ratios. The optical depths and brightness temperatures remain unchanged.}, we fix $\Delta v=10$\,km\,s$^{-1}$ and find $N_{\rm mol}$ from:
	
	\begin{equation}
	N_{\rm mol} = \frac{3.08\times10^{18}n_{{\rm H}_2}X_{\rm mol}\Delta v}{{\rm d}v/{\rm d}r}.
	\end{equation}

	\section{B. Cloud Kinematic Separation}
	
	Our MCMC LVG model fits line ratios, so while it is ideal for identifying the dominant conditions of the sampled gas phases it is poorly adapted for identifying the masses within each phase. A simple but inelegant solution would be to identify the CO line luminosity from each gas phase and then use a canonical $\alpha_{\rm CO}$ factor to estimate the mass of the phase. This is however extremely susceptible to uncertainties in the $\alpha_{\rm CO}$ factor. Instead, we derive an alternative, self-consistent, method, which is used to generate $\alpha_{\rm CO}$ factors for each gas phase and for the galaxy as a whole. We apply a physically motivated upper limit on $N_{{\rm H}_2}$, thereby applying a limit on $\Delta v$ per cloud. 
	
	A single gas cloud representing the galaxy is inadequate, since the column density of H$_2$ can be written as:
	
	\begin{equation}
	N_{{\rm H}_2} =  \frac{3.08\times10^{18}n_{{\rm H}_2}\Delta v}{{\rm d}v/{\rm d}r}.
	\end{equation}
	For the high density gas phase this leads to column densities in excess of $10^{26}$\,cm$^{-2}$! In reality, the dense gas is not in a single cloud but distributed across the galaxy in many, much smaller clouds, so that $\Delta v$ is determined by the large scale motion in the galaxy and does not correspond to the LVG $\Delta v$, which is the line width of the putative emitting cloud.
	
	We therefore separate the emission into multiple identical clouds to obtain more realistic estimates of the mass of the gas phases, without assuming cloud line widths. We start with the general statement that for a cloud of uniform density:

\begin{equation}
M_{\rm cloud} = \frac{4\pi}{3} \left(\frac{\zeta\Delta v}{2 {\rm d}v/{\rm d}r}\right)^3 n_{{\rm H}_2}\mu_{{\rm H}_2},
\end{equation}
where $\zeta=3.08\times10^{18}$ is the parsec to cm conversion factor, $\Delta v$ is the FWHM line width of the cloud, ${\rm d}v/{\rm d}r$ is the velocity gradient of the cloud in km\,s$^{-1}$\,pc$^{-1}$, $n_{{\rm H}_2}$ is the hydrogen number density in cm$^{-3}$ and $\mu_{{\rm H}_2}$ is the helium adjusted ISM mass per H$_2$ molecule.

\bigskip

\noindent Separately, we make the general statement of beam dilution that for a Gaussian surface brightness with peak $T_{\rm b}$ and a Gaussian beam much larger than the size of the cloud:

\begin{equation}
T_{\rm b,d} = \frac{T_{\rm b} \chi^2 1.133}{\Omega_{\rm beam}}\left(\frac{\Delta v}{2 {\rm d}v/{\rm d}r}\right)^2,
\end{equation}
where $\chi=\frac{1}{494}$\,arcsec/pc is the angular scale for NGC\,6240 and $\Omega_{\rm beam}$ is in arcsec$^2$. In this case, $T_{\rm b}$ is a result of RADEX runs, which give optically thick clouds for the HCN and CO $J=1-0$ lines.

\bigskip

\noindent For a Gaussian line, or a line which can be described as a combination of Gaussians, the number of clouds is the ratio of the area ($T\Delta v$) of the observed line to the diluted area per cloud:

\begin{equation}
n_{\rm clouds} = \frac{T_{\rm obs}{\rm FWHM}}{T_{\rm b,d}\Delta v} = \frac{T_{\rm obs}\Omega_{\rm beam}{\rm FWHM}}{1.133\, T_{\rm b}\,\Delta v\, \chi^2 \left(\frac{\Delta v}{2 {\rm d}v/{\rm d}r}\right)^2}.
\end{equation}
Therefore the total mass in the galaxy in this gas phase is:

\begin{align}
M &= M_{\rm cloud}\ n_{\rm clouds}\\[0.5cm] 
&= \frac{4\pi}{3} \left(\frac{\zeta\Delta v}{2 {\rm d}v/{\rm d}r}\right)^3 n_{{\rm H}_2}\mu_{{\rm H}_2} \frac{T_{\rm obs}\Omega_{\rm beam}{\rm FWHM}}{1.133\, T_{\rm b}\,\Delta v\, \chi^2 \left(\frac{\Delta v}{2 {\rm d}v/{\rm d}r}\right)^2} \\[0.5cm]
&= \frac{1.85\,\zeta^3\mu_{{\rm H}_2}}{\chi^2} \frac{T_{\rm obs}\Omega_{\rm beam}}{T_{\rm b}} \frac{n_{{\rm H}_2}}{{\rm d}v/{\rm d}r} {\rm FWHM}\\[0.5cm]
&= 0.123 \frac{T_{\rm obs}\Omega_{\rm beam}}{T_{\rm b}\,\chi^2} \frac{n_{{\rm H}_2}}{{\rm d}v/{\rm d}r} {\rm FWHM}\hspace{1cm} [M_\odot]
\end{align}
This in fact corresponds to the same equation for $\alpha_{\rm mol}$ as that derived by \citet{Papadopoulos2012} by considering iso-velocity surfaces across a galaxy. Also, we find that the mass is independent of the chosen upper limit on $N_{{\rm H}_2}$ which is reassuring given that GMCs are likely to present a range of column densities \citep{VazquezSemadeni1997}.

\section{C: Precursor LVG Models}\label{sec:precursors}

The three phase model presented in Section \ref{subsec:threephasemodel} makes assumptions regarding the gas phases and distribution of molecular species between phases. These assumptions are based on a series of models of increasing complexity that are precursors to our three phase, six species model. These models do not add to the scientific results of the paper, but due to their role in justifying our assumptions we include them here.

\subsection{Single Phase}
We first ran four models with single species and their respective isotopologues: HCN$+$H$^{13}$CN with the detected H$^{13}$CN$(1-0)$ line flux, HCN$+$H$^{13}$CN with the upper limit on the H$^{13}$CN$(1-0)$ line flux, HCO$^++$H$^{13}$CO$^+$ and CS only, with CS data shown in Table \ref{tab:lvgCSInputs}. We also run a combined model with HCN$+$HCO$^++$H$^{13}$CN$+$H$^{13}$CO$^+$. The input parameter ranges for these models are shown in Table \ref{tab:models}. When fixed, molecular abundances were chosen to match those of \citet{Papadopoulos2014}, i.e., $X_{\rm HCN}$ = $2\times10^{-8}$, $X_{{\rm HCO}^+}$ = $8\times10^{-9}$ and $X_{\rm CS}$ = $1\times10^{-9}$.

	\begin{table}
	\centering
	\caption{LVG CS Line Inputs}\label{tab:lvgCSInputs}
	\begin{tabular}{l c c c}\hline
	& $\nu_0$ & $S_{\rm line}$ & References\footnote{1 = \citet{Papadopoulos2014}, 2 = \citet{Wang2011}, 3 = \citet{Scoville2015}}\\
	Line & [GHz] & [Jy\,km\,s$^{-1}$] &\\ \hline\hline
	CS$(2-1)$ & 97.98  & $7.5\pm1.5$ & 1\\
	CS$(3-2)$ & 145.0 & $9\pm2$ & 1\\
	CS$(5-4)$ & 239.1 & $54\pm6$ & 2\\
	CS$(7-6)$ & 342.9 & $7.4\pm1.0$ & 3\\ \hline
	\end{tabular}
	\end{table}

	\begin{table*}
	\centering
	\caption{MCMC model input parameter ranges.}\label{tab:models}
	\begin{tabular}{l c c c c c c c c c}\hline
	& & \multicolumn{8}{c}{All values $\log_{10}$}\\ 
	 & & $T_{\rm k}$ & $n_{{\rm H}_2}$ & ${\rm d}v/{\rm d}r$ & $\frac{[{\rm HC(N/O}^+)]}{[{\rm H}^{13}{\rm C(N/O}^+)]}$ & $X_{\rm HCN}$ & $X_{\rm SiO}$ & $X_{\rm HCO^+}$ & $X_{\rm CS}$ \\
	 Model & ID & [K] & [cm$^{-3}$] & [km\,s$^{-1}$\,pc$^{-1}$] & & & & &\\ \hline\hline
	HCN \& H$^{13}$CN, measured\dotfill&1a& $0.5-3$ & $2-8$ & $0-3$ & $1-3$ & $-7.70$ & $...$ & $...$ & $...$\\
	HCN \& H$^{13}$CN, upper limit\dotfill&1b& $0.5-3$ & $2-8$ & $0-3$ & $1-3$ & $-7.70$ & $...$ & $...$ & $...$\\
	HCO$^+$ \& H$^{13}$CO$^+$\dotfill &2& $0.5-3$ & $2-8$ & $0-3$ & $1-3$ & $...$ & $...$ & $-8.10$ & $...$\\
	CS\dotfill &5& $0.5-3$ & $2-8$ & $0-3$ & $...$ & $...$ & $...$ & $...$ & $-9.00$\\
	HCN, HCO$^+$, H$^{13}$CN \& H$^{13}$CO$^+$\dotfill &4& $0.5-3$ & $2-8$ & $0-3$ & $1-3$ & $-12 - -4$ & $...$ & $-12 - -4$ & $...$\\ \hline
	\end{tabular}
	\end{table*}

The numerical results of the models are given in Table \ref{tab:modelsResults}, where we present both the the means of the MCMC traces and the best fit values. We recover similar temperatures and densities to \citet{Papadopoulos2014}, although the inclusion of our new $^{13}$C isotopologue lines allows us to place tighter constraints on these parameters.

\begin{table*}
	\centering
	\caption{MCMC Results}\label{tab:modelsResults}
	\begin{tabular}{l c c c c c c c c}
	\multicolumn{9}{c}{Posterior mean results and the upper and lower limits on the $68\%$ credible interval.}\\ \hline
	& &\multicolumn{7}{c}{All values $\log_{10}$}\\ 
	& &$T_{\rm k}$ & $n_{{\rm H}_2}$ & ${\rm d}v/{\rm d}r$ & $\frac{[{\rm HC(N/O}^+)]}{[{\rm H}^{13}{\rm C(N/O}^+)]}$ & $X_{\rm HCN}$ & $X_{\rm SiO}$ & $X_{\rm HCO^+}$ \\
	Model & ID & [K] & [cm$^{-3}$] & [km\,s$^{-1}$\,pc$^{-1}$] & & & & \\ \hline\hline
	HCN \& H$^{13}$CN, measured \dotfill &1a& $1.52^{1.75}_{1.00}$ & $4.93^{5.60}_{4.40}$ & $1.28^{2.28}_{0.42}$ & $2.27^{2.46}_{1.90}$ & $...$ & $...$ & $...$\\[1mm]
	HCN \& H$^{13}$CN, upper limit \dotfill &1b& $1.50^{1.75}_{0.95}$ & $4.95^{5.84}_{4.22}$ & $1.39^{1.98}_{0.42}$ & $2.49^{3.00}_{2.34}$ & $...$ & $...$ & $...$\\[1mm]
	HCO$^+$ \& H$^{13}$CO$^+$\dotfill &2& $1.39^{1.60}_{1.13}$ & $4.51^{4.82}_{3.86}$ & $1.00^{1.05}_{0.03}$ & $2.74^{3.00}_{2.68}$ & $...$ & $...$ & $...$\\[1mm]
	CS\dotfill &3& $2.71^{3.0}_{2.65}$ & $3.3^{4.1}_{3.0}$ & $0.68^{0.93}_{0.03}$ & $...$ & $...$ & $...$ & $...$ \\[1mm]
	HCN, HCO$^+$, H$^{13}$CN \& H$^{13}$CO$^+$\dotfill &4& $1.04^{1.13}_{0.85}$ & $5.87^{6.62}_{5.42}$ & $1.81^{2.34}_{1.26}$ & $2.61^{2.82}_{2.42}$ & $-8.32^{-7.52}_{-9.12}$ & $...$ & $-9.04^{-8.4}_{-9.84}$\\[1mm] \hline \\[0.3cm]
	
	\multicolumn{9}{c}{Joint maximum likelihood results.}\\ \hline
	& &\multicolumn{7}{c}{All values $\log_{10}$}\\ 
	& &$T_{\rm k}$ & $n_{{\rm H}_2}$ & ${\rm d}v/{\rm d}r$ & $\frac{[{\rm HC(N/O}^+)]}{[{\rm H}^{13}{\rm C(N/O}^+)]}$ & $X_{\rm HCN}$ & $X_{\rm SiO}$ & $X_{\rm HCO^+}$ \\
	Model & ID & [K] & [cm$^{-3}$] & [km\,s$^{-1}$\,pc$^{-1}$] & & & & \\ \hline\hline
	HCN \& H$^{13}$CN, measured \dotfill &1a& $0.86$ & $6.3$ & $2.7$ & $3.0$ & $...$ & $...$ & $...$\\[0mm]
	HCN \& H$^{13}$CN, upper limit \dotfill &1b& $0.88$ & $6.3$ & $2.7$ & $3.0$ & $...$ & $...$ & $...$\\[0mm]
	HCO$^+$ \& H$^{13}$CO$^+$\dotfill &2& $1.1$ & $4.8$ & $1.1$ & $3.0$ & $...$ & $...$ & $...$\\[0mm]
	CS\dotfill &3& $3.0$ & $3.5$ & $0.1$ & $...$ & $...$ & $...$ & $...$ \\[0mm]
	HCN, HCO$^+$, H$^{13}$CN \& H$^{13}$CO$^+$\dotfill &4& $0.96$ & $6.3$ & $1.7$ & $2.6$ & $-9.0$ & $...$ & $-9.8$\\[0mm] \hline
	\end{tabular}
	\end{table*}

    \subsection{The Effect of $T_{\rm bg}$}
    
       For some very active galaxies with luminous dust emission, background radiation from hot dust behind the gas cloud may significantly affect the line ratios \citep{Papadopoulos2010}. For dense molecular gas tracers, neglecting this may lead to extremely unusual solutions from RADEX fits to the line ratios.
   
    We explored this possibility in NGC\,6240 by testing whether increasing $T_{\rm bg}$ can improve the fits of our HCN $+$ HCO$^+$ single phase model. The model was rerun with $T_{\rm bg}=10$, 20 and 50\,K.
        
\begin{figure*}
    \centering
    \includegraphics[width=0.475\textwidth]{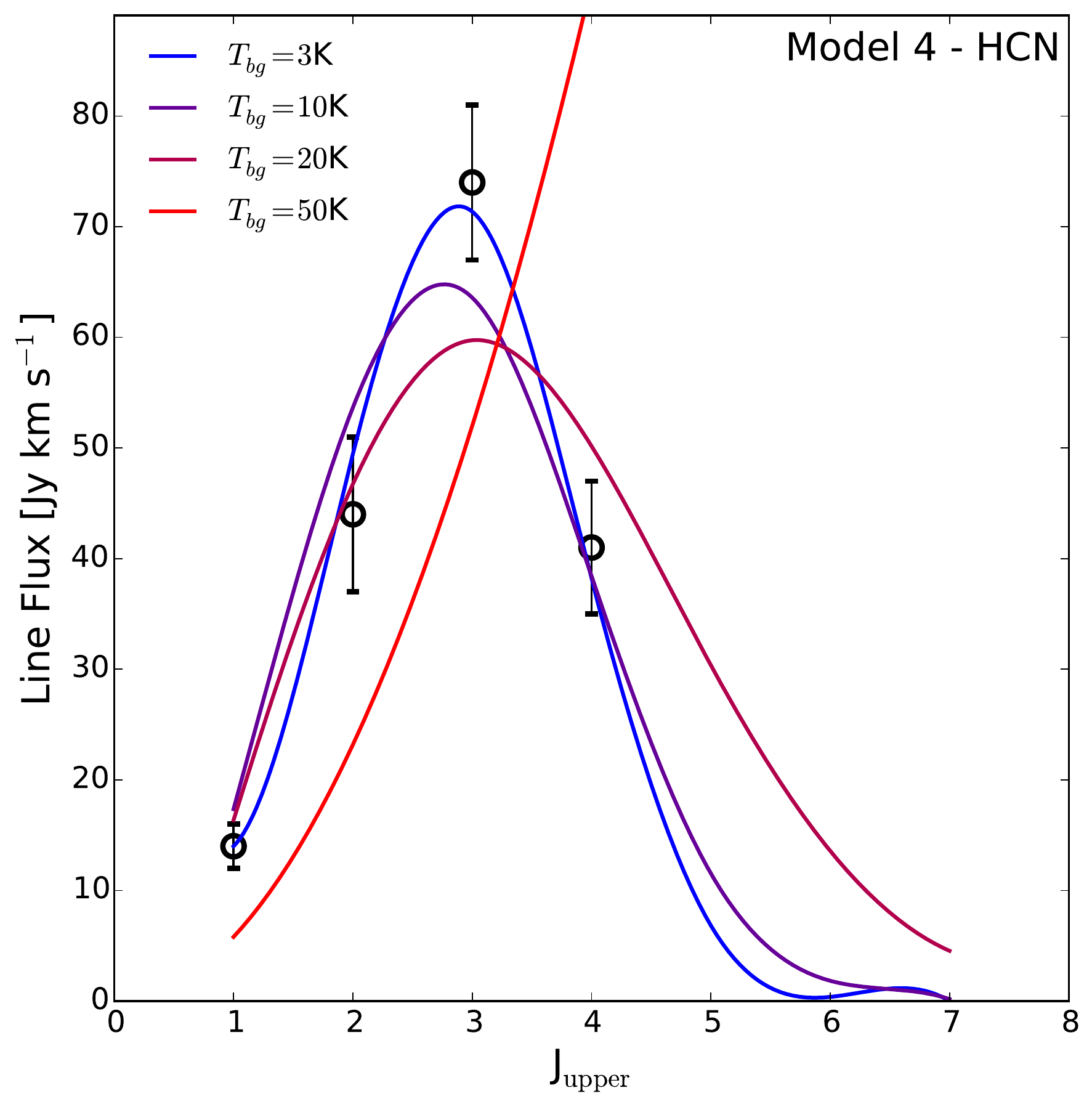}
    \hfill
    \includegraphics[width=0.475\textwidth]{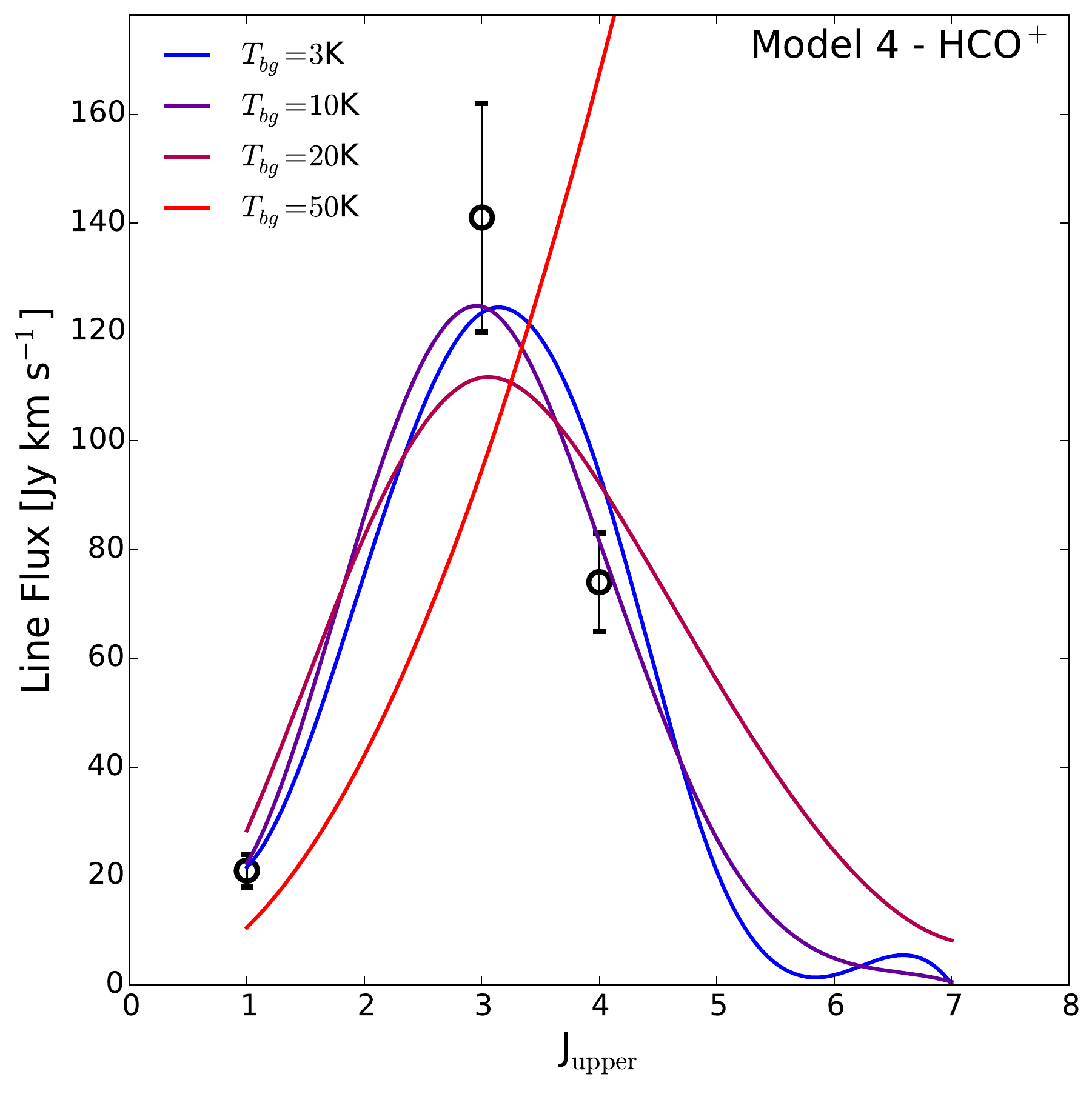}
    \caption{SLED fits for the HCN$+$HCO$^+$ single phase model, with $T_{\rm bg}$ increasing as 3, 10, 20 and 50\,K. The best fits are from the 3\,K background, and the two observationally motivated background temperatures (20 and 50\,K) both produce much worse fits.}\label{fig:hcnhcoTbgCOMP}
    \end{figure*}

   We find that increasing $T_{\rm bg}$ to 10\,K marginally improves the HCO$^+$ SLED fit at the expense of a sightly worse HCN SLED fit. The solutions are constrained to give positive fluxes which indirectly constrains the gas kinetic temperature in these models to be $\gtrsim T_{\rm bg}$. The results push against this limit, so that $T_{\rm k}/T_{\rm dust}\sim T_{\rm k}/T_{\rm bg} \simeq 1$, which is unlikely to be reasonable in the extensively shocked environment of NGC\,6240 \citep{Meijerink2013,Papadopoulos2014}, unless the HCN and HCO$^+$ are located not in shocks but in cold, dense cores. As $T_{\rm bg}$ is increased, the fits become progressively worse, with 50\,K not remotely fitting the SLED and producing $\chi^2>1300$ (cf.\ $4-6$ for $T_{\rm bg}=3-10$\,K). The SLED fits for these models are shown in Figure \ref{fig:hcnhcoTbgCOMP}. Note that the change in shape is largely due to a suppression of the low$-J$ line emission in excess of the background due to the increased background.
   
   The 50\,K case is comparable to the the 52\,K black body dust temperature described by \citet{Tacconi1999}. However, we found no consistent solutions over multiple runs, and pathological fits to the SLEDs for this $T_{\rm bg}$. This is consistent with the inconsistencies between the CO and 1.3\,mm continuum estimates of gas mass commented on by \citet{Tacconi1999}: they find that the dust co-located with the dense gas was most likely significantly cooler (perhaps 20-25\,K), and furthermore that the shock heated gas is thermally decoupled from the cooler dust.
       
    The inability of our results to find good a solution for a 50\,K background is unsurprising, but the poor solution at 20\,K is slightly more interesting. One possible, and we would argue likely, explanation could simply be that the HCN and HCO$^+$ emission are tracing the relatively quiescent regions of shielded, dense gas, i.e., they are performing as classical dense gas tracers. This is consistent with the marginal improvement (worsening) of the SLED fit of HCO$^+$ (HCN) when $T_{\rm bg}$ is increased from 3\,K to 10\,K, as HCO$^+$ is probably more extended and less shielded than HCN, so subject to a slightly stronger background radiation field. It is also consistent with the very high HCN/CN line ratio of 2 found by \citet{Aalto2002}, which is evidence of large reservoirs of UV-shielded gas \citep{Fuente1995,Greaves1996}.
    
    We conclude that there is insufficient evidence to support using a super-CMB background and proceed with the 3\,K CMB.

	\subsection{Two Phase Modelling - HCN+HCO$^+$}\label{subsubsec:twophase}
	
	Here, we extend our model to include a second, non-interacting gas phase with both phases containing HCN and HCO$^+$. This is strongly physically motivated as single dish observations across the galaxy are sampling multiple non-interacting gas phases confused within the beam.
	
	Each of the two phases has their own kinetic temperature, density, and velocity gradient while they share common HCN and HCO$^+$ abundances and a single [HC(N/O$^+$)]/[H$^{13}$C(N/O$^+$)] abundance ratio\footnote{While this is undesirable it is necessary to minimise the number of free parameters we attempt to constrain with our few line ratios. A model was attempted with free abundances in each phase, but this was unable to converge on any solutions.}. The fluxes from the two phases are added, weighted by a free parameter, $f$, such that:
	\begin{equation}
	S = S_1f + S_2(1-f),
	\end{equation}
	where $S_1$ and $S_2$ are the RADEX fluxes for phase 1 and 2 respectively, in a simplified version of the method used in \citep{Zhang2014}. Either of $S_1$ and $S_2$ may be negative, but the combined fluxes $S$ are constrained to be positive, as per observations. Furthermore, there is no line transfer between the two phases: they emulate two phases confused in the imaging beam, not two phases along the same line of sight (unless they are decoupled by a sufficiently large velocity offset). 
		
	\bigskip
	
\begin{figure*}
	\centering
	\includegraphics[width=\textwidth]{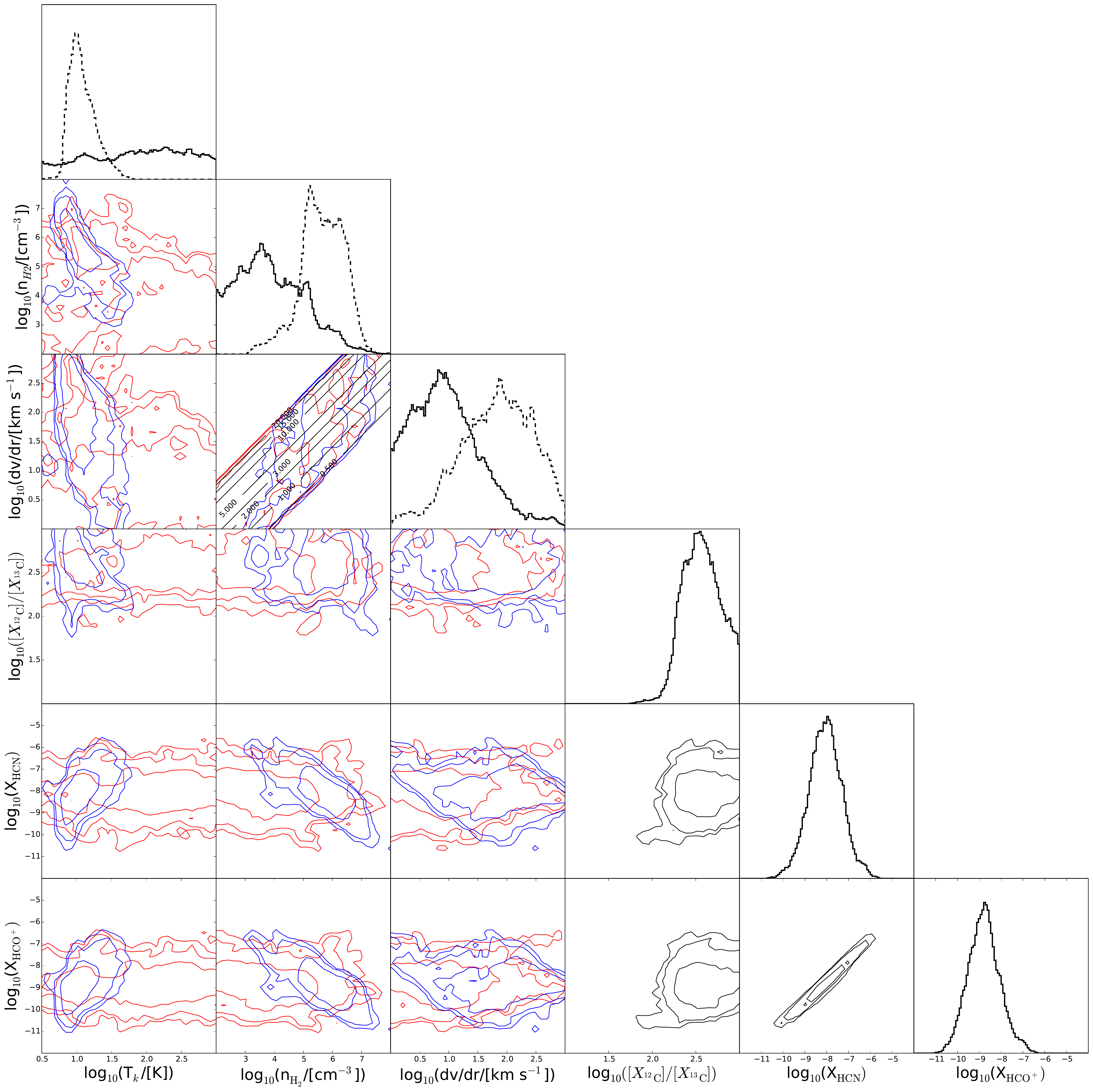}
	\caption{The modified step plot for the two phase HCN$+$HCO$^+$ model. Contours are at the 68\%, 95\% and 99\% credible intervals. Red/solid and blue/dashed contours are phases 1 and 2, with $T_{\rm bg}=3$ and 3\,K respectively. [$^{12}$C]/[$^{13}$C], $X_{\rm HCN}$ and $X_{{\rm HCO}^+}$ are kept the same in the two phases, so have only one set of black contours. The two phases are completely degenerate - unsurprising given the good fits of the single phase models.}\label{fig:2phasestep3}
	\end{figure*}

	The results of this two phase model are shown in Figure \ref{fig:2phasestep3}. It appears from Figure \ref{fig:2phasestep3} that we have an underdetermined model unable to either constrain both phases or clearly exclude one. Even though we are unable to constrain the kinetic temperature, density or velocity gradient, the abundance ratio is well determined. 
	
	The [HC(N/O$^+$)]/[H$^{13}$C(N/O$^+$)] ratio in this model is elevated. It is almost definitely unsuitable to have a single [HC(N/O$^+$)]/[H$^{13}$C(N/O$^+$)] ratio for both phases and both HCN and HCO$^+$; unfortunately, as is shown by the degeneracy of the two phases, we do not have sufficient observations nor are observations of sufficient precision to constrain two phases without making some simplifications. Further observations of the H$^{13}$CN and H$^{13}$CO$^+$ SLEDs, in particular the $J=3-2$ line, would go a long way towards solving this issue.

	\subsection{Two Phase Modelling - HCN+HCO$^+$+CO}\label{subsubsec:hcnhcoco}
	
	The previous two phase model failed due to the paucity of available molecular lines and the large uncertainties on those available. In lieu of additional $^{13}$C isotopologue line observations, which would place much tighter constraints on the models, we explore the addition of CO and $^{13}$CO to the two phase models. We take the line fluxes directly from \citet{Papadopoulos2014}, and do not make any corrections for source size: the $f$ scaling parameter allows the model to account for different beam filling factors due to the different sizes of the HCN and CO emitting regions, while the two phases allow for non-cospatial emission.
	
\begin{figure*}
	\centering
	\includegraphics[width=\textwidth]{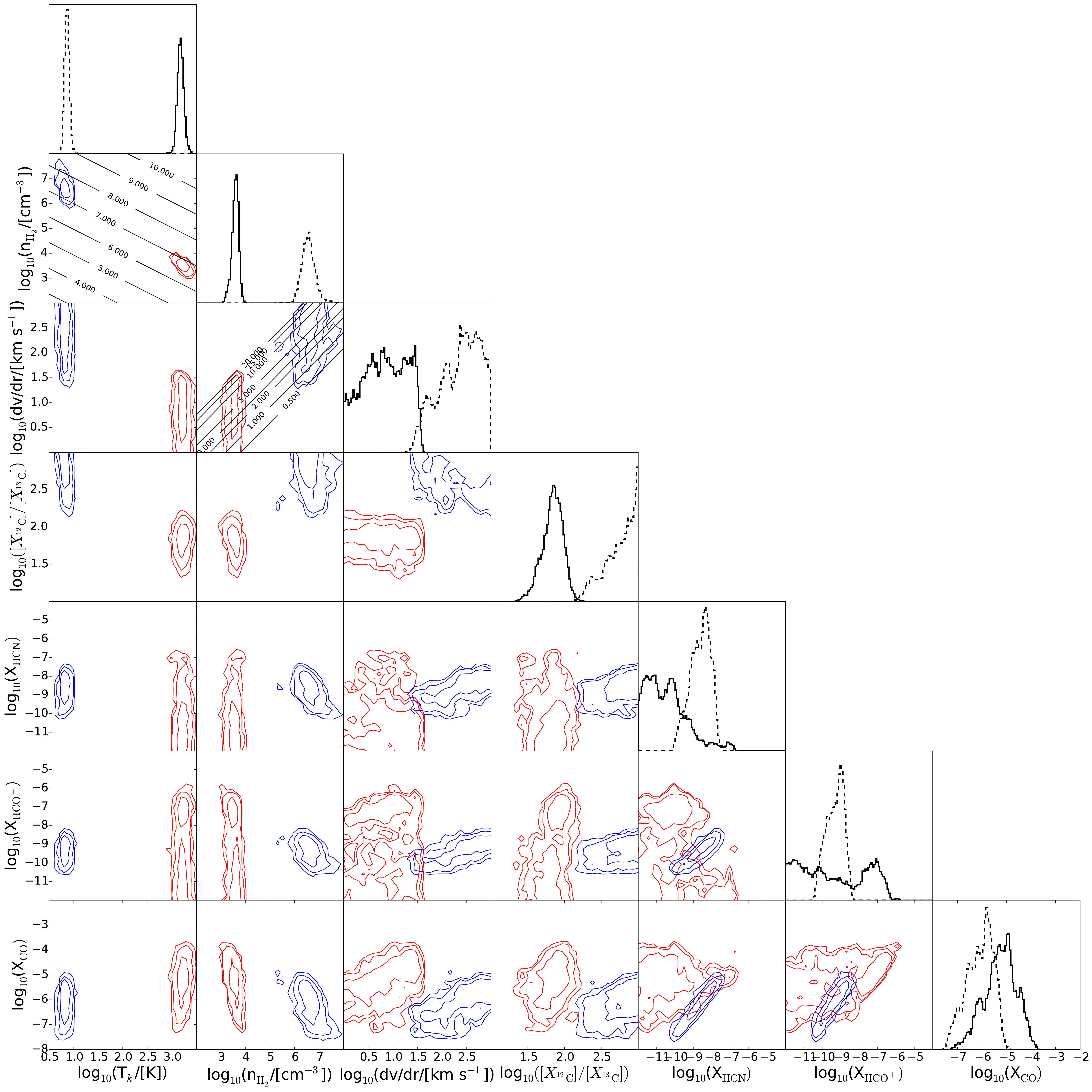}
	\caption{The modified step plot for the two phase HCN$+$HCO$^+$+CO model. Contours are at the 68\%, 95\% and 99\% credible intervals. Red/solid and blue/dashed contours are phases 1 and 2, with $T_{\rm bg}=3$ and 3\,K respectively. Contours of constant pressure have been included in the $T_{\rm k} - n_{{\rm H}_2}$ plot at intervals of one dex. There is a clear, robust and \emph{spontaneous} separation into a hot, diffuse and cold, dense gas phase.}\label{fig:2phaseCO}
	\end{figure*}

\begin{figure*}
	\centering
	\includegraphics[width=0.475\textwidth]{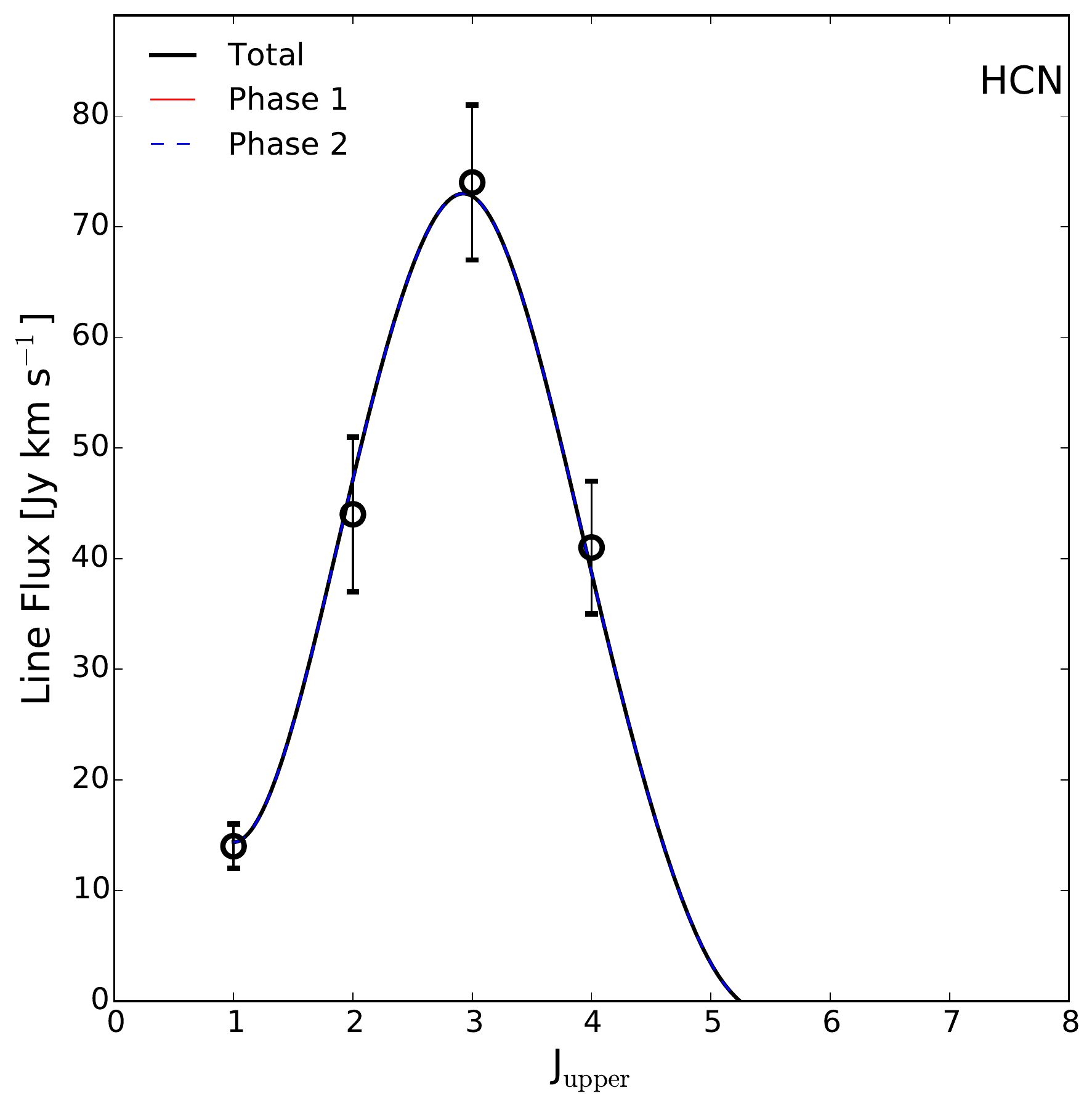}
	\hfill
	\includegraphics[width=0.475\textwidth]{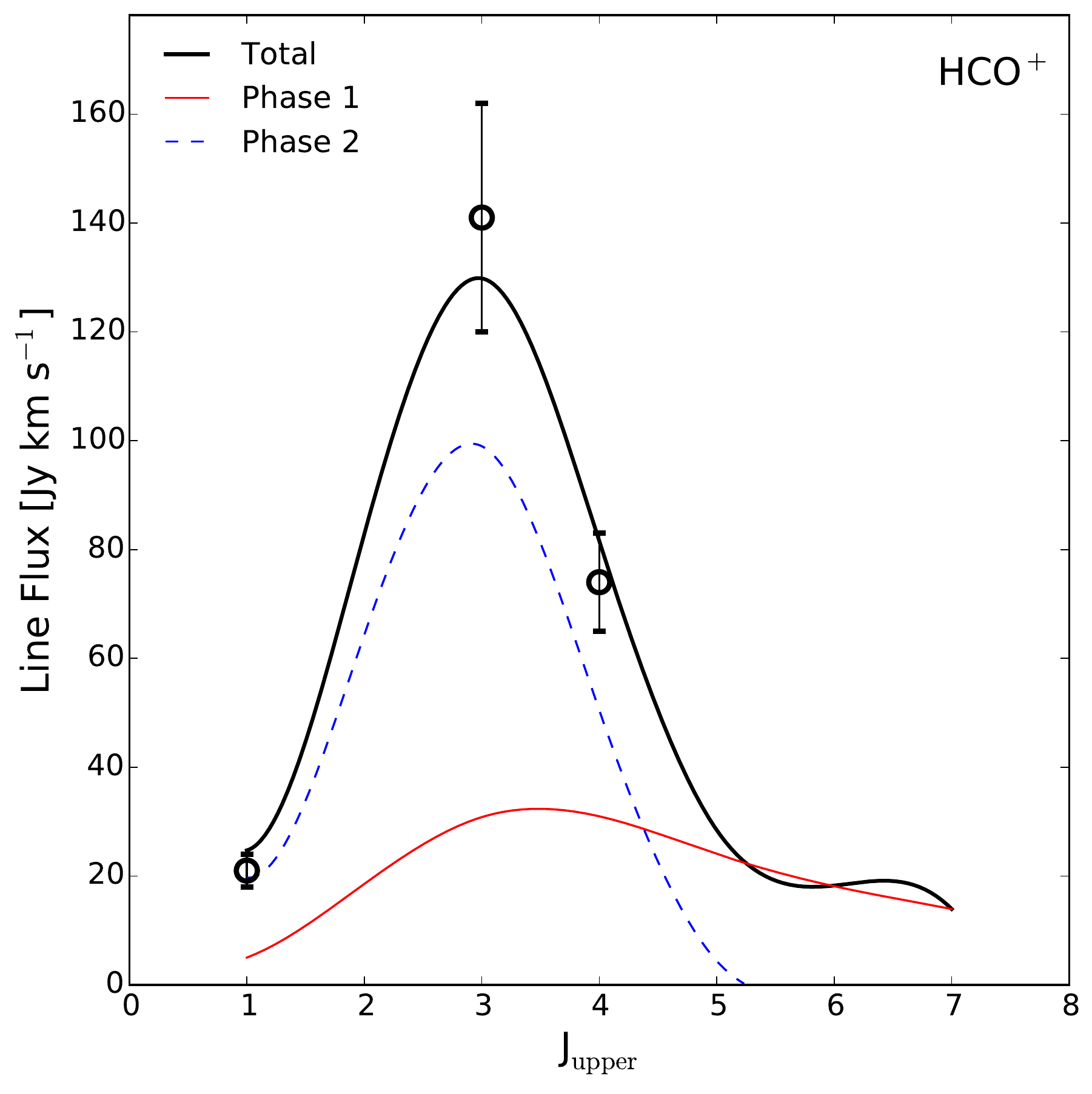}\\
	\includegraphics[width=0.475\textwidth]{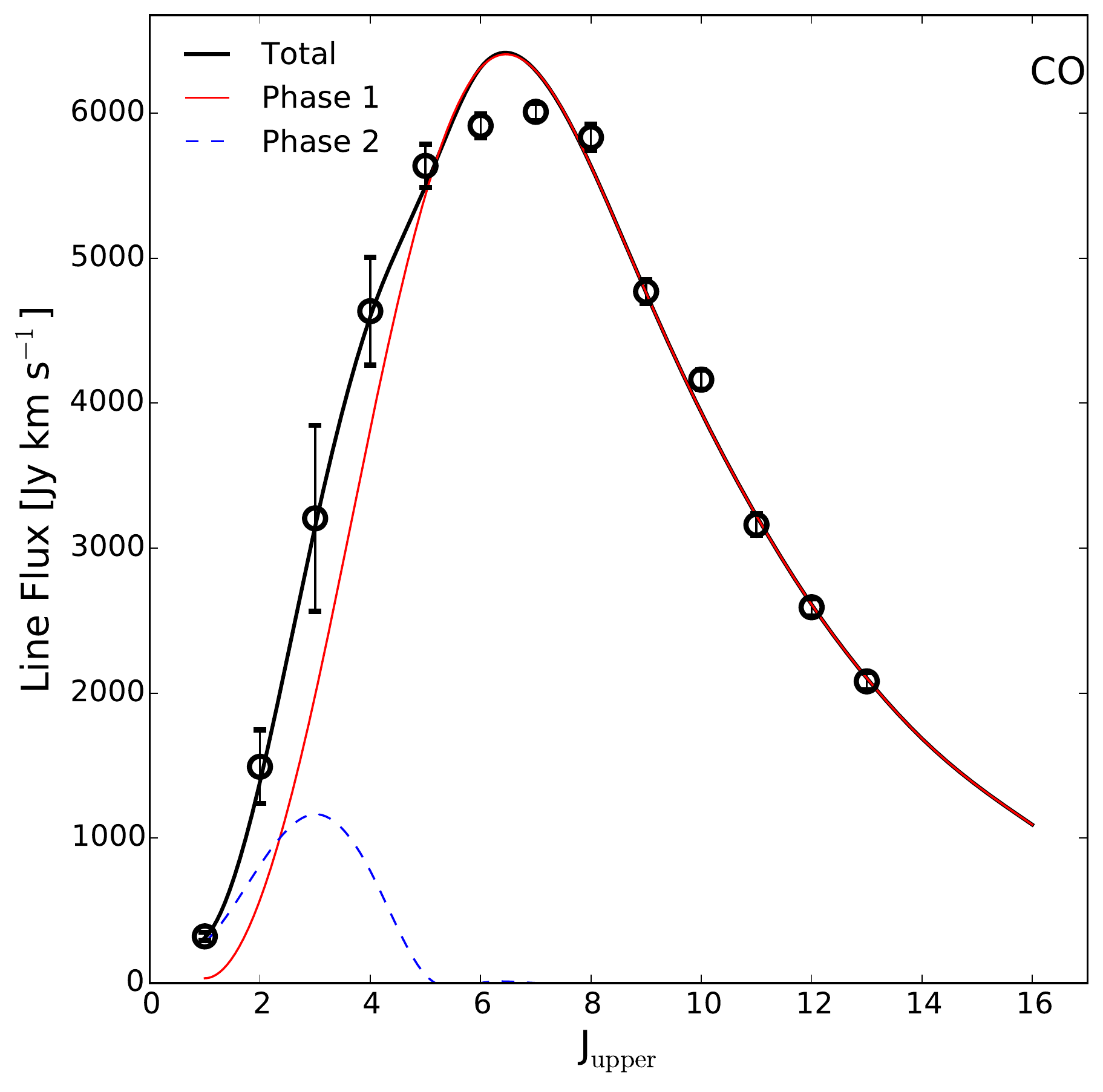}
	\hfill
	\includegraphics[width=0.475\textwidth]{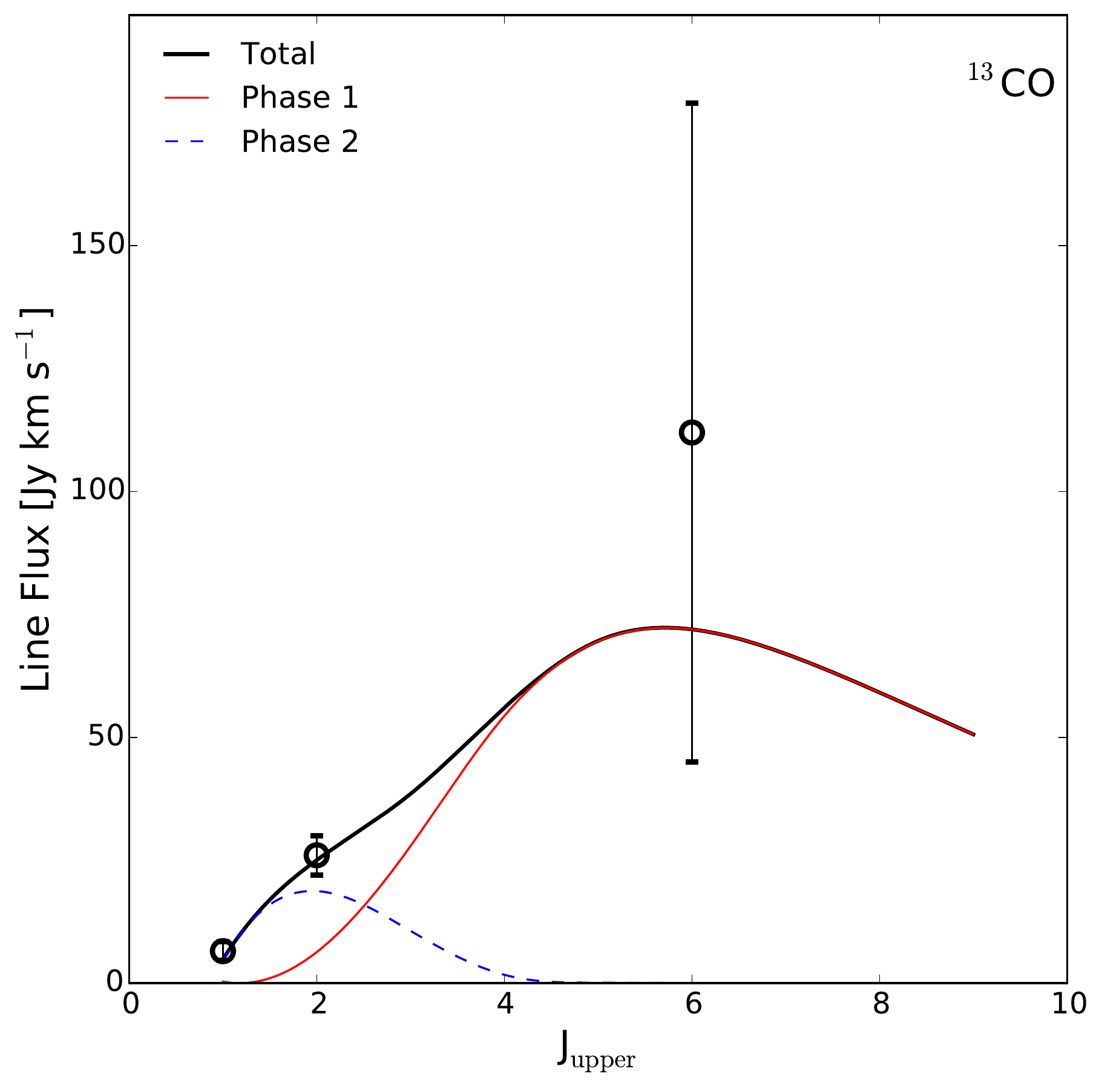}
	\caption{SLED fits for the two phase HCN$+$HCO$^+$+CO model, showing the relative contribution of each gas phase. The same normalisation has been used for the CO and $^{13}$CO SLEDs.}\label{fig:2phaseCOSLED}
	\end{figure*}

	With the additional lines we are able to extend the model complexity so that the free parameters include $X_{\rm CO}$, $X_{\rm HCN}$, $X_{{\rm HCO}^+}$ and $[X_{^{12}{\rm C}}]/[X_{^{13}{\rm C}}]$ in each gas phase. Once again, we have insufficient lines to simultaneously fit free $X_{\rm mol}$ and $[X_{^{12}{\rm C}}]/[X_{^{13}{\rm C}}]$ ratios for every species in both phases, and we use a single $[X_{^{12}{\rm C}}]/[X_{^{13}{\rm C}}]$ ratio in each phase for all species. Even with these restrictions the code can take many more steps to converge, and the traces were carefully inspected to exclude unstable/unconverged runs. Of the runs that converged, all converged onto the same locus, which is presented here. The results are tabulated in Table \ref{tab:2phaseCO} and shown in Figures \ref{fig:2phaseCO} (parameter pdfs) and \ref{fig:2phaseCOSLED} (SLEDs). 
	
	This model provides very tight constraints on the two gas phases, although the tightness of the constraints is no guarantee of accuracy. The model \emph{spontaneously} separates into a hot, diffuse, CO dominated phase and a cold, dense, HCN dominated phase. This result is robust to randomised initial positions over multiple MCMC runs and is similar to the findings of \citet{Kamenetzky2014} who used only CO. $T_{\rm k}$ and $n_{{\rm H}_2}$ are the best constrained parameters, with less than $\pm0.1$\,dex spread in the 68\% credible intervals. In the diffuse phase $X_{\rm HCN}$ falls rapidly, and Figure \ref{fig:2phaseCOSLED} confirms that HCN has no contribution from the diffuse gas phase, so that $X_{\rm HCN}$ in this phase is an upper limit only. In both phases $X_{\rm CO}$ is slightly lower than the canonical value of $2\times10^{-4}$ presenting instead $4\times10^{-5}$ and $6\times10^{-6}$. However, the spread in $X_{\rm CO}$ is very large, especially in the hot phase.
	
	\begin{table*}
	\centering
	\caption{HCN$+$HCO$^+$+CO Two Phase Model Solutions. All values $\log_{10}$.}\label{tab:2phaseCO}
	\begin{tabular}{l c c}\hline
	Parameter & Mean & min $\chi^2$\\ \hline\hline
	$T_{\rm k,1}$\dotfill & $3.2^{3.3}_{3.1}$ & 3.2 \\[1mm]
	$T_{\rm k,2}$\dotfill & $0.9^{0.9}_{0.8}$ & 0.8 \\[1mm]
	$n_{{\rm H}_2,1}$\dotfill & $3.6^{3.8}_{3.5}$ & 3.5 \\[1mm]
	$n_{{\rm H}_2,2}$\dotfill & $6.6^{6.8}_{6.3}$ & 6.5 \\[1mm]
	$({\rm d}v/{\rm d}r)_1$\dotfill & $0.8^{1.5}_{1.1}$ & 1.2 \\[1mm]
	$({\rm d}v/{\rm d}r)_2$\dotfill & $2.4^{3.0}_{2.3}$ & 2.9 \\[1mm]
	$\left(\frac{[{\rm HCN}]}{[{\rm H}^{13}{\rm CN}]} = \frac{[{\rm HCO}^+]}{[{\rm H}^{13}{\rm CO}^+]} = \frac{[{\rm CO}]}{[^{13}{\rm CO}]} \right)_1$\dotfill & $1.8^{2.0}_{1.7}$ & 1.9 \\[1mm]
	$\left(\frac{[{\rm HCN}]}{[{\rm H}^{13}{\rm CN}]} = \frac{[{\rm HCO}^+]}{[{\rm H}^{13}{\rm CO}^+]} = \frac{[{\rm CO}]}{[^{13}{\rm CO}]} \right)_2$\dotfill & $2.7^{3.0}_{2.7}$ & 2.9 \\[1mm]
	$X_{\rm HCN,1}$\dotfill & $-10.8^{-10.4}_{-11.9}$ & $-10.0$ \\[1mm]
	$X_{\rm HCN,2}$\dotfill & $-8.6^{-7.9}_{-9.0}$ & $-7.7$ \\[1mm]
	$X_{{\rm HCO}^+,1}$\dotfill & $-9.3^{-6.7}_{-10.6}$ & $-6.6$ \\[1mm]
	$X_{{\rm HCO}^+,2}$\dotfill & $-9.3^{-8.7}_{-9.6}$ & $-8.6$ \\[1mm]
	$X_{{\rm CO},1}$\dotfill & $-5.3^{-4.5}_{-6.0}$ & $-4.4$ \\[1mm]
	$X_{{\rm CO},2}$\dotfill & $-6.1^{-5.4}_{-6.3}$ & $-5.2$ \\[1mm]
	$f$\dotfill & $-0.9^{-0.7}_{-1.2}$ & $-1.2$ \\\hline
	\end{tabular}
	\end{table*}

	\subsection{The [$X_{^{12}{\rm C}}$]/[$X_{^{13}{\rm C}}$] ratio: evidence for fractionation?}\label{subsubsec:12c13c}
	
	Even with the addition of the CO and $^{13}$CO lines we are unable to constrain two gas phases with free CO, HCN and HCO$^+$; $^{13}$CO, H$^{13}$CN and H$^{13}$CO$^+$ abundances in each phase. In the quest to identify whether the [$^{12}$C]/[$^{13}$C] ratio found by \citet{Papadopoulos2014} is real, or due to biased SLED fitting, we ran a model with $X_{\rm CO}$, $X_{\rm HCN}$ and $X_{{\rm HCO}^+}$ fixed to the best fit parameters of the previous run (see Table \ref{tab:2phaseCO} for values). We then introduce as free parameters [CO]/[$^{13}$CO]$_1$, [CO]/[$^{13}$CO]$_2$, [HCN]/[H$^{13}$CN] and [HCO$^+$]/[H$^{13}$CO$^+$]. I.e., while HCN and HCO$^+$ are present in both phases, each has a single isotopologue ratio used for both phases. This is motivated by us only having the $J=1-0$ line for the H$^{13}$CN and H$^{13}$CO$^+$, which have negligible contributions in the hot phase - in effect we are indirectly excluding H$^{13}$CN and H$^{13}$CO$^+$ from the hot phase.
	
	\bigskip
	
\begin{figure*}
	\centering
	\includegraphics[width=0.9\textwidth]{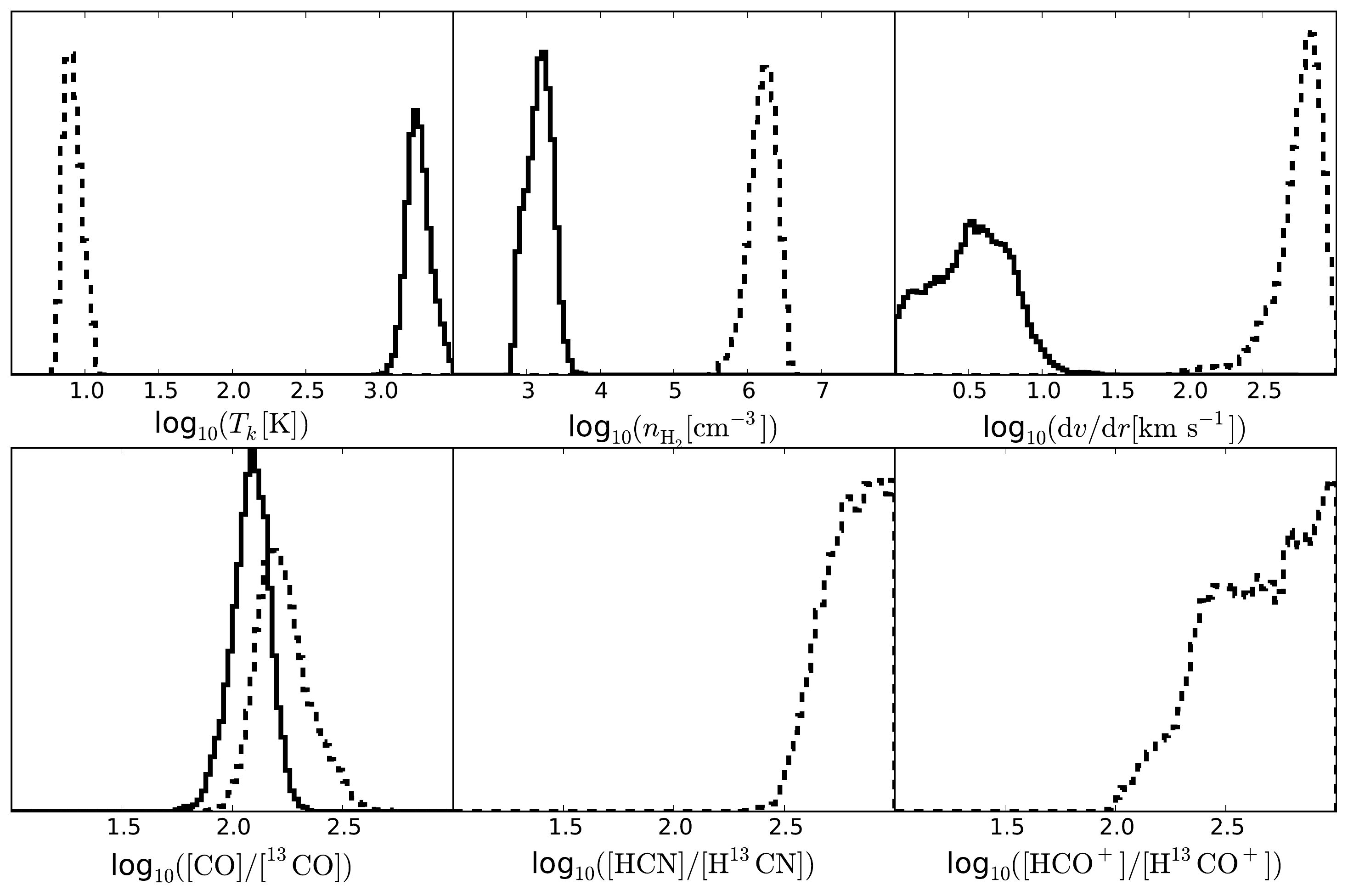}
	\caption{1D marginalised posterior pdfs for the gas phase parameters of the [$X_{^{12}{\rm C}}$]/[$X_{^{13}{\rm C}}$] investigatory model. The gas parameters are very similar to those of the HCN+HCO$^+$+CO 2 phase model. The [CO]/[$^{13}$CO] abundance ratios are similar in both phases, and are in the ``standard'' region for (U)LIRGs, suggesting that the high [HCN]/[H$^{13}$CN] and [HCO$^+$]/[H$^{13}$CO$^+$] ratios are due to isotope fractionation in the cold gas phase.}\label{fig:fixXfree13C}
	\end{figure*}

	The results of the [$^{12}$C]/[$^{13}$C] investigation model are shown in Figure \ref{fig:fixXfree13C} and Table \ref{tab:2phase13CO}.  The model recovers very similar $T_{\rm k}$, $n_{{\rm H}_2}$ and ${\rm d}v/{\rm d}r$ as the previous model. As we have fixed the main isotope abundances this is not surprising, but it is an important confirmation that the isotopologue abundance ratios we derive here are compatible with the less restricted two phase model; by fixing some parameters and freeing others we have not inadvertently moved into a very different locus in parameter space. 
	
	The [HCN]/[H$^{13}$CN] and [HCO$^+$]/[H$^{13}$CO$^+$] ratios are both very high, while the [CO]/[$^{13}$CO] ratio is $\sim 100-200$ in \emph{both} phases. However, the abundance ratios are correlated, so that the spread in ([CO]/[$^{13}$CO]$_2$)/([HCN]/[H$^{13}$CN]) is extremely small, strongly peaking at $2-3$, consistent with \citet{Roueff2015}. These results should be interpreted with caution: the fixing of the molecular abundances could be biasing the results. However, it is expected to bias them towards the values found in the previous model, so [CO]/[$^{13}$CO]$_2$ lying much lower, while [HCN]/[H$^{13}$CN] and [HCO$^+$]/[H$^{13}$CO$^+$] both lie higher, suggests that this is a real effect. This is evidence for a ``standard'' [$^{12}$C]/[$^{13}$C] abundance ratio (for (U)LIRGs) of $\sim 100$ in NGC\,6240. The high ratio found by \citet{Papadopoulos2014} is explained by their attribution of the high$-J$ CO lines to the dense gas phase, whereas our simultaneous fitting and the addition of the H$^{13}$CN and H$^{13}$CO$^+$ lines/limits leads to the model spontaneously fitting these high$-J$ lines to a hot, diffuse phase, as did the two phase CO model of \citet{Kamenetzky2014}. Our high observed [HCN]/[H$^{13}$CN] and [HCO$^+$]/[H$^{13}$CO$^+$] ratios are explained by chemical effects (ICE) in the very cold, dense gas. 
	
	\begin{table*}
	\centering
	\caption{[$^{12}$C]/[$^{13}$C] Study Two Phase Model Solutions. All values $\log_{10}$.}\label{tab:2phase13CO}
	\begin{tabular}{l c c}\hline
	Parameter & Mean & min $\chi^2$\\ \hline\hline
	$T_{\rm k,1}$\dotfill & $3.3^{3.4}_{3.2}$ & 3.3 \\[1mm]
	$T_{\rm k,2}$\dotfill & $0.9^{1.0}_{0.9}$ & 0.8 \\[1mm]
	$n_{{\rm H}_2,1}$\dotfill & $3.2^{3.4}_{3.0}$ & 3.0 \\[1mm]
	$n_{{\rm H}_2,2}$\dotfill & $6.2^{6.4}_{6.1}$ & 6.5 \\[1mm]
	$({\rm d}v/{\rm d}r)_1$\dotfill & $0.5^{0.8}_{0.3}$ & 0.3 \\[1mm]
	$({\rm d}v/{\rm d}r)_2$\dotfill & $2.7^{2.9}_{2.7}$ & 3.0 \\[1mm]
	$[$CO]/[$^{13}$CO]$_1$\dotfill & $2.1^{2.2}_{2.0}$ & 2.2 \\[1mm]
	$[$CO]/[$^{13}$CO]$_2$\dotfill & $2.2^{2.3}_{2.1}$ & 2.3 \\[1mm]
	$[$HCN]/[H$^{13}$CN]\dotfill & $2.8^{3.0}_{2.7}$ & 3.0 \\[1mm]
	$[$HCO$^+$]/[H$^{13}$CO$^+$]\dotfill & $2.6^{3.0}_{2.6}$ & 3.0 \\[1mm]
	$f$\dotfill & $-1.1^{-1.0}_{-1.2}$ & $-1.1$ \\\hline
	\end{tabular}
	\end{table*}

\end{document}